# NEW Fe I LEVEL ENERGIES AND LINE IDENTIFICATIONS FROM STELLAR SPECTRA

Ruth C. Peterson
SETI Institute and Astrophysical Advances, 607 Marion Pl, Palo Alto, CA 94301
e-mail: peterson@ucolick.org

Robert L. Kurucz
Harvard-Smithsonian Center for Astrophysics, 60 Garden Street, Cambridge, MA 02138

*Short title:* New Fe I Identifications from Stellar Spectra

## ABSTRACT

The spectrum of the Fe I atom is critical to many areas of astrophysics and beyond. Measurements of the energies of its high-lying levels remain woefully incomplete, however, despite extensive analysis of ultraviolet laboratory iron absorption spectra, optical laboratory iron emission spectra, and the solar infrared spectrum. In this work we use as sources the high-resolution archival absorption-line ultraviolet and optical spectra of stars, whose warm temperatures favor moderate Fe I excitation. We derive the energy for a particular upper level in Kurucz's semiempirical calculations by adopting a trial value that yields the same wavelength for a given line predicted to be about strong as that of a strong unidentified spectral line observed in the stellar spectra, then checking the new wavelengths of other strong predicted transitions that share the same upper level for coincidence with other strong observed unidentified lines. To date this analysis has provided the upper energies of 66 Fe I levels. Many new level energies are higher than those accessible to laboratory experiments; several exceed the Fe I ionization energy. These levels provide new identifications for over two thousand potentially detectable lines. Almost all of the new levels of odd parity include UV lines that were detected but unclassified in laboratory Fe I absorption spectra, providing an external check on the energy values. We motivate and present the procedure, provide the resulting new level energies and their uncertainties, list all the potentially detectable UV and optical new Fe I line identifications and their gf-values, point out new lines of astrophysical interest, and discuss the prospects for additional Fe I energy-level determinations in the near future.

## 1. INTRODUCTION

Astrophysical research has dramatically gone forward in the last two decades, fueled by rapid advances in key areas. Telescopes and detectors are larger, and make increasingly precise observations of progressively fainter and more distant objects. The analysis of these datasets has surged due to exponentially increasing capabilities of computers and networked systems.

Lagging far behind are the laboratory astrophysics data necessary to interpret this information. These are the fundamental physical parameters that characterize the spectral absorption and emission of the atomic and molecular systems that pervade stars, stellar nebulae, exploding supernovae, and the interstellar and intergalactic medium, from the local environment to the highest redshifts.

Line parameters for the iron atom are a particularly important case. While energy levels, wavelengths, and transition probabilities (gf-values) can often be theoretically derived for light, simple atoms, there is no substitute for determining empirically the energy levels of an atom as complex and as abundant as iron. Because lines of neutral iron dominate the solar absorption spectrum, especially over 1500Å – 2100Å where they outnumber Fe II and Si I lines by a factor of five (Tousey 1988), Fe I has been investigated extensively in the laboratory and from the solar spectrum itself.

Summarizing previous work, Nave & Johansson (1993b) derived energies from laboratory spectra for 86 new Fe I levels, from which Nave & Johansson (1993a) identified and established wavelengths for over 2000 Fe I lines from 1700Å to 5μm. In the UV, Nave & Johansson (1993b) concluded that "many more solar lines in this region are also due to Fe I, and originate from still higher levels that the ones reported here."

A similar situation prevails in the infrared. To reach Fe I levels with energies from 59000 cm$^{-1}$ to 61700 cm$^{-1}$, Johansson et al. (1994) and Schoenfeld et al. (1995) analyzed Fe I supermultiplets in the solar infrared spectrum, identifying nearly 200 lines in three IR windows 35, 6, and 40 cm$^{-1}$ wide. These new IR Fe I lines are weak, usually with depths less than 10% of the solar continuum (Johansson et al. 1994, Fig. 5). A multitude of unidentified IR lines remain, some of which are strong: Table 6 of Hinkle et al. (1995) lists 72 unidentified lines whose depths are 10% or more in the infrared spectrum of the metal-poor K giant Arcturus.

Currently, most of the Fe I lines that remain unidentified fall either in the UV, from 1500Å to 4000Å, or in the infrared, beyond 1μm. Those in the UV are transitions between known low-lying levels to high, still-unmeasured upper levels; those in the infrared are transitions between high levels, one or both of which is unmeasured. Addressing this semiempirically, Kurucz (2011) ran calculations that extrapolate experimentally-determined energy values to unknown energy levels and predict the associated wavelengths. His website currently provides results from such comprehensive calculations for all iron-peak elements. However, because wavelengths are fixed by the difference in energy levels, wavelengths of predicted lines are usually in error, typically by 10Å or more near 2000Å, and by much more in the infrared.

## 2. THE NEED FOR NEW Fe I LINE IDENTIFICATIONS

Several areas of astronomy are severely impacted by unidentified Fe I lines. Among them are the determinations of abundances for trace elements in individual stars from their UV spectra (e.g. Peterson 2011, 2013), with the potential to unravel the nucleosynthesis processes and environments in which the earliest stars and their heavy elements were formed (Sneden et al. 2008). Identified infrared lines and their gf-values are vital (Ruffoni et al. 2013) for infrared spectroscopic iron abundances of luminous red giants in dust-obscured regions like the bulge, bar, and disk of the Milky Way plus the Sagittarius stream (Majewski et al. 2010). All across the spectrum, new line identifications are needed to fill significant poorly-modeled gaps in spectra of stars of solar metallicity and temperature, improving the fidelity of theoretical models of UV (Peterson et al. 2002) and blue (Coelho 2014) spectral energy distributions (SEDs). These in turn are needed to discriminate low metallicity from young age in globular clusters (Dalcanton et al. 2012) and galaxies of moderate redshift (Kelson et al. 2014).

The severity of the problem in the near-UV is seen in Figure 1, adapted from Peterson et al. (2002). This compares observed spectra of *five* near-turnoff stars whose metallicities range from 1/100 solar to slightly supersolar (−2.0 ≤ [Fe/H] ≤ +0.15). Observed stellar near-UV fluxes (heavy lines) are superimposed on theoretical spectra (light lines) that Peterson, Dorman, & Rood (2001) calculated, specifically leaving out the Kurucz predicted lines, whose wavelengths are uncertain. Without them, as temperature drops and metallicity increases, flux is increasingly underestimated in regions not dominated by strong absorption lines. The underestimate reaches a factor of three at solar metallicity in the 2650Å – 2720Å region, as the light grey lines (the calculations) fall increasingly far above the heavy black lines (the observations). This is unfortunate, because the widely-used, high-resolution, 850 – 4700Å UVBLUE theoretical spectral templates (Rodriguez-Merino et al. 2005), incorporated into the Koleva & Vazdekis (2012) assessment of the HST/STIS Next Generation Spectral Library (Gregg et al. 2006), were also calculated without the Kurucz predicted lines. Below 2300Å, matters only become worse (Peterson 2011).

## 3. IDENTIFYING Fe I LINES AND LEVELS FROM STELLAR SPECTRA

To remedy this, in our HST Cycle 21 AR-13263 program we have undertaken the identification of these lines in metal-poor turnoff stars directly from their ultraviolet spectra. We first established that unidentified lines are overwhelmingly due to Fe I. This was seen by running calculations over 1600Å – 8900Å that included only the Kurucz (2011) predicted lines of all neutral species plus Fe II, covering FGK stars in the temperature range 4000K – 6500K. These lines are present in the observed spectra, but have unknown wavelength offsets in the computed spectra.

Peterson then began to empirically establish upper-level energies for Fe I from the same five archival HST E230H echelle spectra of Peterson (2011), by matching modified predictions to the positions of their unidentified absorption lines. The procedures are similar to those followed by Castelli & Kurucz (2010) to identify 109 high levels of Fe II from the optical spectrum of a slowly-rotating B star. The energy difference between the upper and lower levels of a transition fixes its wavelength, and in the UV nearly all lower levels have energies found in the laboratory. A trial energy is chosen by shifting the wavenumber for a specific predicted level to match the predicted and observed wavenumbers of a near-UV line that the Kurucz predictions suggest is as strong as an observed but unidentified spectral line. The wavenumbers of the other strong predicted lines that share the same upper level are checked for coincidence with other strong observed unidentified lines. Matching positions exactly and gf-values approximately for four or more transitions with same upper level confirms its energy. Other lines of the same multiplet will have similar shifts, so their starting guesses become closer.

A critical aspect of this program is the interplay between new identifications and subsequent predictions. Each new level identification and energy value better constrain the Fe I atomic matrix calculations that produce the predicted lines. Once Peterson has found a number of new levels, Kurucz incorporates these into his computation of the Fe I matrix, described just below. The new energy values strongly constrain levels of similar structure whose energies remain unknown, improving their wavelength predictions for Peterson's next Fe I search.

Predicted energy levels and log gf values were computed by Kurucz with his version of the Cowan (1981) code (Kurucz 2009). The calculation included the 61 even configurations $d^6 4s^2$, $d^6 4s 5s$–$10s$, $d^6 4s 4d$–$10d$, $d^6 4s 5g$–$9g$, $d^6 4s 7i$–$9i$, $d^6 4s 9l$, $d^8$, $d^7 4s$–$10s$, $d^7 4d$–$10d$, $d^7 5g$–$9g$, $d^7 7i$–$9i$, $d^9 9l$, $d^5 4s^2$ $5s$–$10s$, $d^5 4s^2 4d$–$10d$, and $d^6 4p^2$ with 18855 levels least-squares fitted to 442 known levels. The 50 odd configurations included $d^6 4s 4p$–$9p$, $d^6 4s 4f$–$9f$, $d^6 4s 6h$–$9h$, $d^6 4s 8k$–$9k$, $d^7 4p$–$10p$, $d^7 4f$–$9f$, $d^7 6h$–$9h$, $d^7 8k$–$9k$, $d^5 4s^2 4p$–$9p$, and $d^5 4s^2 4f$–$9f$ with 18850 levels least-squares fitted to 559 known levels. The calculations were done in LS coupling with all configuration interactions included, with scaled Hartree-Fock starting guesses, and with Hartree-Fock transition integrals. A total of *7512824* lines were saved from the transition array, of which *107602* lines are between known levels and have good wavelengths.

Deriving an energy for a particular Fe I level establishes the identifications and wavelengths for all transitions that share this level and also arise from a known lower level, regardless of its energy and thus its wavelength. Thus a slew of lines from the UV through the IR may be solved in a single level. Because the upper energy remains fixed, the lower energy of this series of transitions increases steadily towards the red. Most of the lower levels of Fe I transitions already have known energies, because these levels are easily populated in laboratory experiments. However, in warm stars at 6000K, the levels of moderate excitation are more easily detected than in the Brown et al. (1988) ~2000C iron furnace. The drawback in stars is that many stellar Fe I lines are blended by lines of other elements.

## 4. STELLAR SPECTRA

Panchromatic stellar spectra are critical to this identification procedure, to increase the number of lines for a given level against which a match can be found, since lines that share the same upper level are spread over wavelengths from 1600Å to 6μm and beyond. Furthermore, as wavelength increases, line profiles become steadily narrower in wavenumber space, reducing the uncertainty in the deduced energy levels. However, at progressively longer wavelengths, Fe I line strengths are diminished by the lower Boltzmann populations as the lower energy level steadily rises towards the red. Consequently, spectra of progressively stronger-lined stars are adopted at redder wavelengths, culminating in stars of solar-metallicity and higher and solar temperatures or lower as wavelengths approached 1μm. In the infrared, Nave & Johansson (1993a) note that "the best source for Fe I is the Sun".

Table 1 summarizes the spectra adopted. Below the stellar names are the model parameters $T_{eff}$, log g, and [Fe/H] used for the calculations. The spectral characteristics include the spectrograph, wavelength coverage, the data reduction procedure or the source from which reduced spectra were downloaded, and the exposure times. The sources include StarCat (Ayres 2010*a*), the UVES and HIRES pipelines, and the UVES ground-based spectral programs of the Next Generation Spectral Library (NGSL; Gregg et al. 2006). All spectra are of intrinsically sharp-lined stars (with rotational velocities $v \sin i < 5$ km s$^{-1}$). Spectral resolution is ~110,000 in the UV and for the solar spectrum, and 100,000 for Arcturus, although turbulence limits the line profiles to an effective resolution of 60,000 to 80,000. Optical spectral resolutions are 50,000 – 60,000. Signal-to-noise ratios S/N exceed 50 for all spectra except below 2000Å. S/N greatly exceeds this for the best-observed optical spectra, those of the Sun and the metal-poor K giant Arcturus. They were obtained with the Fourier Transform Spectrometer (FTS) and coudé feed spectrograph at Kitt Peak; all the others are echelle grating spectra.

In the optical, we have relied very heavily on the Kurucz (2005) solar flux spectrum and the Hinkle et al. (2000) Arcturus atlas. We were able to disentangle an Fe I line coincident with the Li I doublet thanks to the R=140,000, S/N=750 Gemini-S echelle spectrum of HD 217107 (Ghezzi et al. 2009). We also have made use of the R = 80,000, S/N > 200 spectrum of the K5 giant α Tau, obtained for the stellar parameter workshop of Lebzelter et al. (2012). The ~1500K temperature difference between giants and solar-type stars allows us to better discern the plausible values of the lower excitation potential from the difference in the strength of an unknown line in giants versus solar-type stars. For the ground-based near-UV and blue, we have added archival spectra over a wide metallicity range from VLT/UVES (e.g. Bagnulo et al. 2003) and Keck/HIRES with upgraded UV sensitivity (Vogt et al. 1994). The high quality of the latter and our ability to reproduce them synthetically is seen in the figures of Peterson (2013), who derived abundances for light trans-Fe elements from the 3050Å – 4000Å region of the HIRES spectra of metal-poor dwarfs that Boesgaard et al. (2011) obtained for Be II lines near 3130Å.

Conversely, progressively more metal-poor stars were included as wavelength decreases. This is essential to reduce line blending, notably the crowding of the unidentified Fe I lines among themselves. Those from odd-parity levels dominate below 2300Å, and blend so strongly with one another below 1935Å that Brown et al. (1988) could identify only a handful. To minimize this blending, which is significant even at the lowest stellar metallicities available, we added HD 72660, a metallic-lined A star to isolate the very strongest unidentified Fe I lines. Lines from even-parity levels similarly converge near 2500Å, since odd-parity levels from which even-parity levels must arise have their lowest energies near 19500 cm$^{-1}$. This is too high for the furnace of Brown et al., who detected virtually none. In stars of solar temperature, blending is again severe above 1/100 solar metallicity, [Fe/H] = –2.

All but one of the space-based UV spectra were obtained with the E230H echelle grating of the Space Telescope Imaging Spectrograph (STIS). In the mid-UV, for the Hubble Treasury program GO-9455 such spectra were obtained for sharp-lined stars of types sdB, A, F, G, and K at $110,000$ resolution out to 3150Å. Peterson (2008) illustrates how well our calculations match spectra at 3065Å of the metal-poor turnoff stars HD 184499 and HD 157466, with (T$_{eff}$, log g, [Fe/H]) = (5750K, 4.1, –0.5) and (6050K, 4.3, –0.45), as well as the solar-metallicity F5 IV standard Procyon and the Sun itself. The UV spectra near 2000Å were taken primarily for boron abundances (e.g. Thorén & Edvardsson 2000).

Reliable spectral dispersions and stellar radial velocities are critical to the reliability of the values of the derived Fe I energy levels. The latter were established to 0.05 – 0.10 km s$^{-1}$ for each individual spectrum by eye, directly from the match of observed and calculated profiles for previously-identified Fe I lines of similar lower excitation potential to those expected for unidentified Fe I lines in the same wavelength region. In all cases this absorbs whatever velocity shift is contributed to these lines by stellar convective motions. For the Sun, a shift of 0.25 km s$^{-1}$ was applied, to partially restore the removal of its gravitational redshift from the Kurucz (2005) spectrum. For reference, an 0.1 km s$^{-1}$ shift in velocity corresponds to wavenumber shifts of $0.017 cm^{-1}$, $0.008 cm^{-1}$, and $0.004 cm^{-1}$ at air wavelengths of 2000Å, 4000Å, and 8000Å.

The internal reliability of spectral dispersion solutions varies somewhat; bench-mounted FTS and coude feed spectrographs have fewer complications than do echelles on moving platforms and in varying thermal environments (Chaffee & Schroeder 1976). The latter include HIRES, UVES, and STIS. Ayres (2010b) has incorporated improvements in the STIS E230H near-UV echelle dispersion solutions, reducing the three-sigma internal error in a given position from 0.3 km s$^{-1}$ for spectra with pipeline reductions to 0.1 km s$^{-1}$ for StarCat spectra. Values typical of the ground-based echelles *also* fall within this range, judging from the variations in Fe I profile coincidence as well as from overlapping segments of adjacent orders.

## 5. SPECTRAL SYNTHESIS METHODS

Peterson generates stellar spectra using an updated version of the Kurucz (1993) program SYNTHE. Input is a list of molecular and atomic line transitions with wavelengths, energy levels, and laboratory and computed gf-values (revised to match the stellar line strengths), and a static, one-dimensional model stellar photosphere of effective temperature T$_{eff}$, gravity log g, overall metallicity [Fe/H], and microturbulent velocity v$_t$. These parameters are derived exclusively from the spectra, not colors. T$_{eff}$ is derived as noted just below, [Fe/H] from iron lines, and v$_t$ and log g from trends with line strength or ionization, plus the breadth of wings of strong lines. As reported by Peterson, Dorman, & Rood (2001), a consistent determination of the stellar temperature emerges from all available diagnostics. These always include demanding the same abundances be deduced from low- and high-excitation lines of the same species, and that the wings of the profiles of Balmer lines be reproduced. If space-based ultraviolet spectra are on hand, we also match mid-UV flux levels and the continuum slope of the mid-UV spectrum. The Balmer wings agree with other $T_{eff}$ diagnostics only when convective overshoot is turned off. We thus use Castelli & Kurucz (2003) ODFNEW models, which also adopt an improved solar iron abundance and continuum and line opacities.

The input line list was improved by comparing calculations to echelle spectra of a wide variety of standard stars. This includes lines identified in the laboratory, but not lines whose wavelengths or identifications are unknown. Moving from weak-lined to stronger-lined stars, each spectrum is calculated, gf-values are adjusted singly for atomic lines and as a function of band and energy for molecular lines, and both are iterated until all spectra match. These LTE calculations fit both optical and mid-UV spectra of a wide range of metal-poor and solar-metallicity standard stars (Peterson 2005, 2008, 2011, 2013).

Any nonLTE effects are expected to be small for the range in temperature, metallicity, and gravity of the stars considered here (Table 1), according to Lind et al. (2012). Their Fig. 2 shows that the nonLTE effect expected on our Fe I abundances never exceeds 0.1 dex. Similarly, from their Fig. 6, the nonLTE effect on $T_{eff}$ values derived from Fe I excitation, such as ours, is 30K or less. Lind et al. (2012) find good agreement between their results and those of Mashonkina et al. (2011); both show smaller nonLTE effects than do previous studies. Lind et al. attribute this to the recent inclusion of the numerous high-lying Fe I levels of the Kurucz (2007) expanded Fe I calculations, providing "a realistic coupling to the next ionization state."

## 6. ILLUSTRATIONS OF THE LINE IDENTIFICATION PROCEDURE

Figures 2 – 5 illustrate the identification of a single level via four of its least-blended transitions. Progressing from the ultraviolet redward, each figure includes a line whose upper level is 5.0 (4F)6p 3G, at 59357.03 cm$^{-1}$. These lines are at 47380.79 cm$^{-1}$ (2108.89Å; Fig. 2), 33005.99 cm$^{-1}$ (3028.87Å; Fig. 3), 22311.10 cm$^{-1}$ (4480.82Å; Fig. 4), and 11396.01 cm$^{-1}$ (8772.53Å, Fig. 5). Each plot compares observed and calculated spectra for several stars. Their identifications are given at left, along with the effective temperature $T_{eff}$, gravity log g, metallicity [Fe/H], and microturbulent velocity $v_t$ of the atmospheric model, derived as described above. Observed spectra are in blue. Calculations that include the new line identifications are in red; those lacking them, in black.

The ultraviolet region near 2110Å in Figure 2 shows four strong newly-identified lines and illustrates the high quality of five archival STIS E230H spectra. In Figure 3, the region near 3030Å shows another four new identifications, all moderately strong. The high S/N HD 140283 spectrum brings out very weak lines that grow substantially below. Nonetheless, none of the four new identifications is strong enough to be detected in that spectrum, nor are any seen in the lower-resolution STIS E230M spectrum of HD 94028. Two new weak lines are seen in the stronger-lined, ground-based spectra of Figure 4. The star HD 184499 at the bottom of Fig. 3 is now found at the top of Fig. 4, and spectra of the Sun, the super-metal-rich giant μ Leo, and the metal-poor giant Arcturus appear at the bottom. The other two stars are a solar-metallicity dwarf and a metal-rich turnoff star. Figure 5 shows one new detection in the near-IR region near 8770Å. Weak-lined stars are dropped, and the cool giant α Tau is substituted for μ Leo due to the excellent quality of its spectrum (Lebzelter et al. 2012).

In each figure the new identifications stand out: the red line is deeper than the black. The calculated and observed positions of the four 5.0 (4F)6p 3G lines all coincide, confirming our result for its energy value. Each newly-identified line is well reproduced in all stars. Moreover, the strengths of the new lines match regardless of stellar temperature. The two reddest 5.0 (4F)6p 3G lines illustrate this well. The line at 4480.82Å in Fig. 4 is stronger in Arcturus than in the Sun, but the line at 8772.53Å in Fig. 5 is stronger in the Sun than in Arcturus. The former has a value of 37045.93 cm$^{-1}$ for its lower excitation potential, while the latter has a high value of 47960.94 cm$^{-1}$. Taken together, the figures also illustrate the general trend for both known and unidentified lines to become weaker towards the red.

## 7. RESULTS TO DATE

In this way we have now matched four or more transitions in 66 levels with energies up to 67716 cm$^{-1}$. In so doing we have identified more than two thousand individual lines over 1600Å – 5.4μm that are strong enough to be detected in warm and cool stars of moderate to high metallicity, of which more than a third are in the infrared. There are 1414 lines from 2000Å to 9000Å with log gf > –3, and 154 lines blueward to 1680Å with log gf > –4. From 9000Å to 5.4μ are over 700 lines with log gf > -2. The transitions whose energies are most easily established are those with strong lines in well-observed, uncrowded regions (e.g. Figs 2, 4). These form the majority of the levels we have successfully identified to date.

Thanks to the UV spectra, and to the high quality of the solar and Arcturus optical spectra, this approach reaches higher Fe I energies than any previous work, reaching a maximum at 67716 cm$^{-1}$. Five levels have energies higher even than the Fe I ionization potential of 63737.7 cm$^{-1}$ (Schoenfeld et al. 1995). Table 2 lists separately for the newly-identified even and odd levels the full and abbreviated labels and J-value of each new level, and the associated energy and its uncertainty in wavenumbers.

Table 3 provides a wavelength-ordered list of the newly-identified UV and optical lines. Wavelengths are given in vacuum below 2000Å and in air above. For each line sufficiently strong and unblended, we estimate a gf-value good to ±0.2 dex above 2617Å. Blueward, gf-value uncertainties rise to ±0.4 dex, as blends are poorly understood due to the lack of high-resolution spectra for stars with -2 ≤ [Fe/H] ≤ -1 (Table 1). Even larger uncertainties apply in the 2150Å – 2380Å region, where HD 140283 is the only star with high-resolution spectra (Table 1), and below 1950Å, where line blending sharply increases and signal-to-noise drops (Peterson 2017*l*). For every line for which the gf-value can be assessed, Table 3 includes an entry for dgf, the log gf-value from spectral matching minus the predicted log gf-value. Table 3 also provides theoretical line damping constants. Γ_R is the logarithm of the radiative damping constant. Γ_s is the logarithm of the Stark damping constant / electron number density per cm$^3$. Γ_w is the logarithm of the van der Waals damping constant / neutral hydrogen number density per cm$^3$.

We have estimated the uncertainty in each individual energy determination from visual inspection of the goodness of fit of every reasonably-unblended line. Our values range from 0.01 to 0.1 cm$^{-1}$, generally higher than the Nave & Johansson (1993a) uncertainty of 0.01 cm$^{-1}$. Since line profiles are broader at short wavelengths, as noted above, our uncertainties depend strongly on the distribution in wavelength of the subset of lines sufficiently unblended to constrain the energy.

Confirmation of our energy values is available for 41 levels of odd parity. These have two to five strong UV lines whose wavenumbers Brown et al. (1988) determined but could not classify. For many such lines we were able to assign an energy and J value; 31 of our 41 levels have these assignments, and our results always agree. The difference in the average of the 41 level energies between our work and theirs is 0.04 ± 0.03 cm$^{-1}$. The latter is similar to the 0.042 cm$^{-1}$ rms value Schoenfeld et al. (1995) found in identifying infrared Fe I lines from the solar spectrum.

Our work also is uncovering individual line identifications of astrophysical interest. An example is the new Fe I identification at 6707.786Å, a blend with the principal component of the Li I doublet at 6707.761Å, upon which most stellar lithium abundance determinations depend. By analyzing the Ghezzi et al. (2009) high-quality bHROS spectrum of HD 217107, which has a very low lithium abundance, we were able to establish a gf-value for this line. Its size is large enough to mildly affect the lithium abundances of solar-type stars of solar metallicity, and thus its presence is of importance for problems such as detecting lithium trends with stellar metallicity.

## 8. FUTURE PROSPECTS

Due to the very high data quality of the solar spectra, thousands of potentially detectable lines are available for many of the remaining unidentified levels, for levels of both even and odd parity. Over two thousand of these are in the infrared. From existing solar spectra we expect to identify hundreds more Fe I levels, and are beginning with new f, g, h, and i levels. In the UV, should additional spectra become available, hundreds of additional unknown Fe I levels could be found that have few strong lines outside the UV. These include levels with moderately weak lines scattered throughout the UV, and high-lying levels of odd parity with several strong lines below 1930Å but few beyond.

Each set of new identifications is submitted for publication at the same time that Kurucz updates all the Fe I material on his website, including the new energy levels, line identifications, predicted gf-values, and gf-values derived from the stellar spectra. In this way, the entire community engaged in solving the astrophysical problems in Sec. 1 and 2 is able to freely access and use these improvements immediately.

We thank Richard Monier for suggesting HD 72660 as a target and providing the far UV data, V. Smith and L. Ghezzi for providing the high-resolution HD 217107 Gemini-S spectrum, J. X. Prochaska for his reductions of the Keck HIRES data, and D. Silva and R. Hanuschik for the reduced UVES NGSL spectra. T. Ayres provided helpful information regarding STIS echelle dispersion solutions. Support for this work under program number HST-AR-13263 was provided by NASA through a grant from the Space Telescope Science Institute, which is operated by the association of Universities for Research in Astronomy, Inc. under NASA contract NAS 5-26555. Ground-based spectra are largely based on observations made with ESO Telescopes at the Paranal Observatory with the UVES spectrograph under programs 065.L-0507(A), 072.B-0585(A), and 266.D-5655(A), and with the Keck Observatory HIRES spectrograph, under programs H6aH (PI A. Boesgaard), N01H, N12H, and N13H (PI D. Latham), U17H and U63H (PI J. Prochaska), U35H (PI A. Wolfe), and U44H (PI M. Rich), as well as under programs GS-2066A-C-5 and GS-2006B-Q-47 at the Gemini Observatory, which is operated by the Association of Universities for Research in Astronomy, Inc., under cooperative agreement with the NSF on behalf of the Gemini partnership. This research has made use of the Keck Observatory Archive (KOA), which is operated by the W. M. Keck Observatory and the NASA Exoplanet Science Institute (NExScI), under contract with the National Aeronautics and Space Administration. Space-based spectra are based on observations made with the NASA/ESA Hubble Space Telescope under GO programs 7348, 7402, 8197, 9146, 9455, 9491, and 9804. These data were obtained from the HST and StarCat archives hosted by the Mikulski Archive for Space Telescopes (MAST). IRAF is distributed by the National Optical Astronomy Observatories, which are operated by the Association of Universities for Research in Astronomy, Inc., under cooperative agreement with the National Science Foundation.

TABLE 1

Stellar Parameters and Spectra

| Star, Model | Wavelength (Å) | Instrument | Program | Reduction | T (ks) |
|---|---|---|---|---|---|
| Sun<br>5775 4.40 +0.00 | 2960 – 13000 | NSO FTS | | Kurucz (2005) | |
| HD 29139 (α Tau)<br>3950 1.10 +0.00 | 4900 – 9750 | 2m NARVAL<br>at Pic du Midi | U. Heiter | Lebzelter et al. (2012) | |
| HD 72660<br>9525 4.00 +0.35 | 1630 – 1902<br>2129 – 2888<br>3022 – 5845 | STIS 230H<br>STIS 230H<br>HIRES | GO 9146<br>GO 9455<br>U17H | Monier<br>StarCat uvsum 52690<br>Prochaska | 1.65<br>1.64<br>0.15 |
| HD 76932<br>5900 4.10 -1.00 | 1880 – 2150<br>3022 – 4975 | STIS 230H<br>UVES | GO 9804<br>266.D-5655(A) | StarCat 53054-53056<br>Pipeline | 23.86<br>0.34 |
| HD 85503 (μ Leo)<br>4650 2.70 +0.40 | 5582 – 5665<br>5578 – 8560 | HIRES<br>HIRES | U44H<br>U63H | Pipeline + IRAF<br>Pipeline + IRAF | 0.01<br>0.06 |
| HD 94028<br>6050 4.30 -1.40 | 1880 – 2150<br>2278 – 3120<br>3050 – 4989 | STIS 230H<br>STIS 230M<br>UVES | GO 8197<br>GO 7402<br>072.B-0585(A) | IRAF<br>IRAF<br>NGSL | 33.05<br>0.60<br>0.75 |
| HD 124897<br>(Arcturus, α Boo)<br>4275 1.30 -0.55 | 3727 – 9300 | Coudé Feed<br><br>KPNO 0.9m | Hinkle et al.<br><br>Table 3 | Hinkle et al. (2000) | |
| HD 140283<br>5400 3.60 -2.60 | 1950 – 2300<br>2378 – 2891<br>2885 – 3147<br>3080 – 5953 | STIS E230H<br>STIS E230H<br>STIS 230H<br>HIRES | GO 7348<br>GO 9455<br>GO 9491<br>U35H | StarCat uvsum2126<br>IRAF<br>StarCat 52831-52844<br>Prochaska | 18.32<br>5.28<br>62.57<br>0.60 |
| HD 157466<br>6050 4.30 -0.45 | 2378 – 3158<br>3085 – 3996 | STIS 230H<br>HIRES | GO 9455<br>U35H | IRAF<br>Prochaska | 11.11<br>1.26 |
| HD 160617<br>6000 3.80 -1.80 | 1880 – 2150<br>3057 – 3873<br>4400 – 6780 | STIS 230H<br>UVES<br>HIRES | GO 8197<br>65.L-0507(A)<br>H6aH | StarCat 51480-51787<br>Pipeline<br>Extracted | 39.39<br>3.00<br>0.42 |
| HD 165341<br>5300 4.50 +0.00 | 3736 – 10425 | UVES | 71.B-0529(A) | Pipeline | 0.05 |
| HD 184499<br>5750 4.10 -0.50 | 2378 – 3159<br>3847 – 4986 | STIS 230H<br>HIRES | GO 9455<br>N01/N12/N13H | IRAF<br>Pipeline + IRAF | 11.26<br>0.18 |
| HD 211998<br>5300 3.30 -2.60 | 1880 – 2150<br>3040 – 10400 | STIS 230H<br>UVES | GO 9804<br>266.D-5655(A) | IRAF<br>Pipeline | 29.40<br>0.60 |
| HD 217107<br>5600 4.20 +0.30 | 3750 – 10252<br>5730 – 7230 | UVES<br>bHROS,<br>Gemini-S | 076.B-0055(A)<br>2600A-C-5,<br>2006B-Q-47 | Pipeline<br>Ghezzi | 0.33<br>0.60 |

## TABLE 2
### New Fe I Levels and Energies
*See also http://kurucz.harvard.edu/atoms/2600/2600.readme*

| Expanded Label | Label | J | E (cm$^{-1}$) | σ (cm$^{-1}$) |
|---|---|---|---|---|
| *23 Even Levels:* | | | | |
| 3d6 4s(6D)4d e7F | 4s6D4d e7F | 0 | 51143.92 | 0.03 |
| 3d7(4F)4d 5D | (4F)4d 5D | 0 | 54304.21 | 0.02 |
| 3d6 4s(6D)4d 5D | 4s6D4d 5D | 0 | 58428.17 | 0.03 |
| 3d6 4s(4D)4d 5P | 4s4D4d 5P | 1 | 58628.41 | 0.03 |
| 3d7(4P)5s 3P | (4P)5s 3P | 1 | 59300.54 | 0.03 |
| 3d6 4s(4D)4d 3D | 4s4D4d 3D | 2 | 58779.59 | 0.02 |
| 3d7(4F)5d 5F | (4F)5d 5F | 2 | 59366.79 | 0.02 |
| 3d6 4s(4D)4d 3P | 4s4D4d 3P | 2 | 60087.26 | 0.03 |
| 3d7(2F)4s 1F | (2F)4s 1F | 3 | 38602.26 | 0.02 |
| 3d7(4F)5d 5P | (4F)5d 5P | 3 | 58616.11 | 0.02 |
| 3d7(4F)5d 5F | (4F)5d 5F | 3 | 59196.87 | 0.02 |
| 3d6 4s(4D)4d 3G | 4s4D4d 3G | 3 | 59294.38 | 0.02 |
| 3d7(4F)5d 5F5D3G | 5d 5F5D3G | 3 | 59636.36 | 0.02 |
| 3d7(2G)5s 3G | (2G)5s 3G | 3 | 61724.84 | 0.01 |
| 3d6 4s(6D)6d 3+[4+] | s6d 3+[4+] | 4 | 59532.97 | 0.02 |
| 3d7(2G)5s 3G | (2G)5s 3G | 4 | 61340.46 | 0.01 |
| 3d7(2G)5s 1G | (2G)5s 1G | 4 | 61935.47 | 0.01 |
| 3d6 4s(3H)5s 5H | 4s3H5s 5H | 4 | 64531.78 | 0.03 |
| 3d7(2G)5s 3G | (2G)5s 3G | 5 | 61198.49 | 0.01 |
| 3d7(2H)5s 1H | (2H)5s 1H | 5 | 66293.98 | 0.01 |
| 3d7(2G)4d 3I | (2G)4d 3I | 5 | 67687.99 | 0.01 |
| 3d6 4s(3H)5s 5H | 4s3H5s 5H | 6 | 64300.51 | 0.02 |
| 3d7(2G)4d 1I | (2G)4d 1I | 6 | 67716.75 | 0.01 |
| *43 Odd Levels:* | | | | |
| d7(4F)5p 5D | (4F)5p 5D | 0 | 54720.67 | 0.02 |
| d7(2P)4p 1S | (2P)4p 1S | 0 | 55179.91 | 0.08 |
| d6(3P)4s4p(3P) 1P | 3Psp3P 1P | 1 | 50675.08 | 0.05 |
| d7(4F)6p 5D | (4F)6p 5D | 1 | 59703.05 | 0.05 |
| d6(5D)4s(4F)7p 5D | 4s4F7p 5D | 1 | 60169.33 | 0.03 |
| d6(5D)4s(6D)7p 5F | 4s6D7p 5F | 1 | 60336.16 | 0.03 |
| d7(2F)4p 3D | (2F)4p 3D | 1 | 60375.65 | 0.03 |
| d6(3F)4s4p(1P) | 3Fsp1P 3D | 1 | 61075.16 | 0.03 |
| d6(3P)4s4p(1P) | 3Psp1P 3P | 1 | 61155.62 | 0.06 |
| d6(5D)4s(6D)7p 5D | 4s6D7p 5D | 2 | 60237.81 | 0.02 |
| d6(3P)4s4p(1P) 3P | 3Psp1P 3P | 2 | 60585.09 | 0.04 |
| d7(4F)7p 5D | (4F)7p 5D | 2 | 61866.45 | 0.05 |
| d7(4F)6p 5F | (4F)6p 5F | 3 | 59418.83 | 0.03 |
| d7(4F)6p 5G5D3D | 6p 5G5D3D | 3 | 59503.40 | 0.04 |
| d6(3P)4s4p(3P) 1F | 3Dsp3P 1F | 3 | 59794.85 | 0.03 |
| d6(5D)4s(6D)7p 5D | 4s6D7p 5D | 3 | 59875.89 | 0.04 |
| d7(4F)6p 3G | (4F)6p 3G | 3 | 60013.27 | 0.05 |

| Expanded Label | Label | J | E (cm$^{-1}$) | σ (cm$^{-1}$) |
|---|---|---|---|---|
| d6(5D)4s(6D)7p 5F | 4s6D7p 5F | 3 | 60055.93 | 0.05 |
| d7(4F)7p 3D | (4F)7p 3D | 3 | 61351.66 | 0.06 |
| d7(4F)7p 5D | (4F)7p 5D | 3 | 61770.94 | 0.04 |
| d7(4F)7p 3G | (4F)7p 3G | 3 | 62016.99 | 0.10 |
| d7(4F)7p 5G | (4F)7p 5G | 3 | 62287.54 | 0.10 |
| d7(4F)8p 3D3G3F | 8p 3D3G3F | 3 | 62509.75 | 0.04 |
| d7(4F)6p 5D | (4F)6p 5D | 4 | 58729.80 | 0.08 |
| d6(5D)4s(6D)7p 7D | 4s6D7p 7D | 4 | 59317.86 | 0.04 |
| d7(4F)6p 5G | (4F)6p 5G | 4 | 59377.30 | 0.02 |
| d6(5D)4s(6D)7p 5D | 4s6D7p 5D | 4 | 59496.62 | 0.05 |
| d6(5D)4s(6D)7p 7F | 4s6D7p 7F | 4 | 59595.12 | 0.03 |
| d7(4F)6p 3G | (4F)6p 3G | 4 | 59731.29 | 0.05 |
| d6(5D)4s(6D)7p 5F | 4s6D7p 5F | 4 | 59804.54 | 0.02 |
| d7(4F)7p 3F | (4F)7p 3F | 4 | 61113.38 | 0.05 |
| d7(4F)7p 5D | (4F)7p 5D | 4 | 61173.80 | 0.03 |
| d7(4F)7p 3G | (4F)7p 3G | 4 | 61648.30 | 0.08 |
| d7(4F)7p 5F | (4F)7p 5F | 4 | 61678.26 | 0.05 |
| d7(4F)8p 3G5G5F | 8p 3G5G5F | 4 | 62683.77 | 0.05 |
| d7(4F)6p 5F | (4F)6p 5F | 5 | 58609.56 | 0.03 |
| d7(4F)6p 5G | (4F)6p 5G | 5 | 59021.31 | 0.06 |
| d7(4F)6p 3G | (4F)6p 3G | 5 | 59357.03 | 0.02 |
| d7(4F)7p 3G | (4F)7p 3G | 5 | 61140.62 | 0.08 |
| d7(4F)7p 5F | (4F)7p 5F | 5 | 61155.95 | 0.05 |
| d7(4F)7p 5G | (4F)7p 5G | 5 | 61693.44 | 0.04 |
| d6(3G)4s4p(1P) 3G | 3Gsp1P 3G | 5 | 62107.29 | 0.05 |

Note. The three largest eigenvector components for each level can be found in the log files from the least squares fits, b2600e.log and b2600o.log, on the Kurucz website kurucz.harvard.edu/atoms/2600. All of the levels are mixed. The identification is a label, not a definitive assignment, especially for levels that are highly mixed.

# TABLE 3

## Newly Classified Lines of Fe I

| Wavelength (nm) | log gf | dgf | E_even (cm⁻¹) | J_e | Label_e | E_odd (cm⁻¹) | J_o | Label_o | Γ_R | Γ_S | Γ_W |
|---|---|---|---|---|---|---|---|---|---|---|---|
| 160.5458 | -3.197 | … | 0.000 | 4 | 4s2 a5D | 62287.54 | 3 | (4F)7p 5G | 7.76 | -3.71 | -6.99 |
| 160.5966 | -3.591 | … | 415.933 | 3 | 4s2 a5D | 62683.77 | 4 | 8p 3G5G5F | 8.30 | -3.23 | -7.02 |
| 161.0117 | -2.672 | … | 0.000 | 4 | 4s2 a5D | 62107.29 | 5 | 3Gsp1P 3G | 8.32 | -2.99 | -7.03 |
| 161.0466 | -3.777 | … | 415.933 | 3 | 4s2 a5D | 62509.75 | 3 | 8p 3D3G3F | 8.28 | -3.80 | -7.03 |
| 161.2461 | -2.708 | … | 0.000 | 4 | 4s2 a5D | 62016.99 | 3 | (4F)7p 3G | 8.12 | -3.24 | -7.09 |
| 161.6250 | -2.065 | … | 415.933 | 3 | 4s2 a5D | 62287.54 | 3 | (4F)7p 5G | 7.76 | -3.71 | -6.99 |
| 161.7973 | -2.875 | … | 704.007 | 2 | 4s2 a5D | 62509.75 | 3 | 8p 3D3G3F | 8.28 | -3.80 | -7.03 |
| 161.8884 | -2.605 | … | 0.000 | 4 | 4s2 a5D | 61170.94 | 3 | (4F) 7p 5D | 7.39 | -2.63 | -6.96 |
| 162.0918 | -2.814 | … | 0.000 | 4 | 4s2 a5D | 61693.44 | 5 | (4F) 7p 5G | 7.69 | -3.92 | -7.08 |
| 162.1317 | -1.956 | … | 0.000 | 4 | 4s2 a5D | 61678.26 | 4 | (4F)7p 5F | 7.60 | -3.47 | -7.02 |
| 162.2105 | -3.979 | … | 0.000 | 4 | 4s2 a5D | 61648.30 | 4 | (4F)7p 3G | 7.95 | -4.20 | -7.12 |
| 162.3349 | -2.683 | … | 415.933 | 3 | 4s2 a5D | 62016.99 | 3 | (4F)7p 3G | 8.12 | -3.24 | -7.09 |
| 162.3811 | -3.265 | … | 704.007 | 2 | 4s2 a5D | 62287.54 | 3 | (4F)7p 5G | 7.76 | -3.71 | -6.99 |
| 162.7326 | -3.827 | … | 415.933 | 3 | 4s2 a5D | 61866.45 | 2 | (4F)7p 3D | 7.75 | -3.45 | -7.06 |
| 162.9859 | -2.608 | … | 415.933 | 3 | 4s2 a5D | 61170.94 | 3 | (4F)7p 5D | 7.39 | -2.63 | -6.96 |
| 163.2325 | -2.719 | … | 415.933 | 3 | 4s2 a5D | 61678.26 | 4 | (4F)7p 5F | 7.60 | -3.47 | -7.02 |
| 163.4687 | -1.737 | -0.60 | 0.000 | 4 | 4s2 a5D | 61173.80 | 4 | (4F)7p 5D | 7.40 | -3.66 | -7.03 |
| 163.4990 | -3.146 | … | 704.007 | 2 | 4s2 a5D | 61866.45 | 2 | (4F)7p 5D | 7.70 | -3.59 | -7.06 |
| 163.5164 | -1.975 | -0.20 | 0.000 | 4 | 4s2 a5D | 61155.95 | 5 | (4F)7p 5F | 7.10 | -4.19 | -7.06 |

Note. Table 3 is published in its entirety in the electronic edition of The Astrophysical Journal. *The complete listing appears in the addendum below, and the digital version is also available for download at http://kurucz.harvard.edu/atoms/2600/gf2600pk.lab.* A portion is shown here for guidance regarding its form and content.

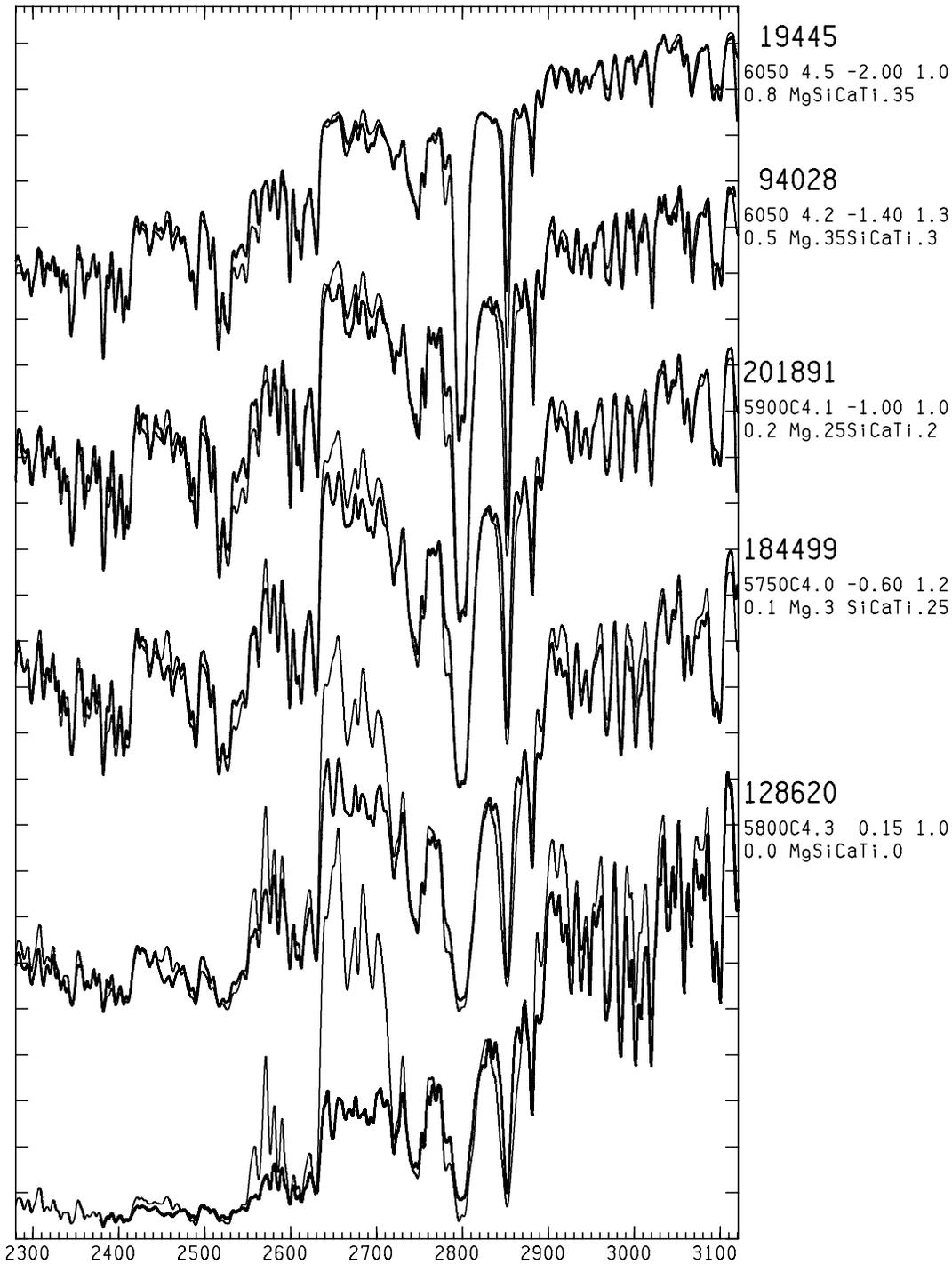

19445
6050 4.5 -2.00 1.0
0.8 MgSiCaTi.35

94028
6050 4.2 -1.40 1.3
0.5 Mg.35SiCaTi.3

201891
5900C4.1 -1.00 1.0
0.2 Mg.25SiCaTi.2

184499
5750C4.0 -0.60 1.2
0.1 Mg.3 SiCaTi.25

128620
5800C4.3  0.15 1.0
0.0 MgSiCaTi.0

Figure 1 – Comparisons are shown of spectral calculations (light lines) to observations (heavy lines) for four metal-poor stars plus α Cen A. Their HD numbers and model parameters appear at the right. Each stellar comparison is vertically offset; Y-axis ticks represent 10% of full scale. Wavelengths in Å appear at the bottom. Adapted from Peterson et al. (2002).

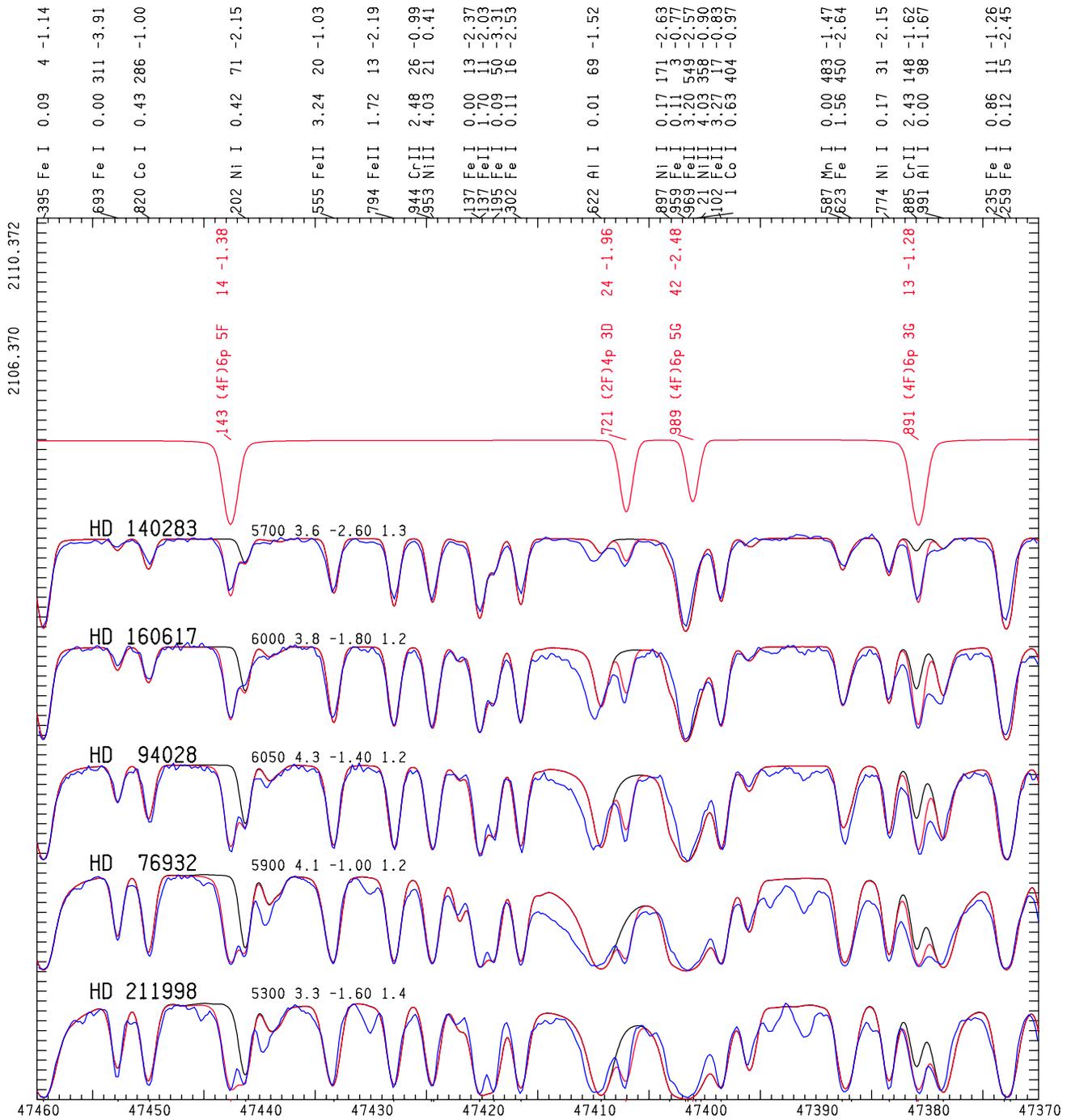

Figure 2 – Comparisons are shown in the 2110Å region between the observed and calculated spectra for five metal-poor stars, vertically offset for clarity. Each tick on the y-axis represents one-tenth of full scale. The stellar identification is on the left with the parameters $T_{eff}$, log g, [Fe/H], and $v_t$, the temperature, gravity, overall metallicity and microturbulent velocity adopted for the model atmosphere. Wavenumbers in $cm^{-1}$ appear at the bottom. The wavelength range in Å of the plot is given at upper left. Strong lines are identified at the top. First are the digits following the decimal place of the line center wavelength in Å (in air > 2000Å). Next are the species giving rise to the line, the lower excitation of the line in eV, an indicator of its strength (stronger lines have smaller numbers), and its log gf-value. Blue lines are observations, which are HST STIS spectra below 3050Å and ground-based in the optical. Black and red lines are spectral calculations. The black line lacks the newly-identified Fe I lines; the red line includes them. Only the new lines were included in the top calculation.

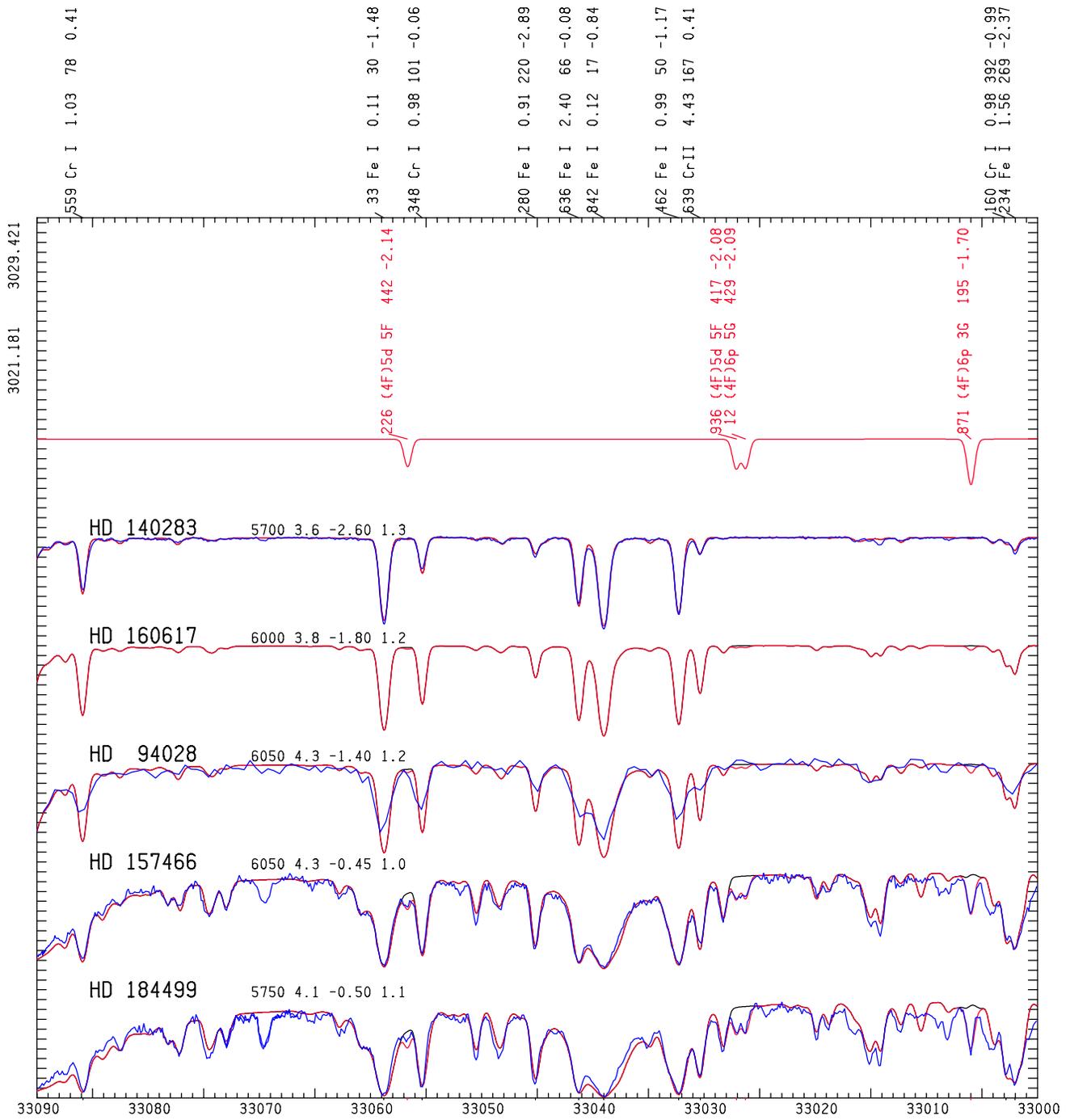

Figure 3 – Comparisons like those in Figure 2 are shown for stars in the 3025Å region.

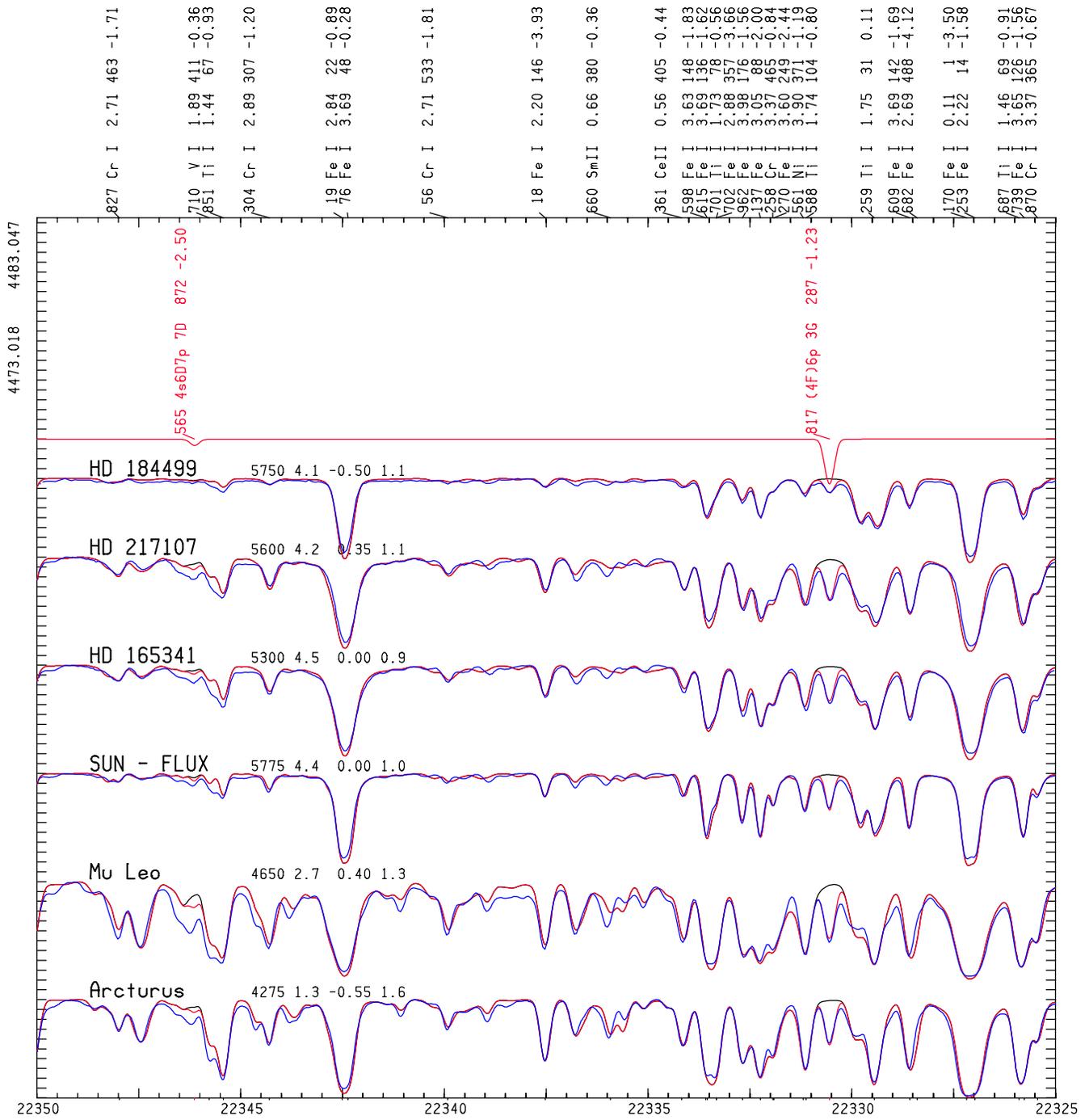

Figure 4 – Comparisons like those in Figure 2 are shown for five stars in the 4480Å region.

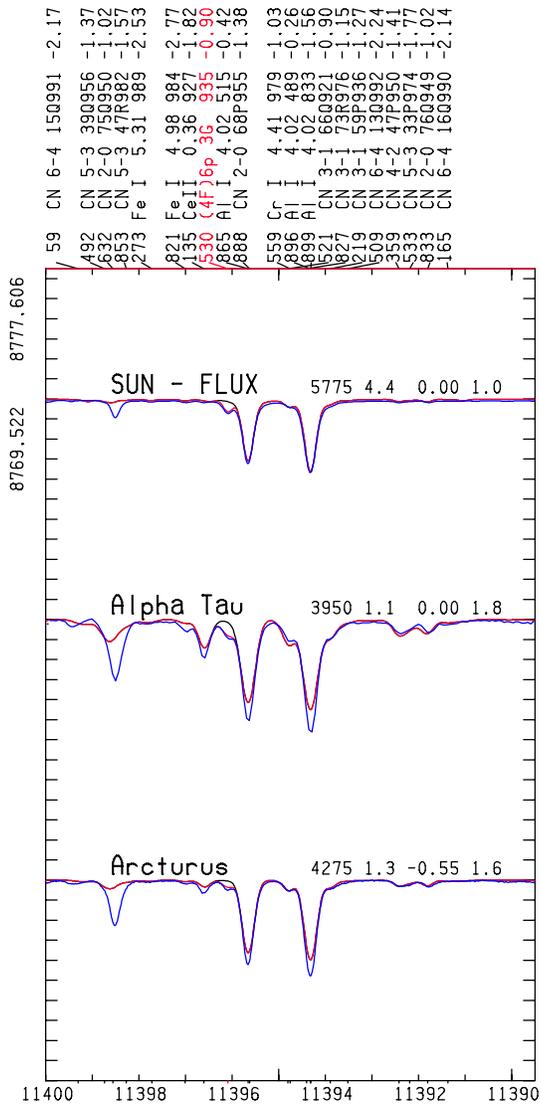

Figure 5 – Comparisons like those in Figure 2 are shown for three stars in the 8770Å region. The single newly-identified Fe I line in this region appears at 8772.53Å.

# Addendum: Full Table 3: Newly Classified Lines of Fe I

## *Format and Content Specification for Table 3*
### *http://kurucz.harvard.edu/atoms/2600/gf2600pk.lab*

```
Title: New Fe I Level Energies and Line identifications from Stellar Spectra
Authors: Peterson, R.C. and Kurucz, R.L.
Table 3: Newly Classified Lines of Fe I
================================================================================
Byte-by-byte Description of file: mrt.tb3
--------------------------------------------------------------------------------
   Bytes Format Units   Label     Explanations
--------------------------------------------------------------------------------
   1- 11 F11.4 nm      WL        Wavelength in air; vacuum below 200 nm.
  17- 18 F7.3  ---     gf        Calculated log gf
  19- 24 F6.2  ---     dgf       gf from spectral match minus predicted gf
  25- 36 F12.3 cm-1    Eeven     Even level energy
  37- 41 F5.1  ---     Jeven     Even level J
  42     1X    ---
  43- 52 A10   ---     Label_e   Even level label
  53- 64 F12.3 cm-1    Eodd      Odd level energy
  65- 69 F5.1  ---     Jodd      Odd level J
  70     1X    ---
  71- 80 A10   ---     Label_o   Odd level label
  81- 86 F6.2  ---     GammaR    Log radiative damping constant
  87- 92 F6.2  ---     GammaS    log Stark damping constant/electron density
cm-3
  93-108 F6.2  ---     GammaW    log van der Waals damping constant/hydrogen
density cm-3
--------------------------------------------------------------------------------
```

## Table 3: Newly Classified Lines of Fe I
*http://kurucz.harvard.edu/atoms/2600/gf2600pk.lab*

| | | | | | | | | | | | | | |
|---|---|---|---|---|---|---|---|---|---|---|---|---|---|
| 160.5458 | -3.197 | ... | 0.000 | 4.0 | 4s2 | a5D | 62287.540 | 3.0 | (4F)7p | 5G | 7.78 | -4.05 | -7.00 |
| 160.5966 | -3.591 | ... | 415.933 | 3.0 | 4s2 | a5D | 62683.770 | 4.0 | 8p | 3G5G5F | 8.29 | -3.24 | -7.02 |
| 161.0117 | -2.672 | ... | 0.000 | 4.0 | 4s2 | a5D | 62107.290 | 5.0 | 3Gsp1P | 3G | 8.33 | -2.99 | -7.04 |
| 161.0466 | -3.777 | ... | 415.933 | 3.0 | 4s2 | a5D | 62509.750 | 3.0 | 8p | 3D3G3F | 8.28 | -3.79 | -7.03 |
| 161.2461 | -2.708 | ... | 0.000 | 4.0 | 4s2 | a5D | 62016.990 | 3.0 | (4F)7p | 3G | 8.12 | -3.24 | -7.09 |
| 161.6250 | -2.065 | ... | 415.933 | 3.0 | 4s2 | a5D | 62287.540 | 3.0 | (4F)7p | 5G | 7.78 | -4.05 | -7.00 |
| 161.7973 | -2.875 | ... | 704.007 | 2.0 | 4s2 | a5D | 62509.750 | 3.0 | 8p | 3D3G3F | 8.28 | -3.79 | -7.03 |
| 161.8884 | -2.605 | ... | 0.000 | 4.0 | 4s2 | a5D | 61770.940 | 3.0 | (4F)7p | 5D | 7.39 | -2.63 | -6.96 |
| 162.0918 | -2.814 | ... | 0.000 | 4.0 | 4s2 | a5D | 61693.440 | 5.0 | (4F)7p | 5G | 7.69 | -3.92 | -7.08 |
| 162.1317 | -1.956 | ... | 0.000 | 4.0 | 4s2 | a5D | 61678.260 | 4.0 | (4F)7p | 5F | 7.62 | -3.54 | -7.02 |
| 162.2105 | -3.979 | ... | 0.000 | 4.0 | 4s2 | a5D | 61648.300 | 4.0 | (4F)7p | 3G | 7.95 | -4.20 | -7.12 |
| 162.3349 | -2.683 | ... | 415.933 | 3.0 | 4s2 | a5D | 62016.990 | 3.0 | (4F)7p | 3G | 8.12 | -3.24 | -7.09 |
| 162.3811 | -3.265 | ... | 704.007 | 2.0 | 4s2 | a5D | 62287.540 | 3.0 | (4F)7p | 5G | 7.78 | -4.05 | -7.00 |
| 162.7326 | -3.827 | ... | 415.933 | 3.0 | 4s2 | a5D | 61866.450 | 2.0 | (4F)7p | 3D | 7.75 | -3.45 | -7.06 |
| 162.9859 | -2.608 | ... | 415.933 | 3.0 | 4s2 | a5D | 61770.940 | 3.0 | (4F)7p | 5D | 7.39 | -2.63 | -6.96 |
| 163.2325 | -2.719 | ... | 415.933 | 3.0 | 4s2 | a5D | 61678.260 | 4.0 | (4F)7p | 5F | 7.62 | -3.54 | -7.02 |
| 163.4687 | -1.737 | -0.60 | 0.000 | 4.0 | 4s2 | a5D | 61173.800 | 4.0 | (4F)7p | 5D | 7.40 | -3.66 | -7.03 |
| 163.4990 | -3.146 | ... | 704.007 | 2.0 | 4s2 | a5D | 61866.450 | 2.0 | (4F)7p | 3D | 7.75 | -3.45 | -7.06 |
| 163.5164 | -1.975 | -0.20 | 0.000 | 4.0 | 4s2 | a5D | 61155.950 | 5.0 | (4F)7p | 5F | 7.07 | -4.17 | -7.06 |
| 163.6303 | -3.397 | ... | 0.000 | 4.0 | 4s2 | a5D | 61113.380 | 4.0 | (4F)7p | 3F | 7.84 | -3.82 | -7.11 |
| 163.7547 | -2.332 | ... | 704.007 | 2.0 | 4s2 | a5D | 61770.940 | 3.0 | (4F)7p | 5D | 7.39 | -2.63 | -6.96 |
| 163.9927 | -2.531 | ... | 888.132 | 1.0 | 4s2 | a5D | 61866.450 | 2.0 | (4F)7p | 3D | 7.75 | -3.45 | -7.06 |
| 164.1073 | -2.752 | ... | 415.933 | 3.0 | 4s2 | a5D | 61351.660 | 3.0 | (4F)7p | 5D | 7.90 | -4.31 | -7.07 |
| 164.5877 | -2.061 | -0.40 | 415.933 | 3.0 | 4s2 | a5D | 61173.800 | 4.0 | (4F)7p | 5D | 7.40 | -3.66 | -7.03 |
| 164.7516 | -3.225 | ... | 415.933 | 3.0 | 4s2 | a5D | 61113.380 | 4.0 | (4F)7p | 3F | 7.84 | -3.82 | -7.11 |
| 164.8868 | -2.441 | ... | 704.007 | 2.0 | 4s2 | a5D | 61351.660 | 3.0 | (4F)7p | 5D | 7.90 | -4.31 | -7.07 |
| 165.4216 | -3.014 | ... | 704.007 | 2.0 | 4s2 | a5D | 61155.620 | 1.0 | 3Psp1P | 3P | 8.52 | -4.00 | -7.19 |
| 165.9269 | -2.933 | ... | 888.132 | 1.0 | 4s2 | a5D | 61155.620 | 1.0 | 3Psp1P | 3P | 8.52 | -4.00 | -7.19 |
| 166.1981 | -3.060 | ... | 415.933 | 3.0 | 4s2 | a5D | 60585.090 | 2.0 | 3Psp1P | 3P | 8.58 | -4.76 | -7.25 |
| 166.5115 | -2.829 | ... | 0.000 | 4.0 | 4s2 | a5D | 60055.930 | 3.0 | 4s6D7p | 5F | 7.21 | -4.07 | -7.07 |
| 166.6298 | -3.900 | ... | 0.000 | 4.0 | 4s2 | a5D | 60013.270 | 3.0 | (4F)6p | 3G | 8.00 | -3.41 | -7.26 |
| 167.0121 | -2.381 | ... | 0.000 | 4.0 | 4s2 | a5D | 59875.890 | 3.0 | 4s6D7p | 5D | 7.67 | -3.29 | -7.13 |
| 167.1629 | -2.294 | ... | 415.933 | 3.0 | 4s2 | a5D | 60237.810 | 2.0 | 4s6D7p | 5D | 7.53 | -3.90 | -7.06 |
| 167.2114 | -3.666 | ... | 0.000 | 4.0 | 4s2 | a5D | 59804.540 | 4.0 | 4s6D7p | 5F | 7.10 | -4.56 | -7.07 |
| 167.2385 | -2.358 | -0.10 | 0.000 | 4.0 | 4s2 | a5D | 59794.850 | 3.0 | 3Dsp3P | 1F | 7.86 | -4.18 | -7.25 |
| 167.4164 | -3.126 | ... | 415.933 | 3.0 | 4s2 | a5D | 59731.290 | 4.0 | (4F)6p | 3G | 8.03 | -3.71 | -7.26 |
| 167.5838 | -3.152 | ... | 704.007 | 2.0 | 4s2 | a5D | 60375.650 | 1.0 | (2F)4p | 3D | 8.52 | -4.33 | -7.37 |
| 167.6727 | -3.306 | ... | 415.933 | 3.0 | 4s2 | a5D | 60055.930 | 3.0 | 4s6D7p | 5F | 7.21 | -4.07 | -7.07 |
| 167.6948 | -2.757 | +0.70 | 704.007 | 2.0 | 4s2 | a5D | 60336.160 | 1.0 | 4s6D7p | 5F | 7.42 | -4.13 | -7.07 |
| 167.7927 | -2.357 | ... | 415.933 | 3.0 | 4s2 | a5D | 60013.270 | 3.0 | (4F)6p | 3G | 8.00 | -3.41 | -7.26 |
| 167.9718 | -2.006 | ... | 704.007 | 2.0 | 4s2 | a5D | 60237.810 | 2.0 | 4s6D7p | 5D | 7.53 | -3.90 | -7.06 |
| 168.0576 | -1.615 | ... | 0.000 | 4.0 | 4s2 | a5D | 59503.400 | 3.0 | 6p | 5G5D3D | 8.15 | -3.43 | -7.17 |
| 168.0768 | -1.614 | +0.10 | 0.000 | 4.0 | 4s2 | a5D | 59496.620 | 4.0 | 4s6D7p | 5D | 7.42 | -3.41 | -7.08 |
| 168.1025 | -2.791 | ... | 888.132 | 1.0 | 4s2 | a5D | 60375.650 | 1.0 | (2F)4p | 3D | 8.52 | -4.33 | -7.37 |
| 168.1652 | -1.700 | ... | 704.007 | 2.0 | 4s2 | a5D | 60169.330 | 1.0 | 4s4F7p | 5D | 7.56 | -3.76 | -7.07 |
| 168.1804 | -1.580 | +0.10 | 415.933 | 3.0 | 4s2 | a5D | 59875.890 | 3.0 | 4s6D7p | 5D | 7.67 | -3.29 | -7.13 |
| 168.2142 | -1.998 | ... | 888.132 | 1.0 | 4s2 | a5D | 60336.160 | 1.0 | 4s6D7p | 5F | 7.42 | -4.13 | -7.07 |
| 168.2968 | -2.900 | ... | 0.000 | 4.0 | 4s2 | a5D | 59418.830 | 3.0 | (4F)6p | 5F | 7.59 | -4.10 | -7.19 |
| 168.3825 | -1.682 | ... | 415.933 | 3.0 | 4s2 | a5D | 59804.540 | 4.0 | 4s6D7p | 5F | 7.10 | -4.56 | -7.07 |
| 168.4099 | -2.105 | ... | 415.933 | 3.0 | 4s2 | a5D | 59794.850 | 3.0 | 3Dsp3P | 1F | 7.86 | -4.18 | -7.25 |
| 168.4145 | -3.658 | ... | 0.000 | 4.0 | 4s2 | a5D | 59377.300 | 4.0 | (4F)6p | 5G | 7.82 | -3.96 | -7.22 |
| 168.4690 | -2.786 | ... | 978.074 | 0.0 | 4s2 | a5D | 60336.160 | 1.0 | 4s6D7p | 5F | 7.42 | -4.13 | -7.07 |
| 168.4720 | -3.803 | ... | 0.000 | 4.0 | 4s2 | a5D | 59357.030 | 5.0 | (4F)6p | 3G | 8.26 | -4.27 | -7.28 |
| 168.4865 | -1.692 | -0.10 | 704.007 | 2.0 | 4s2 | a5D | 60055.930 | 3.0 | 4s6D7p | 5F | 7.21 | -4.07 | -7.07 |
| 168.4929 | -1.702 | +0.00 | 888.132 | 1.0 | 4s2 | a5D | 60237.810 | 2.0 | 4s6D7p | 5D | 7.53 | -3.90 | -7.06 |
| 168.5833 | -2.005 | +0.50 | 0.000 | 4.0 | 4s2 | a5D | 59317.860 | 4.0 | 4s6D7p | 7D | 7.22 | -4.13 | -7.11 |
| 168.9439 | -3.750 | ... | 978.074 | 0.0 | 4s2 | a5D | 60169.330 | 1.0 | 4s4F7p | 5D | 7.56 | -3.76 | -7.07 |
| 168.9783 | -2.184 | +0.50 | 415.933 | 3.0 | 4s2 | a5D | 59595.120 | 4.0 | 4s6D7p | 7F | 6.88 | -4.11 | -7.10 |
| 169.4832 | -3.053 | ... | 415.933 | 3.0 | 4s2 | a5D | 59418.830 | 3.0 | (4F)6p | 5F | 7.59 | -4.10 | -7.19 |
| 169.6026 | -3.756 | ... | 415.933 | 3.0 | 4s2 | a5D | 59377.300 | 4.0 | (4F)6p | 5G | 7.82 | -3.96 | -7.22 |
| 170.0249 | -3.409 | ... | 888.132 | 1.0 | 4s2 | a5D | 59703.050 | 1.0 | (4F)6p | 5D | 7.40 | -3.86 | -7.16 |
| 170.0698 | -3.576 | ... | 704.007 | 2.0 | 4s2 | a5D | 59503.400 | 3.0 | 6p | 5G5D3D | 8.15 | -3.43 | -7.17 |
| 170.2713 | -2.324 | ... | 0.000 | 4.0 | 4s2 | a5D | 58729.800 | 4.0 | (4F)6p | 5D | 7.68 | -4.39 | -7.23 |

| | | | | | | | | | | | |
|---|---|---|---|---|---|---|---|---|---|---|---|
| 170.2853 | -2.595 | ... | 978.074 | 0.0 | 4s2 a5D | 59703.050 | 1.0 | (4F)6p 5D | 7.40 | -3.86 | -7.16 |
| 170.3147 | -3.113 | ... | 704.007 | 2.0 | 4s2 a5D | 59418.830 | 3.0 | (4F)6p 5F | 7.59 | -4.10 | -7.19 |
| 170.6206 | -1.703 | ... | 0.000 | 4.0 | 4s2 a5D | 58609.560 | 5.0 | (4F)6p 5F | 7.19 | -3.24 | -7.23 |
| 171.4858 | -1.377 | ... | 415.933 | 3.0 | 4s2 a5D | 58729.800 | 4.0 | (4F)6p 5D | 7.68 | -4.39 | -7.23 |
| 179.3545 | -3.900 | ... | 6928.268 | 5.0 | (4F)4s a5F | 62683.770 | 4.0 | 8p 3G5G5F | 8.29 | -3.24 | -7.02 |
| 180.8089 | -3.339 | ... | 7376.764 | 4.0 | (4F)4s a5F | 62683.770 | 4.0 | 8p 3G5G5F | 8.29 | -3.24 | -7.02 |
| 181.2283 | -2.880 | ... | 6928.268 | 5.0 | (4F)4s a5F | 62107.290 | 5.0 | 3Gsp1P 3G | 8.33 | -2.99 | -7.04 |
| 181.3796 | -2.933 | ... | 7376.764 | 4.0 | (4F)4s a5F | 62509.750 | 3.0 | 8p 3D3G3F | 8.28 | -3.79 | -7.03 |
| 181.9647 | -3.179 | ... | 7728.060 | 3.0 | (4F)4s a5F | 62683.770 | 4.0 | 8p 3G5G5F | 8.29 | -3.24 | -7.02 |
| 182.1136 | -2.283 | ... | 7376.764 | 4.0 | (4F)4s a5F | 62287.540 | 3.0 | (4F)7p 5G | 7.78 | -4.05 | -7.00 |
| 182.7134 | -3.736 | ... | 7376.764 | 4.0 | (4F)4s a5F | 62107.290 | 5.0 | 3Gsp1P 3G | 8.33 | -2.99 | -7.04 |
| 183.0153 | -3.237 | ... | 7376.764 | 4.0 | (4F)4s a5F | 62016.990 | 3.0 | (4F)7p 3G | 8.12 | -3.24 | -7.09 |
| 183.2862 | -2.653 | ... | 7728.060 | 3.0 | (4F)4s a5F | 62287.540 | 3.0 | (4F)7p 5G | 7.78 | -4.05 | -7.00 |
| 183.4056 | -3.346 | ... | 7985.785 | 2.0 | (4F)4s a5F | 62509.750 | 3.0 | 8p 3D3G3F | 8.28 | -3.79 | -7.03 |
| 183.8432 | -1.626 | +0.00 | 7376.764 | 4.0 | (4F)4s a5F | 61770.940 | 3.0 | (4F)7p 5D | 7.39 | -2.63 | -6.96 |
| 184.1055 | -1.845 | ... | 7376.764 | 4.0 | (4F)4s a5F | 61693.440 | 5.0 | (4F)7p 5G | 7.69 | -3.92 | -7.08 |
| 184.1561 | -1.992 | ... | 7985.785 | 2.0 | (4F)4s a5F | 62287.540 | 3.0 | (4F)7p 5G | 7.78 | -4.05 | -7.00 |
| 184.1570 | -1.783 | ... | 7376.764 | 4.0 | (4F)4s a5F | 61678.260 | 4.0 | (4F)7p 5F | 7.62 | -3.54 | -7.02 |
| 184.1996 | -2.789 | ... | 7728.060 | 3.0 | (4F)4s a5F | 62016.990 | 3.0 | (4F)7p 3G | 8.12 | -3.24 | -7.09 |
| 184.3470 | -1.271 | ... | 6928.268 | 5.0 | (4F)4s a5F | 61173.800 | 4.0 | (4F)7p 5D | 7.40 | -3.66 | -7.03 |
| 184.4077 | -1.632 | +0.40 | 6928.268 | 5.0 | (4F)4s a5F | 61155.950 | 5.0 | (4F)7p 5D | 7.07 | -4.17 | -7.06 |
| 184.4598 | -3.292 | ... | 6928.268 | 5.0 | (4F)4s a5F | 61140.620 | 5.0 | (4F)7p 3G | 7.98 | -3.98 | -7.12 |
| 184.7118 | -1.997 | ... | 7728.060 | 3.0 | (4F)4s a5F | 61866.450 | 2.0 | (4F)7p 3D | 7.75 | -3.45 | -7.06 |
| 185.0383 | -2.497 | ... | 7728.060 | 3.0 | (4F)4s a5F | 61770.940 | 3.0 | (4F)7p 5D | 7.39 | -2.63 | -6.96 |
| 185.0782 | -2.780 | ... | 7985.785 | 2.0 | (4F)4s a5F | 62016.990 | 3.0 | (4F)7p 3G | 8.12 | -3.24 | -7.09 |
| 185.2713 | -2.127 | ... | 7376.764 | 4.0 | (4F)4s a5F | 61351.660 | 3.0 | (4F)7p 5F | 7.90 | -4.31 | -7.07 |
| 185.3561 | -3.745 | ... | 7728.060 | 3.0 | (4F)4s a5F | 61678.260 | 4.0 | (4F)7p 5F | 7.62 | -3.54 | -7.02 |
| 185.4591 | -2.696 | ... | 7728.060 | 3.0 | (4F)4s a5F | 61648.300 | 4.0 | (4F)7p 3G | 7.95 | -4.20 | -7.12 |
| 185.5953 | -2.824 | ... | 7985.785 | 2.0 | (4F)4s a5F | 61866.450 | 2.0 | (4F)7p 3D | 7.75 | -3.45 | -7.06 |
| 185.7613 | -1.762 | +0.00 | 888.132 | 1.0 | 4s2 a5D | 54720.670 | 0.0 | (4F)5p 5D | 8.08 | -4.64 | -7.43 |
| 185.8838 | -2.385 | ... | 7376.764 | 4.0 | (4F)4s a5F | 61173.800 | 4.0 | (4F)7p 5D | 7.40 | -3.66 | -7.03 |
| 185.9249 | -3.279 | ... | 7985.785 | 2.0 | (4F)4s a5F | 61770.940 | 3.0 | (4F)7p 5D | 7.39 | -2.63 | -6.96 |
| 185.9455 | -3.471 | ... | 7376.764 | 4.0 | (4F)4s a5F | 61155.950 | 5.0 | (4F)7p 5D | 7.07 | -4.17 | -7.06 |
| 185.9986 | -2.800 | ... | 7376.764 | 4.0 | (4F)4s a5F | 61140.620 | 5.0 | (4F)7p 3G | 7.98 | -3.98 | -7.12 |
| 186.0928 | -2.771 | ... | 7376.764 | 4.0 | (4F)4s a5F | 61113.380 | 4.0 | (4F)7p 3F | 7.84 | -3.82 | -7.11 |
| 186.1791 | -3.685 | ... | 8154.714 | 1.0 | (4F)4s a5F | 61866.450 | 2.0 | (4F)7p 3D | 7.75 | -3.45 | -7.06 |
| 186.4851 | -3.426 | ... | 7728.060 | 3.0 | (4F)4s a5F | 61351.660 | 3.0 | (4F)7p 5F | 7.90 | -4.31 | -7.07 |
| 187.1057 | -3.455 | ... | 7728.060 | 3.0 | (4F)4s a5F | 61173.800 | 4.0 | (4F)7p 5D | 7.40 | -3.66 | -7.03 |
| 188.3616 | -3.764 | ... | 7985.785 | 2.0 | (4F)4s a5F | 61075.160 | 1.0 | 3Fsp1P 3D | 8.23 | -2.91 | -7.06 |
| 188.6760 | -3.561 | ... | 8154.714 | 1.0 | (4F)4s a5F | 61155.620 | 1.0 | 3Psp1P 3P | 8.52 | -4.00 | -7.19 |
| 189.1207 | -2.430 | +0.00 | 6928.268 | 5.0 | (4F)4s a5F | 59804.540 | 4.0 | 4s6D7p 5F | 7.10 | -4.56 | -7.07 |
| 189.3831 | -3.255 | ... | 6928.268 | 5.0 | (4F)4s a5F | 59731.290 | 4.0 | (4F)6p 3G | 8.03 | -3.71 | -7.26 |
| 189.8284 | -2.110 | ... | 7376.764 | 4.0 | (4F)4s a5F | 60055.930 | 3.0 | 4s6D7p 5F | 7.21 | -4.07 | -7.07 |
| 189.8727 | -2.914 | +0.50 | 6928.268 | 5.0 | (4F)4s a5F | 59595.120 | 4.0 | 4s6D7p 7F | 6.88 | -4.11 | -7.10 |
| 189.9822 | -2.912 | ... | 7376.764 | 4.0 | (4F)4s a5F | 60013.270 | 3.0 | (4F)6p 3G | 8.00 | -3.41 | -7.26 |
| 190.2285 | -1.790 | -0.20 | 6928.268 | 5.0 | (4F)4s a5F | 59496.620 | 4.0 | 4s6D7p 5D | 7.42 | -3.41 | -7.08 |
| 190.4408 | -2.088 | -0.10 | 7728.060 | 3.0 | (4F)4s a5F | 60237.810 | 2.0 | 4s6D7p 5D | 7.53 | -3.90 | -7.06 |
| 190.4794 | -2.401 | ... | 7376.764 | 4.0 | (4F)4s a5F | 59875.890 | 3.0 | 4s6D7p 5D | 7.67 | -3.29 | -7.13 |
| 190.7386 | -2.645 | ... | 7376.764 | 4.0 | (4F)4s a5F | 59804.540 | 4.0 | 4s6D7p 5F | 7.10 | -4.56 | -7.07 |
| 190.7738 | -3.260 | ... | 7376.764 | 4.0 | (4F)4s a5F | 59794.850 | 3.0 | 3Dsp3P 1F | 7.86 | -4.18 | -7.25 |
| 190.8766 | -3.854 | ... | 7985.785 | 2.0 | (4F)4s a5F | 60375.650 | 1.0 | (2F)4p 3D | 8.52 | -4.33 | -7.37 |
| 190.8776 | -2.271 | +0.30 | 6928.268 | 5.0 | (4F)4s a5F | 59317.860 | 4.0 | 4s6D7p 7D | 7.22 | -4.13 | -7.11 |
| 191.0206 | -2.529 | ... | 7985.785 | 2.0 | (4F)4s a5F | 60336.160 | 1.0 | 4s6D7p 5F | 7.42 | -4.13 | -7.07 |
| 191.1028 | -3.972 | ... | 7728.060 | 3.0 | (4F)4s a5F | 60055.930 | 3.0 | 4s6D7p 5F | 7.21 | -4.07 | -7.07 |
| 191.2587 | -3.008 | -0.80 | 7728.060 | 3.0 | (4F)4s a5F | 60013.270 | 3.0 | (4F)6p 3G | 8.00 | -3.41 | -7.26 |
| 191.3801 | -3.041 | ... | 7985.785 | 2.0 | (4F)4s a5F | 60237.810 | 2.0 | 4s6D7p 5D | 7.53 | -3.90 | -7.06 |
| 191.5035 | -2.984 | +0.30 | 7376.764 | 4.0 | (4F)4s a5F | 59595.120 | 4.0 | 4s6D7p 7F | 6.88 | -4.11 | -7.10 |
| 191.6313 | -2.185 | ... | 7985.785 | 2.0 | (4F)4s a5F | 60169.330 | 1.0 | 4s4F7p 5D | 7.56 | -3.76 | -7.07 |
| 191.6390 | -3.072 | ... | 8154.714 | 1.0 | (4F)4s a5F | 60336.160 | 1.0 | 4s6D7p 5F | 7.42 | -4.13 | -7.07 |
| 191.7625 | -2.066 | ... | 7728.060 | 3.0 | (4F)4s a5F | 59875.890 | 3.0 | 4s6D7p 5D | 7.67 | -3.29 | -7.13 |
| 191.8405 | -2.706 | -0.30 | 7376.764 | 4.0 | (4F)4s a5F | 59503.400 | 3.0 | 6p 5G5D3D | 8.15 | -3.43 | -7.17 |
| 191.8655 | -1.971 | ... | 7376.764 | 4.0 | (4F)4s a5F | 59496.620 | 4.0 | 4s6D7p 5D | 7.42 | -3.41 | -7.08 |
| 192.0253 | -2.619 | -0.10 | 7728.060 | 3.0 | (4F)4s a5F | 59804.540 | 4.0 | 4s6D7p 5F | 7.10 | -4.56 | -7.07 |
| 192.0486 | -3.131 | +0.10 | 7985.785 | 2.0 | (4F)4s a5F | 60055.930 | 3.0 | 4s6D7p 5F | 7.21 | -4.07 | -7.07 |
| 192.0610 | -2.760 | ... | 7728.060 | 3.0 | (4F)4s a5F | 59794.850 | 3.0 | 3Dsp3P 1F | 7.86 | -4.18 | -7.25 |
| 192.1522 | -2.872 | ... | 7376.764 | 4.0 | (4F)4s a5F | 59418.830 | 3.0 | (4F)6p 5F | 7.59 | -4.10 | -7.19 |
| 192.2061 | -2.285 | ... | 7985.785 | 2.0 | (4F)4s a5F | 60013.270 | 3.0 | (4F)6p 3G | 8.00 | -3.41 | -7.26 |
| 192.2537 | -1.975 | ... | 8154.714 | 1.0 | (4F)4s a5F | 60169.330 | 1.0 | 4s4F7p 5D | 7.56 | -3.76 | -7.07 |

```
192.2957 -2.131 -0.30    7728.060  3.0 (4F)4s a5F   59731.290  4.0 (4F)6p 3G    8.03 -3.71 -7.26
192.3807 -2.325 -0.30    7376.764  4.0 (4F)4s a5F   59357.030  5.0 (4F)6p 3G    8.26 -4.27 -7.28
192.5258 -2.001 +0.10    7376.764  4.0 (4F)4s a5F   59317.860  4.0 4s6D7p 7D    7.22 -4.13 -7.11
192.7150 -2.648 -0.30    7985.785  2.0 (4F)4s a5F   59875.890  3.0 4s6D7p 5D    7.67 -3.29 -7.13
192.8006 -3.498  ...     7728.060  3.0 (4F)4s a5F   59595.120  4.0 4s6D7p 7F    6.88 -4.11 -7.10
193.0445 -1.104  ...     6928.268  5.0 (4F)4s a5F   58729.800  4.0 (4F)6p 5D    7.68 -4.39 -7.23
193.1421 -2.314  ...     7728.060  3.0 (4F)4s a5F   59503.400  3.0 6p 5G5D3D    8.15 -3.43 -7.17
193.1674 -2.505  ...     7728.060  3.0 (4F)4s a5F   59496.620  4.0 4s6D7p 5D    7.42 -3.41 -7.08
193.3590 -2.041 -0.40    7985.785  2.0 (4F)4s a5F   59703.050  1.0 (4F)6p 5D    7.40 -3.86 -7.16
193.4581 -1.996 -0.40    7728.060  3.0 (4F)4s a5F   59418.830  3.0 (4F)6p 5F    7.59 -4.10 -7.19
193.4936 -1.333 -0.50    6928.268  5.0 (4F)4s a5F   58609.560  5.0 (4F)6p 5F    7.19 -3.24 -7.23
193.6137 -1.768  ...     7728.060  3.0 (4F)4s a5F   59377.300  4.0 (4F)6p 5G    7.82 -3.96 -7.22
193.6313 -1.564  ...     7376.764  4.0 (4F)4s a5F   59021.310  5.0 (4F)6p 5G    7.98 -4.37 -7.26
193.8368 -2.890  ...     7728.060  3.0 (4F)4s a5F   59317.860  4.0 4s6D7p 7D    7.22 -4.13 -7.11
193.9927 -3.420  ...     8154.714  1.0 (4F)4s a5F   59703.050  1.0 (4F)6p 5D    7.40 -3.86 -7.16
194.1084 -2.993 -0.40    7985.785  2.0 (4F)4s a5F   59503.400  3.0 6p 5G5D3D    8.15 -3.43 -7.17
194.7305 -3.413 +0.00    7376.764  4.0 (4F)4s a5F   58729.800  4.0 (4F)6p 5D    7.68 -4.39 -7.23
195.1875 -3.548 -0.20    7376.764  4.0 (4F)4s a5F   58609.560  5.0 (4F)6p 5F    7.19 -3.24 -7.23
196.0717 -3.330 -0.60    7728.060  3.0 (4F)4s a5F   58729.800  4.0 (4F)6p 5D    7.68 -4.39 -7.23
197.2094 -0.726 -0.50   11976.239  4.0 (4F)4s a3F   62683.770  4.0 8p 3G5G5F    8.29 -3.24 -7.02
197.8885 -1.625 -0.60   11976.239  4.0 (4F)4s a3F   62509.750  3.0 8p 3D3G3F    8.28 -3.79 -7.03
199.4772 -0.682 -0.40   11976.239  4.0 (4F)4s a3F   62107.290  5.0 3Gsp1P 3G    8.33 -2.99 -7.04
199.5099 -1.070 -0.80   12560.934  3.0 (4F)4s a3F   62683.770  4.0 8p 3G5G5F    8.29 -3.24 -7.02
200.1401 -0.744 -0.20   12560.934  3.0 (4F)4s a3F   62509.750  3.0 8p 3D3G3F    8.28 -3.79 -7.03
200.7596 -2.151 -0.10   11976.239  4.0 (4F)4s a3F   61770.940  3.0 (4F)7p 5D    7.39 -2.63 -6.96
201.0726 -1.518 -0.20   11976.239  4.0 (4F)4s a3F   61693.440  5.0 (4F)7p 5G    7.69 -3.92 -7.08
201.1341 -1.421 -0.30   11976.239  4.0 (4F)4s a3F   61678.260  4.0 (4F)7p 5F    7.62 -3.54 -7.02
202.1345 -2.714  ...    12560.934  3.0 (4F)4s a3F   62016.990  3.0 (4F)7p 3G    8.12 -3.24 -7.09
202.4647 -1.312 -0.60   11976.239  4.0 (4F)4s a3F   61351.660  3.0 (4F)7p 3D    7.90 -4.31 -7.07
202.6964 -1.509 -0.40   12968.554  2.0 (4F)4s a3F   62287.540  3.0 (4F)7p 5G    7.78 -4.05 -7.00
202.7518 -2.486 -0.20   12560.934  3.0 (4F)4s a3F   61866.450  2.0 (4F)7p 3D    7.75 -3.45 -7.06
203.3339 -2.103  ...    11976.239  4.0 (4F)4s a3F   61140.620  5.0 (4F)7p 3G    7.98 -3.98 -7.12
203.4466 -1.357 -0.20   11976.239  4.0 (4F)4s a3F   61113.380  4.0 (4F)7p 3F    7.84 -3.82 -7.11
203.6529 -1.366 -0.50   12560.934  3.0 (4F)4s a3F   61648.300  4.0 (4F)7p 3G    7.95 -4.20 -7.12
203.8146 -1.132 -0.70   12968.554  2.0 (4F)4s a3F   62016.990  3.0 (4F)7p 3G    8.12 -3.24 -7.09
204.4422 -2.495 +0.00   12968.554  2.0 (4F)4s a3F   61866.450  2.0 (4F)7p 3D    7.75 -3.45 -7.06
204.8913 -2.281 -1.20   12560.934  3.0 (4F)4s a3F   61351.660  3.0 (4F)7p 3D    7.90 -4.31 -7.07
205.8970 -2.351 -0.10   12560.934  3.0 (4F)4s a3F   61113.380  4.0 (4F)7p 3F    7.84 -3.82 -7.11
207.4584 -2.017  ...    12968.554  2.0 (4F)4s a3F   61155.620  1.0 3Psp1P 3P    8.52 -4.00 -7.19
207.8054 -1.465 -0.70   12968.554  2.0 (4F)4s a3F   61075.160  1.0 3Fsp1P 3D    8.23 -2.91 -7.06
208.1064 -2.171 -0.50   11976.239  4.0 (4F)4s a3F   60013.270  3.0 (4F)6p 3G    8.00 -3.41 -7.26
208.1622 -2.769 +0.70   12560.934  3.0 (4F)4s a3F   60585.090  2.0 3Psp1P 3P    8.58 -4.76 -7.25
208.7034 -2.093 -0.60   11976.239  4.0 (4F)4s a3F   59875.890  3.0 4s6D7p 5D    7.67 -3.29 -7.13
209.0571 -1.698 -0.80   11976.239  4.0 (4F)4s a3F   59794.850  3.0 3Dsp3P 1F    7.86 -4.18 -7.25
209.3354 -2.966 +0.20   11976.239  4.0 (4F)4s a3F   59731.290  4.0 (4F)6p 3G    8.03 -3.71 -7.26
209.9444 -2.510 +0.20   12968.554  2.0 (4F)4s a3F   60585.090  2.0 3Psp1P 3P    8.58 -4.76 -7.25
210.3393 -1.113 -0.50   11976.239  4.0 (4F)4s a3F   59503.400  3.0 6p 5G5D3D    8.15 -3.43 -7.17
210.7143 -2.082 +0.70   11976.239  4.0 (4F)4s a3F   59418.830  3.0 (4F)6p 5F    7.59 -4.10 -7.19
210.8721 -1.564 -0.40   12968.554  2.0 (4F)4s a3F   60375.650  1.0 (2F)4p 3D    8.52 -4.33 -7.37
210.8989 -2.483  ...    11976.239  4.0 (4F)4s a3F   59377.300  4.0 (4F)6p 5G    7.82 -3.96 -7.22
210.9891 -0.776 -0.50   11976.239  4.0 (4F)4s a3F   59357.030  5.0 (4F)6p 3G    8.26 -4.27 -7.28
211.0479 -2.832 +0.70   12968.554  2.0 (4F)4s a3F   60336.160  1.0 4s6D7p 5F    7.42 -4.13 -7.07
211.1637 -2.820 +0.10   11976.239  4.0 (4F)4s a3F   59317.860  4.0 4s6D7p 7D    7.22 -4.13 -7.11
211.6453 -2.249 -0.50   12560.934  3.0 (4F)4s a3F   59794.850  3.0 3Dsp3P 1F    7.86 -4.18 -7.25
211.9305 -1.431 -0.40   12560.934  3.0 (4F)4s a3F   59731.290  4.0 (4F)6p 3G    8.03 -3.71 -7.26
212.4950 -1.032  ...    11976.239  4.0 (4F)4s a3F   59021.310  5.0 (4F)6p 5G    7.98 -4.37 -7.26
212.4966 -1.943 -0.40   12968.554  2.0 (4F)4s a3F   60013.270  3.0 (4F)6p 3G    8.00 -3.41 -7.26
212.5442 -2.898 +0.90   12560.934  3.0 (4F)4s a3F   59595.120  4.0 4s6D7p 7F    6.88 -4.11 -7.10
212.9595 -2.136 -0.50   12560.934  3.0 (4F)4s a3F   59503.400  3.0 6p 5G5D3D    8.15 -3.43 -7.17
212.9903 -2.327 -0.50   12560.934  3.0 (4F)4s a3F   59496.620  4.0 4s6D7p 5D    7.42 -3.41 -7.08
213.1190 -2.991 -0.10   12968.554  2.0 (4F)4s a3F   59875.890  3.0 4s6D7p 5D    7.67 -3.29 -7.13
213.3439 -2.059  ...    12560.934  3.0 (4F)4s a3F   59418.830  3.0 (4F)6p 5F    7.59 -4.10 -7.19
213.5332 -1.407  ...    12560.934  3.0 (4F)4s a3F   59377.300  4.0 (4F)6p 5G    7.82 -3.96 -7.22
213.8046 -2.329  ...    12560.934  3.0 (4F)4s a3F   59317.860  4.0 4s6D7p 7D    7.22 -4.13 -7.11
213.9073 -2.709 +0.20   12968.554  2.0 (4F)4s a3F   59703.050  1.0 (4F)6p 5D    7.40 -3.86 -7.16
214.6816 -1.309 -0.60    8154.714  1.0 (4F)4s a5F   54720.670  0.0 (4F)5p 5D    8.08 -4.64 -7.43
216.5282 -2.004 -0.50   12560.934  3.0 (4F)4s a3F   58729.800  4.0 (4F)6p 5D    7.68 -4.39 -7.23
223.4574 -2.192  ...    17550.181  3.0 (4P)4s a5P   62287.540  3.0 (4F)7p 5G    7.78 -4.05 -7.00
```

```
224.8171 -2.588  ...  17550.181  3.0 (4P)4s a5P  62016.990  3.0 (4F)7p 3G   8.12 -3.24 -7.09
225.5809 -2.638  ...  17550.181  3.0 (4P)4s a5P  61866.450  2.0 (4F)7p 3D   7.75 -3.45 -7.06
225.7147 -2.474  ...  17726.988  2.0 (4P)4s a5P  62016.990  3.0 (4F)7p 3G   8.12 -3.24 -7.09
226.0682 -2.607  ...  17550.181  3.0 (4P)4s a5P  61770.940  3.0 (4F)7p 5D   7.39 -2.63 -6.96
226.4846 -2.436  ...  17726.988  2.0 (4P)4s a5P  61866.450  2.0 (4F)7p 3D   7.75 -3.45 -7.06
226.5251 -1.949 +0.00 18378.186  2.0 4s2 a3P    62509.750  3.0 8p 3D3G3F   8.28 -3.79 -7.03
226.9758 -2.225  ...  17726.988  2.0 (4P)4s a5P  61770.940  3.0 (4F)7p 5D   7.39 -2.63 -6.96
227.5176 -2.490  ...  17927.382  1.0 (4P)4s a5P  61866.450  2.0 (4F)7p 3D   7.75 -3.45 -7.06
229.1574 -2.500  ...  17726.988  2.0 (4P)4s a5P  61351.660  3.0 (4F)7p 3D   7.90 -4.31 -7.07
229.1630 -1.847  ...  17550.181  3.0 (4P)4s a5P  61173.800  4.0 (4F)7p 5D   7.40 -3.66 -7.03
229.8763 -2.522  ...  18378.186  2.0 4s2 a3P    61866.450  2.0 (4F)7p 3D   7.75 -3.45 -7.06
230.1920 -2.889  ...  17726.988  2.0 (4P)4s a5P  61155.620  1.0 3Psp1P 3P   8.52 -4.00 -7.19
230.3823 -2.485  ...  18378.186  2.0 4s2 a3P    61770.940  3.0 (4F)7p 5D   7.39 -2.63 -6.96
232.1479 -1.835  ...  19621.006  5.0 4s2 a3H    62683.770  4.0 8p 3G5G5F   8.29 -3.24 -7.02
232.6303 -1.304  ...  18378.186  2.0 4s2 a3P    61351.660  3.0 (4F)7p 3D   7.90 -4.31 -7.07
233.0531 -2.380  ...  19788.251  4.0 4s2 a3H    62683.770  4.0 8p 3G5G5F   8.29 -3.24 -7.02
233.2566 -2.855  ...  17726.988  2.0 (4P)4s a5P  60585.090  2.0 3Psp1P 3P   8.58 -4.76 -7.25
233.6965 -1.458  ...  18378.186  2.0 4s2 a3P    61155.620  1.0 3Psp1P 3P   8.52 -4.00 -7.19
234.0025 -1.288  ...  19788.251  4.0 4s2 a3H    62509.750  3.0 8p 3D3G3F   8.28 -3.79 -7.03
234.0264 -0.551  ...  19390.168  6.0 4s2 a3H    62107.290  5.0 3Gsp1P 3G   8.33 -2.99 -7.04
234.1369 -2.341  ...  18378.186  2.0 4s2 a3P    61075.160  1.0 3Fsp1P 3D   8.23 -2.91 -7.06
234.1881 -2.407  ...  17550.181  3.0 (4P)4s a5P  60237.810  2.0 4s6D7p 5D   7.53 -3.90 -7.06
235.1622 -2.878  ...  17726.988  2.0 (4P)4s a5P  60237.810  2.0 4s6D7p 5D   7.53 -3.90 -7.06
235.1903 -2.735  ...  17550.181  3.0 (4P)4s a5P  60055.930  3.0 4s6D7p 5F   7.21 -4.07 -7.07
235.2261 -1.417  ...  19788.251  4.0 4s2 a3H    62287.540  3.0 (4F)7p 5G   7.78 -4.05 -7.00
235.4266 -2.949  ...  17550.181  3.0 (4P)4s a5P  60013.270  3.0 (4F)6p 3G   8.00 -3.41 -7.26
236.1908 -2.145  ...  17550.181  3.0 (4P)4s a5P  59875.890  3.0 4s6D7p 5D   7.67 -3.29 -7.13
236.2761 -2.802  ...  17927.382  1.0 (4P)4s a5P  60237.810  2.0 4s6D7p 5D   7.53 -3.90 -7.06
236.3161 -1.453  ...  19390.168  6.0 4s2 a3H    61693.440  5.0 (4F)7p 5G   7.69 -3.92 -7.08
236.6440 -2.594  ...  17550.181  3.0 (4P)4s a5P  59794.850  3.0 3Dsp3P 1F   7.86 -4.18 -7.25
236.6592 -2.319  ...  17927.382  1.0 (4P)4s a5P  60169.330  1.0 4s4F7p 5D   7.56 -3.76 -7.07
236.7332 -1.189  ...  19788.251  4.0 4s2 a3H    62016.990  3.0 (4F)7p 3G   8.12 -3.24 -7.09
236.8557 -0.689  ...  18378.186  2.0 4s2 a3P    60585.090  2.0 3Psp1P 3P   8.58 -4.76 -7.25
237.6128 -2.378  ...  19621.006  4.0 4s2 a3H    61693.440  5.0 (4F)7p 5G   7.69 -3.92 -7.08
237.6986 -2.143  ...  19621.006  5.0 4s2 a3H    61678.260  4.0 (4F)7p 5F   7.62 -3.54 -7.02
237.7811 -0.493 -0.20 20641.110  4.0 4s2 b3F    62683.770  4.0 8p 3G5G5F   8.29 -3.24 -7.02
237.8680 -1.624  ...  19621.006  5.0 4s2 a3H    61648.300  4.0 (4F)7p 3G   7.95 -4.20 -7.12
238.0370 -2.214  ...  18378.186  2.0 4s2 a3P    60375.650  1.0 (2F)4p 3D   8.52 -4.33 -7.37
238.1208 -2.913  ...  19788.251  4.0 4s2 a3H    61770.940  3.0 (4F)7p 5D   7.39 -2.63 -6.96
238.1584 -2.765  ...  17726.988  2.0 (4P)4s a5P  59703.050  1.0 (4F)6p 5D   7.40 -3.86 -7.16
238.2881 -2.340  ...  17550.181  3.0 (4P)4s a5P  59503.400  3.0 6p 5G5D3D   8.15 -3.43 -7.17
238.3266 -2.579  ...  17550.181  3.0 (4P)4s a5P  59496.620  4.0 4s6D7p 5D   7.42 -3.41 -7.08
238.7695 -1.656 +0.20 20641.110  4.0 4s2 b3F    62509.750  3.0 8p 3D3G3F   8.28 -3.79 -7.03
238.8185 -1.622  ...  19788.251  4.0 4s2 a3H    61648.300  4.0 (4F)7p 3G   7.95 -4.20 -7.12
238.8209 -1.880  ...  18378.186  2.0 4s2 a3P    60237.810  2.0 4s6D7p 5D   7.53 -3.90 -7.06
239.1084 -2.319 +0.60 20874.482  3.0 4s2 b3F    62683.770  4.0 8p 3G5G5F   8.29 -3.24 -7.02
239.2966 -1.855  ...  17726.988  2.0 (4P)4s a5P  59503.400  3.0 6p 5G5D3D   8.15 -3.43 -7.17
239.3009 -2.378  ...  17927.382  1.0 (4P)4s a5P  59703.050  1.0 (4F)6p 5D   7.40 -3.86 -7.16
239.4454 -2.691  ...  19390.168  6.0 4s2 a3H    61140.620  5.0 (4F)7p 3G   7.98 -3.98 -7.12
239.8632 -2.980  ...  18378.186  2.0 4s2 a3P    60055.930  3.0 4s6D7p 5F   7.21 -4.07 -7.07
240.0435 -2.833  ...  20641.110  4.0 4s2 b3F    62287.540  3.0 (4F)7p 5G   7.78 -4.05 -7.00
240.1034 -1.979 -0.50 61198.480  5.0 (2G)5s 3G  19562.439  4.0 5Dsp3P z7D  7.63 -3.84 -7.13
240.1079 -0.676 -0.20 20874.482  3.0 4s2 b3F    62509.750  3.0 8p 3D3G3F   8.28 -3.79 -7.03
240.1090 -1.963  ...  18378.186  2.0 4s2 a3P    60013.270  3.0 (4F)6p 3G   8.00 -3.41 -7.26
240.2933 -1.127 -0.30 19552.478  1.0 4s2 a3P    61155.620  1.0 3Psp1P 3P   8.52 -4.00 -7.19
240.5231 -2.347  ...  19788.251  4.0 4s2 a3H    61351.660  3.0 (4F)7p 3D   7.90 -4.31 -7.07
240.6879 -2.282  ...  19621.006  5.0 4s2 a3H    61155.950  5.0 (4F)7p 5F   7.07 -4.17 -7.06
240.7590 -1.379 +0.40 19552.478  1.0 4s2 a3P    61075.160  1.0 3Fsp1P 3D   8.23 -2.91 -7.06
240.7768 -0.628 -0.60 19621.006  5.0 4s2 a3H    61140.620  5.0 (4F)7p 3G   7.98 -3.98 -7.12
240.9039 -1.406 -0.90 18378.186  2.0 4s2 a3P    59875.890  3.0 4s6D7p 5D   7.67 -3.29 -7.13
241.0604 -1.172  ...  21038.987  2.0 4s2 b3F    62509.750  3.0 8p 3D3G3F   8.28 -3.79 -7.03
241.0871 -0.961 -0.30 20641.110  4.0 4s2 b3F    62107.290  5.0 3Gsp1P 3G   8.33 -2.99 -7.04
241.3753 -1.007 -0.40 18378.186  2.0 4s2 a3P    59794.850  3.0 3Dsp3P 1F   7.86 -4.18 -7.25
241.6133 -2.361  ...  20641.110  4.0 4s2 b3F    62016.990  3.0 (4F)7p 3G   8.12 -3.24 -7.09
241.7507 -2.047  ...  19788.251  4.0 4s2 a3H    61140.620  5.0 (4F)7p 3G   7.98 -3.98 -7.12
241.9100 -1.407  ...  19788.251  4.0 4s2 a3H    61113.380  4.0 (4F)7p 3F   7.84 -3.82 -7.11
242.3591 -1.844 +0.00 21038.987  2.0 4s2 b3F    62287.540  3.0 (4F)7p 5G   7.78 -4.05 -7.00
242.7649 -1.749 -0.20 17550.181  3.0 (4P)4s a5P  58729.800  4.0 (4F)6p 5D   7.68 -4.39 -7.23
```

```
242.9839 -2.954  ...   20874.482  3.0 4s2 b3F   62016.990  3.0 (4F)7p 3G   8.12 -3.24 -7.09
243.0588 -1.810 -0.10  20641.110  4.0 4s2 b3F   61770.940  3.0 (4F)7p 5D   7.39 -2.63 -6.96
243.0861 -0.360 -0.80  18378.186  2.0 4s2 a3P   59503.400  3.0 6p 5G5D3D   8.15 -3.43 -7.17
243.1299 -0.620 -0.20  20037.816  0.0 4s2 a3P   61155.620  1.0 3Psp1P 3P  8.52 -4.00 -7.19
243.5177 -1.669  ...   20641.110  4.0 4s2 b3F   61693.440  5.0 (4F)7p 5G   7.69 -3.92 -7.08
243.5870 -1.151 +0.60  18378.186  2.0 4s2 a3P   59418.830  3.0 (4F)6p 5F  7.59 -4.10 -7.19
243.6066 -2.986  ...   20037.816  0.0 4s2 a3P   61075.160  1.0 3Fsp1P 3D  8.23 -2.91 -7.06
243.6078 -1.565 +0.40  20641.110  4.0 4s2 b3F   61678.260  4.0 (4F)7p 5F  7.62 -3.54 -7.02
243.6347 -0.536 +0.10  19552.478  1.0 4s2 a3P   60585.090  2.0 3Psp1P 3P  8.58 -4.76 -7.25
243.7857 -2.935  ...   20641.110  4.0 4s2 b3F   61648.300  4.0 (4F)7p 3G   7.95 -4.20 -7.12
243.8763 -1.530  ...   20874.482  3.0 4s2 b3F   61866.450  3.0 (4F)7p 3D   7.75 -3.45 -7.06
243.9594 -1.328  ...   21038.987  2.0 4s2 b3F   62016.990  3.0 (4F)7p 3G   8.12 -3.24 -7.09
244.0187 -0.809 -0.40  21715.732  5.0 (2G)4s a3G 62683.770  4.0 8p 3G5G5F  8.29 -3.24 -7.02
244.4459 -2.331  ...   20874.482  3.0 4s2 b3F   61770.940  3.0 (4F)7p 5D   7.39 -2.63 -6.96
244.8590 -2.671  ...   21038.987  2.0 4s2 b3F   61866.450  2.0 (4F)7p 3D   7.75 -3.45 -7.06
244.8847 -0.673 -0.30  19552.478  1.0 4s2 a3P   60375.650  1.0 (2F)4p 3D   8.52 -4.33 -7.37
245.0011 -1.900  ...   20874.482  3.0 4s2 b3F   61678.260  4.0 (4F)7p 5F   7.62 -3.54 -7.02
245.1219 -2.246  ...   19552.478  1.0 4s2 a3P   60336.160  1.0 4s6d7p 5F   7.42 -4.13 -7.07
245.1812 -1.665 +0.00  20874.482  3.0 4s2 b3F   61648.300  4.0 (4F)7p 3G   7.95 -4.20 -7.12
245.5622 -0.782  ...   20641.110  4.0 4s2 b3F   61351.660  3.0 (4F)7p 3D   7.90 -4.31 -7.07
245.7186 -1.645 -0.30  21999.130  4.0 (2G)4s a3G 62683.770  4.0 8p 3G5G5F  8.29 -3.24 -7.02
246.1288 -2.834  ...   19552.478  1.0 4s2 a3P   60169.330  1.0 4s4F7p 5D   7.56 -3.76 -7.07
246.7742 -1.173  ...   21999.130  4.0 (2G)4s a3G 62509.750  3.0 8p 3D3G3F  8.28 -3.79 -7.03
246.8419 -2.289 -0.30  20641.110  4.0 4s2 b3F   61140.620  5.0 (4F)7p 3G   7.98 -3.98 -7.12
246.9781 -1.457 -0.10  20874.482  3.0 4s2 b3F   61351.660  3.0 (4F)7p 3D   7.90 -4.31 -7.07
247.0081 -2.012  ...   20641.110  4.0 4s2 b3F   61113.380  4.0 (4F)7p 3F   7.84 -3.82 -7.11
247.2398 -2.632  ...   22249.429  3.0 (2G)4s a3G 62683.770  4.0 8p 3G5G5F  8.29 -3.24 -7.02
247.5017 -0.566 -0.10  21715.732  5.0 (2G)4s a3G 62107.290  5.0 3Gsp1P 3G  8.33 -2.99 -7.04
247.8314 -0.856 -0.30  20037.816  0.0 4s2 a3P   60375.650  1.0 (2F)4p 3D   8.52 -4.33 -7.37
247.9861 -2.860  ...   21038.987  2.0 4s2 b3F   61351.660  3.0 (4F)7p 3D   7.90 -4.31 -7.07
248.0742 -2.320 +0.50  20037.816  0.0 4s2 a3P   60336.160  1.0 4s6d7p 5F   7.42 -4.13 -7.07
248.1354 -2.328 +0.00  21999.130  4.0 (2G)4s a3G 62287.540  3.0 (4F)7p 5G   7.78 -4.05 -7.00
248.3085 -1.540  ...   22249.429  3.0 (2G)4s a3G 62509.750  3.0 8p 3D3G3F  8.28 -3.79 -7.03
248.5265 -0.784 +0.00  19788.251  4.0 4s2 a3H   60013.270  3.0 (4F)6p 3G   8.00 -3.41 -7.26
248.7831 -2.044  ...   19621.006  5.0 4s2 a3H   59804.540  4.0 4s6d7p 5F   7.10 -4.56 -7.07
248.7921 -2.715  ...   59532.970  4.0 s6d 3+[4+] 19350.891  5.0 5Dsp3P z7D  7.67 -3.91 -7.15
248.9873 -2.067  ...   19552.478  1.0 4s2 a3P   59703.050  1.0 (4F)6p 5D   7.40 -3.86 -7.16
249.1056 -2.615  ...   20037.816  0.0 4s2 a3P   60169.330  1.0 4s4F7p 5D   7.56 -3.76 -7.07
249.1980 -1.694  ...   21038.987  2.0 4s2 b3F   61155.620  1.0 3Psp1P 3P  8.52 -4.00 -7.19
249.2374 -0.408 +0.00  19621.006  5.0 4s2 a3H   59731.290  4.0 (4F)6p 3G   8.03 -3.71 -7.26
249.2506 -1.848 -0.10  21999.130  4.0 (2G)4s a3G 62107.290  5.0 3Gsp1P 3G  8.33 -2.99 -7.04
249.3782 -1.836 -0.30  19788.251  4.0 4s2 a3H   59875.890  3.0 4s6d7p 5D   7.67 -3.29 -7.13
249.6867 -1.418 +0.00  22249.429  3.0 (2G)4s a3G 62287.540  3.0 (4F)7p 5G   7.78 -4.05 -7.00
249.6988 -0.569 +0.00  21038.987  2.0 4s2 b3F   61075.160  1.0 3Fsp1P 3D  8.23 -2.91 -7.06
249.8131 -1.943  ...   21999.130  4.0 (2G)4s a3G 62016.990  3.0 (4F)7p 3G   8.12 -3.24 -7.09
249.8834 -2.125  ...   19788.251  4.0 4s2 a3H   59794.850  3.0 3Dsp3P 1F  7.86 -4.18 -7.25
250.0640 -1.201 -0.10  21715.732  5.0 (2G)4s a3G 61693.440  5.0 (4F)7p 5G   7.69 -3.92 -7.08
250.0865 -1.927  ...   19621.006  5.0 4s2 a3H   59595.120  4.0 4s6d7p 7F   6.88 -4.11 -7.10
250.1089 -2.114  ...   59532.970  4.0 s6d 3+[4+] 19562.439  4.0 5Dsp3P z7D  7.67 -3.91 -7.15
250.1319  0.071 -0.50  19390.168  6.0 4s2 a3H   59357.030  5.0 (4F)6p 3G   8.26 -4.27 -7.28
250.1590 -2.222  ...   21715.732  5.0 (2G)4s a3G 61678.260  4.0 (4F)7p 5F   7.62 -3.54 -7.02
250.2811 -2.042  ...   19788.251  4.0 4s2 a3H   59731.290  4.0 (4F)6p 3G   8.03 -3.71 -7.26
250.3467 -1.817  ...   21715.732  5.0 (2G)4s a3G 61648.300  4.0 (4F)7p 3G   7.95 -4.20 -7.12
250.6810 -2.688  ...   59636.360  3.0 5d 5F5D3G  19757.032  3.0 5Dsp3P z7D  7.73 -3.56 -7.25
250.7043 -1.410  ...   19621.006  5.0 4s2 a3H   59496.620  4.0 4s6d7p 5D   7.42 -3.41 -7.08
251.3856 -0.953  ...   22249.429  3.0 (2G)4s a3G 62016.990  3.0 (4F)7p 3G   8.12 -3.24 -7.09
251.4568 -0.488 -0.50  19621.006  5.0 4s2 a3H   59377.300  4.0 (4F)6p 5G   7.82 -3.96 -7.22
251.5851 -2.743  ...   19621.006  5.0 4s2 a3H   59357.030  5.0 (4F)6p 3G   8.26 -4.27 -7.28
251.7173 -1.529 -0.40  19788.251  4.0 4s2 a3H   59503.400  3.0 6p 5G5D3D   8.15 -3.43 -7.17
251.7461 -0.511 -0.50  20874.482  3.0 4s2 b3F   60585.090  2.0 3Psp1P 3P  8.58 -4.76 -7.25
251.8333 -1.378  ...   19621.006  5.0 4s2 a3H   59317.860  4.0 4s6d7p 7D   7.22 -4.13 -7.11
251.8495 -2.440  ...   21999.130  4.0 (2G)4s a3G 61693.440  5.0 (4F)7p 5G   7.69 -3.92 -7.08
251.9458 -2.253  ...   21999.130  4.0 (2G)4s a3G 61678.260  4.0 (4F)7p 5F   7.62 -3.54 -7.02
252.0341 -2.169  ...   20037.816  0.0 4s2 a3P   59703.050  1.0 (4F)6p 5D   7.40 -3.86 -7.16
252.1362 -1.095  ...   21999.130  4.0 (2G)4s a3G 61648.300  4.0 (4F)7p 3G   7.95 -4.20 -7.12
252.2509 -0.211  ...   19390.168  6.0 4s2 a3H   59021.310  5.0 (4F)6p 5D   7.68 -4.37 -7.26
252.3871 -2.819  ...   59366.790  2.0 (4F)5d 5F  19757.032  3.0 5Dsp3P z7D  7.57 -3.98 -7.25
252.7934 -0.929  ...   21038.987  2.0 4s2 b3F   60585.090  2.0 3Psp1P 3P  8.58 -4.76 -7.25
```

```
252.8494 -2.841  ...   59294.380 3.0 4s4D4d 3G  19757.032 3.0 5Dsp3P z7D  8.29 -3.62 -7.37
253.4721 -2.565 +0.50  21715.732 5.0 (2G)4s a3G 61155.950 5.0 (4F)7p 5F   7.07 -4.17 -7.06
253.5707 -1.080 +0.00  21715.732 5.0 (2G)4s a3G 61140.620 5.0 (4F)7p 3G   7.98 -3.98 -7.12
253.6355 -2.079  ...   20641.110 4.0 4s2 b3F    60055.930 3.0 4s6D7p 5F   7.21 -4.07 -7.07
253.7289 -1.813  ...   19621.006 5.0 4s2 a3H    59021.310 5.0 (4F)6p 5G   7.98 -4.37 -7.26
253.7381 -2.120  ...   22249.429 3.0 (2G)4s a3G 61648.300 4.0 (4F)7p 3G   7.95 -4.20 -7.12
253.7460 -2.407  ...   21715.732 5.0 (2G)4s a3G 61113.380 4.0 (4F)7p 3F   7.84 -3.82 -7.11
253.9103 -1.541  ...   20641.110 4.0 4s2 b3F    60013.270 3.0 (4F)6p 3G   8.00 -3.41 -7.26
254.1394 -0.513 -0.60  21038.987 2.0 4s2 b3F    60375.650 1.0 (2F)4p 3D   8.52 -4.33 -7.37
254.3948 -2.297  ...   21038.987 2.0 4s2 b3F    60336.160 1.0 4s6D7p 5F   7.42 -4.13 -7.07
254.7994 -2.581  ...   20641.110 4.0 4s2 b3F    59875.890 3.0 4s6D7p 5D   7.67 -3.29 -7.13
254.8106 -1.567  ...   19788.251 4.0 4s2 a3H    59021.310 5.0 (4F)6p 5G   7.98 -4.37 -7.26
254.8994 -2.715  ...   19390.168 6.0 4s2 a3H    58609.560 5.0 (4F)6p 5F   7.19 -3.24 -7.23
255.0332 -2.152 -0.80  21038.987 2.0 4s2 b3F    60237.810 2.0 4s6D7p 5D   7.53 -3.90 -7.06
255.1463 -2.472 +0.50  20874.482 3.0 4s2 b3F    60055.930 3.0 4s6D7p 5F   7.21 -4.07 -7.07
255.3268 -1.650  ...   20641.110 4.0 4s2 b3F    59794.850 3.0 3Dsp3P 1F   7.86 -4.18 -7.25
255.4068 -2.033  ...   21999.130 4.0 (2G)4s a3G 61140.620 5.0 (4F)7p 3G   7.98 -3.98 -7.12
255.4244 -1.046 -0.40  20874.482 3.0 4s2 b3F    60013.270 3.0 (4F)6p 3G   8.00 -3.41 -7.26
255.5846 -2.617 +0.10  21999.130 4.0 (2G)4s a3G 61113.380 4.0 (4F)7p 3F   7.84 -3.82 -7.11
255.6203 -1.245 -0.20  19621.006 5.0 4s2 a3H    58729.800 4.0 (4F)6p 5D   7.68 -4.39 -7.23
255.7420 -1.052 -0.20  20641.110 4.0 4s2 b3F    59731.290 4.0 (4F)6p 3G   8.03 -3.71 -7.26
255.9811 -1.795  ...   58616.110 3.0 (4F)5d 5P  19562.439 4.0 5Dsp3P z7D  8.00 -4.14 -7.38
256.1487 -1.802  ...   22838.323 2.0 (4P)4s b3P 61866.450 2.0 (4F)7p 3D   7.75 -3.45 -7.06
256.3242 -2.128  ...   20874.482 3.0 4s2 b3F    59875.890 3.0 4s6D7p 5D   7.67 -3.29 -7.13
256.5026 -2.291 +0.50  21038.987 2.0 4s2 b3F    60013.270 3.0 (4F)6p 3G   8.00 -3.41 -7.26
256.6361 -2.529  ...   20641.110 4.0 4s2 b3F    59595.120 4.0 4s6D7p 7F   6.88 -4.11 -7.10
256.7771 -2.152  ...   22838.323 2.0 (4P)4s b3P 61770.940 3.0 (4F)6p 3D   7.39 -2.63 -6.96
256.7940 -2.900  ...   20874.482 3.0 4s2 b3F    59804.540 4.0 4s6D7p 5F   7.10 -4.56 -7.07
256.8579 -1.298  ...   20874.482 3.0 4s2 b3F    59794.850 3.0 3Dsp3P 1F   7.86 -4.18 -7.25
256.9914 -0.315 -0.30  23783.619 5.0 4s2 b3G    62683.770 4.0 8p 3G5G5F   8.29 -3.24 -7.02
257.2100 -2.595  ...   58779.590 2.0 4s4D4d 3D  19912.495 2.0 5Dsp3P z7D  8.14 -4.56 -7.43
257.2418 -1.315 -0.80  20641.110 4.0 4s2 b3F    59503.400 3.0 6p 5G5D3D   8.15 -3.43 -7.17
257.2631 -1.865  ...   58616.110 3.0 (4F)5d 5P  19757.032 3.0 5Dsp3P z7D  8.00 -4.14 -7.38
257.2781 -1.524 +0.00  20874.482 3.0 4s2 b3F    59731.290 4.0 (4F)6p 3G   8.03 -3.71 -7.26
257.2867 -1.862 -0.60  20641.110 4.0 4s2 b3F    59496.620 4.0 4s6D7p 5D   7.42 -3.41 -7.08
257.9210 -2.571  ...   58779.590 2.0 4s4D4d 3D  20019.635 1.0 5Dsp3P z7D  8.14 -4.56 -7.43
257.9483 -1.863  ...   21038.987 2.0 4s2 b3F    59794.850 3.0 3Dsp3P 1F   7.86 -4.18 -7.25
258.0793 -1.802 -0.20  20641.110 4.0 4s2 b3F    59377.300 4.0 (4F)6p 5G   7.82 -3.96 -7.22
258.2144 -1.080 -0.60  20641.110 4.0 4s2 b3F    59357.030 5.0 (4F)6p 3G   8.26 -4.27 -7.28
258.2965 -2.358  ...   58616.110 3.0 (4F)5d 5P  19912.495 2.0 5Dsp3P z7D  8.00 -4.14 -7.38
258.7960 -2.023  ...   20874.482 3.0 4s2 b3F    59503.400 3.0 6p 5G5D3D   8.15 -3.43 -7.17
258.8414 -2.612  ...   20874.482 3.0 4s2 b3F    59496.620 4.0 4s6D7p 5D   7.42 -3.41 -7.08
259.2253 -1.185  ...   24118.819 4.0 4s2 b3G    62683.770 4.0 8p 3G5G5F   8.29 -3.24 -7.02
259.3639 -1.173  ...   20874.482 3.0 4s2 b3F    59418.830 3.0 (4F)6p 5F   7.59 -4.10 -7.19
259.5727 -1.224 -0.10  22838.323 2.0 (4P)4s b3P 61351.660 3.0 (4F)7p 3D   7.90 -4.31 -7.07
259.6436 -1.860  ...   20874.482 3.0 4s2 b3F    59377.300 4.0 (4F)6p 5G   7.82 -3.96 -7.22
259.9029 -2.523  ...   21038.987 2.0 4s2 b3F    59503.400 3.0 6p 5G5D3D   8.15 -3.43 -7.17
260.0451 -2.993  ...   20874.482 3.0 4s2 b3F    59317.860 4.0 4s6D7p 7D   7.22 -4.13 -7.11
260.4004 -0.688 +0.10  24118.819 4.0 4s2 b3G    62509.750 3.0 8p 3D3G3F   8.28 -3.79 -7.03
260.4732 -1.874  ...   20641.110 4.0 4s2 b3F    59021.310 5.0 (4F)6p 5G   7.98 -4.37 -7.26
260.4756 -2.461  ...   21038.987 2.0 4s2 b3F    59418.830 3.0 (4F)6p 5F   7.59 -4.10 -7.19
260.7123 -2.272  ...   24338.767 3.0 4s2 b3G    62683.770 4.0 8p 3G5G5F   8.29 -3.24 -7.02
260.7758 -2.867  ...   22249.429 3.0 (2G)4s a3G 60585.090 2.0 3Psp1P 3P   8.58 -4.76 -7.25
260.8574  0.007 -0.10  23783.619 5.0 4s2 b3G    62107.290 5.0 3Gsp1P 3G   8.33 -2.99 -7.04
260.9008 -2.242 -0.30  22838.323 2.0 (4P)4s b3P 61155.620 1.0 3Psp1P 3D   8.52 -4.00 -7.19
261.4499 -2.492  ...   22838.323 2.0 (4P)4s b3P 61075.160 1.0 3Fsp1P 3D   8.23 -2.91 -7.06
261.6896 -1.665 -0.70  61198.480 5.0 (2G)5s 3G  22996.674 4.0 5Dsp3P z7F  7.63 -3.84 -7.13
261.8804 -2.712 +0.10  24335.766 2.0 (2P)4s c3P 62509.750 3.0 8p 3D3G3F   8.28 -3.79 -7.03
261.9009 -0.910 +0.10  24338.767 3.0 4s2 b3G    62509.750 3.0 8p 3D3G3F   8.28 -3.79 -7.03
261.9165 -1.611  ...   24118.819 4.0 4s2 b3G    62287.540 3.0 (4F)7p 5G   7.78 -4.05 -7.00
262.1938 -1.438  ...   22946.816 1.0 (4P)4s b3P 61075.160 1.0 3Fsp1P 3D   8.23 -2.91 -7.06
262.3623 -1.812  ...   23051.750 0.0 (4P)4s b3P 61155.620 1.0 3Psp1P 3P   8.52 -4.00 -7.19
262.4668 -2.114 +0.00  20641.110 4.0 4s2 b3F    58729.800 4.0 (4F)6p 5D   7.68 -4.39 -7.23
262.6868 -2.664  ...   21999.130 4.0 (2G)4s a3G 60055.930 3.0 4s6D7p 5F   7.21 -4.07 -7.07
262.9175 -1.518 +0.00  23051.750 0.0 (4P)4s b3P 61075.160 1.0 3Fsp1P 3D   8.23 -2.91 -7.06
262.9718 -1.326 -0.50  21715.732 5.0 (2G)4s a3G 59731.290 4.0 (4F)6p 3G   8.03 -3.71 -7.26
262.9816 -1.260 -0.60  21999.130 4.0 (2G)4s a3G 60013.270 3.0 (4F)6p 3G   8.00 -3.41 -7.26
263.1593 -1.265  ...   24118.819 4.0 4s2 b3G    62107.290 5.0 3Gsp1P 3G   8.33 -2.99 -7.04
```

```
263.4346 -0.890 +0.00   24338.767   3.0 4s2 b3G    62287.540   3.0 (4F)7p 5G    7.78 -4.05 -7.00
263.5296 -2.395  ...     24574.655   4.0 (2G)4s a1G  62509.750   3.0 8p 3D3G3F    8.28 -3.79 -7.03
263.7053 -0.598 -0.10   23783.619   5.0 4s2 b3G    61693.440   5.0 (4F)7p 5G    7.69 -3.92 -7.08
263.7864 -1.224 -0.30   24118.819   4.0 4s2 b3G    62016.990   3.0 (4F)7p 3G    8.12 -3.24 -7.09
263.8109 -1.087 -0.20   23783.619   5.0 4s2 b3G    61678.260   4.0 (4F)7p 5F    7.62 -3.54 -7.02
263.9172 -2.814  ...     21715.732   5.0 (2G)4s a3G  59595.120   4.0 4s6D7p 7F    6.88 -4.11 -7.10
263.9355 -1.851 -0.30   24118.819   4.0 (2G)4s a3G  59875.890   3.0 4s6D7p 5D    7.67 -3.29 -7.13
264.0197 -1.166 +0.00   23783.619   5.0 4s2 b3G    61648.300   4.0 (4F)7p 3G    7.95 -4.20 -7.12
264.5014 -2.569  ...     21999.130   4.0 (2G)4s a3G  59794.850   3.0 3Dsp3P 1F    7.86 -4.18 -7.25
264.6053 -2.142 -0.50   21715.732   5.0 (2G)4s a3G  59496.620   4.0 4s6D7p 5D    7.42 -3.41 -7.08
264.7247 -1.568 -0.20   22249.429   3.0 (2G)4s a3G  60013.270   3.0 (4F)6p 3G    8.00 -3.41 -7.26
264.8445 -1.047 -0.50   22838.323   2.0 (4P)4s b3P  60585.090   2.0 3Psp1P 3P    8.58 -4.76 -7.25
264.9470 -1.881  ...     21999.130   4.0 (2G)4s a3G  59731.290   4.0 (4F)6p 3G    8.03 -3.71 -7.26
265.3263 -0.390 -0.30   24338.767   3.0 4s2 b3G    62016.990   3.0 (4F)7p 3G    8.12 -3.24 -7.09
265.4437 -1.688 -0.20   21715.732   5.0 (2G)4s a3G  59377.300   4.0 (4F)6p 5G    7.82 -3.96 -7.22
265.5866 -2.134 -0.40   21715.732   5.0 (2G)4s a3G  59357.030   5.0 (4F)6p 3G    8.26 -4.27 -7.28
265.6080 -2.063  ...     22946.816   1.0 (4P)4s b3P  60585.090   2.0 3Psp1P 3P    8.58 -4.76 -7.25
265.6913 -2.633 -0.30   22249.429   3.0 (2G)4s a3G  59875.890   3.0 4s6D7p 5D    7.67 -3.29 -7.13
266.0579 -1.919 -0.20   24118.819   4.0 4s2 b3G    61693.440   5.0 (4F)7p 5G    7.69 -3.92 -7.08
266.1655 -1.283 +0.00   24118.819   4.0 4s2 b3G    61678.260   4.0 (4F)7p 5F    7.62 -3.54 -7.02
266.2649 -2.487 -0.30   22249.429   3.0 (2G)4s a3G  59794.850   3.0 3Dsp3P 1F    7.86 -4.18 -7.25
266.3556 -2.956  ...     24574.655   4.0 (2G)4s a1G  62107.290   5.0 3Gsp1P 3G    8.33 -2.99 -7.04
266.3694 -1.282  ...     24335.766   2.0 (2P)4s c3P  61866.450   2.0 (4F)7p 3D    7.75 -3.45 -7.06
266.3780 -0.437 -0.10   24118.819   4.0 4s2 b3G    61648.300   4.0 (4F)7p 3G    7.95 -4.20 -7.12
266.3907 -2.014  ...     24338.767   3.0 4s2 b3G    61866.450   2.0 (4F)7p 3D    7.75 -3.45 -7.06
266.5570 -2.077 -0.30   21999.130   4.0 (2G)4s a3G  59503.400   3.0 6p 5G5D3D    8.15 -3.43 -7.17
266.6797 -2.994  ...     61198.480   5.0 (2G)5s 3G   23711.456   4.0 5Dsp3P z7P   7.63 -3.84 -7.13
266.9980 -2.745  ...     24574.655   4.0 (2G)4s a1G  62016.990   3.0 (4F)7p 3G    8.12 -3.24 -7.09
267.0491 -2.812  ...     24335.766   2.0 (2P)4s c3P  61770.940   3.0 (4F)7p 5D    7.39 -2.63 -6.96
267.0705 -2.975  ...     24338.767   3.0 4s2 b3G    61770.940   3.0 (4F)7p 5D    7.39 -2.63 -6.96
267.1595 -1.798 -0.80   21999.130   4.0 (2G)4s a3G  59418.830   3.0 (4F)6p 5F    7.59 -4.10 -7.19
267.3039 -2.705  ...     22838.323   2.0 (4P)4s b3P  60237.810   2.0 4s6D7p 5D    7.53 -3.90 -7.06
267.4981 -1.881 +0.70   23783.619   5.0 4s2 b3G    61155.950   5.0 (4F)7p 3D    7.07 -4.17 -7.06
267.6015 -2.404 +0.00   21999.130   4.0 (2G)4s a3G  59357.030   5.0 (4F)6p 3G    8.26 -4.27 -7.28
267.6079 -0.382 -0.20   23783.619   5.0 4s2 b3G    61140.620   5.0 (4F)7p 3G    7.98 -3.98 -7.12
267.7334 -2.580  ...     24338.767   3.0 4s2 b3G    61678.260   4.0 (4F)7p 5F    7.62 -3.54 -7.02
267.8032 -0.580 +0.00   23783.619   5.0 4s2 b3G    61113.380   4.0 (4F)7p 3F    7.84 -3.82 -7.11
267.8453 -2.049 -0.80   23051.750   0.0 (4P)4s b3P  60375.650   1.0 (2F)4p 3D    8.52 -4.33 -7.37
267.9484 -1.615 -0.10   24338.767   3.0 4s2 b3G    61648.300   4.0 (4F)7p 3G    7.95 -4.20 -7.12
267.9768 -1.766  ...     21715.732   5.0 (2G)4s a3G  59021.310   5.0 (4F)6p 5G    7.98 -4.37 -7.26
268.3481 -2.919  ...     22249.429   3.0 (2G)4s a3G  59503.400   3.0 6p 5G5D3D    8.15 -3.43 -7.17
268.5004 -1.665 -0.50   24118.819   4.0 4s2 b3G    61351.660   3.0 (4F)7p 3D    7.90 -4.31 -7.07
268.9185 -2.292  ...     22838.323   2.0 (4P)4s b3P  60013.270   3.0 (4F)6p 3G    8.00 -3.41 -7.26
269.5023 -1.370 -0.40   24772.018   1.0 (2P)4s c3P  61866.450   2.0 (4F)7p 3D    7.75 -3.45 -7.06
269.6534 -2.502  ...     24574.655   4.0 (2G)4s a1G  61648.300   4.0 (4F)7p 3G    7.95 -4.20 -7.12
269.9161 -2.655  ...     22838.323   2.0 (4P)4s b3P  59875.890   3.0 4s6D7p 5D    7.67 -3.29 -7.13
270.0282 -2.372 -0.40   21999.130   4.0 (2G)4s a3G  59021.310   5.0 (4F)6p 5G    7.98 -4.37 -7.26
270.0310 -1.654 -0.90   24118.819   4.0 4s2 b3G    61140.620   5.0 (4F)7p 3G    7.98 -3.98 -7.12
270.0741 -2.914  ...     24335.766   2.0 (2P)4s c3P  61351.660   3.0 (4F)7p 3D    7.90 -4.31 -7.07
270.0874 -2.031 -0.20   21715.732   5.0 (2G)4s a3G  58729.800   4.0 (4F)6p 5D    7.68 -4.39 -7.23
270.0960 -2.317  ...     24338.767   3.0 4s2 b3G    61351.660   3.0 (4F)7p 3D    7.90 -4.31 -7.07
270.2298 -1.078 -0.20   24118.819   4.0 4s2 b3G    61113.380   4.0 (4F)7p 3F    7.84 -3.82 -7.11
270.5080 -1.896 -0.40   22838.323   2.0 (4P)4s b3P  59794.850   3.0 3Dsp3P 1F    7.86 -4.18 -7.25
270.9609 -2.031  ...     60087.260   2.0 4s4D4d 3P   23192.500   2.0 5Dsp3P z7P   7.83 -3.79 -7.18
271.3458 -1.613 +0.00   60087.260   2.0 4s4D4d 3P   23244.838   1.0 5Dsp3P z7F   7.83 -3.79 -7.18
271.5121 -1.177  ...     24335.766   2.0 (2P)4s c3P  61155.620   1.0 3Psp1P 3P    8.52 -4.00 -7.19
271.7082 -2.484 -0.80   19927.382   1.0 (4P)4s a5P  54720.670   0.0 (4F)5p 5D    8.08 -4.64 -7.43
271.8285 -2.894  ...     24574.655   4.0 (2G)4s a1G  61351.660   3.0 (4F)7p 3D    7.90 -4.31 -7.07
271.8462 -2.409  ...     24338.767   3.0 4s2 b3G    61113.380   4.0 (4F)7p 3F    7.84 -3.82 -7.11
272.1714 -2.763 -0.30   21999.130   4.0 (2G)4s a3G  58729.800   4.0 (4F)6p 5D    7.68 -4.39 -7.23
272.4947 -2.294  ...     59532.970   4.0 s6d 3+[4+]  22845.869   5.0 5Dsp3P z7F   7.67 -3.91 -7.15
272.6584 -1.712 -0.60   22838.323   2.0 (4P)4s b3P  59503.400   3.0 6p 5G5D3D    8.15 -3.43 -7.17
272.8473 -2.438 -0.40   59636.360   3.0 5d 5F5D3G   59875.890   4.0 5Dsp3P z7F   7.73 -3.56 -7.25
273.2887 -2.384  ...     22838.323   2.0 (4P)4s b3P  59418.830   3.0 (4F)6p 5F    7.59 -4.10 -7.19
273.7009 -2.245  ...     59636.360   3.0 5d 5F5D3G   23110.939   3.0 5Dsp3P z7F   7.73 -3.56 -7.25
274.2011 -1.576  ...     26224.969   3.0 a2D)4s a3D  62683.770   4.0 8p 3G5G5F    8.29 -3.24 -7.02
274.4779 -2.018 -0.10   59532.970   4.0 s6d 3+[4+]  23110.939   3.0 5Dsp3P z7F   7.67 -3.91 -7.15
274.7678 -1.244 -0.10   24772.018   1.0 (2P)4s c3P  61155.620   1.0 3Psp1P 3P    8.52 -4.00 -7.19
```

```
275.5162 -2.646  ...   26224.969  3.0 a2D)4s a3D  62509.750  3.0 8p 3D3G3F  8.28 -3.79 -7.03
275.7857 -0.967 +0.20  24335.766  2.0 (2P)4s c3P  60585.090  2.0 3Psp1P 3P  8.58 -4.76 -7.25
275.8085 -2.919  ...   24338.767  3.0 4s2 b3G     60585.090  2.0 3Psp1P 3P  8.58 -4.76 -7.25
276.1600 -2.910  ...   59196.870  3.0 (4F)5d 5F   22996.674  4.0 5Dsp3P z7F  7.63 -3.95 -7.28
277.2028 -0.917 -0.10  25091.599  0.0 (2P)4s c3P  61155.620  1.0 3Psp1P 3P  8.52 -4.00 -7.19
277.3885 -1.981  ...   24335.766  2.0 (2P)4s c3P  60375.650  1.0 (2F)4p 3D  8.52 -4.33 -7.37
277.5345 -2.564  ...   23783.619  5.0 4s2 b3G     59804.540  4.0 4s6D7p 5F  7.10 -4.56 -7.07
277.6852 -2.448  ...   26105.908  6.0 (2H)4s b3H  62107.290  5.0 3Gsp1P 3G  8.33 -2.99 -7.04
277.8227 -2.380  ...   25091.599  0.0 (2P)4s c3P  61075.160  1.0 3Fsp1P 3D  8.23 -2.91 -7.06
278.1001 -1.015 -0.40  23783.619  5.0 4s2 b3G     59731.290  4.0 (4F)6p 3G  8.03 -3.71 -7.26
278.1818 -2.482  ...   24118.819  4.0 4s2 b3G     60055.930  3.0 4s6D7p 5F  7.21 -4.07 -7.07
278.4535 -2.126  ...   24335.766  2.0 (2P)4s c3P  60237.810  2.0 4s6D7p 5D  7.53 -3.90 -7.06
278.4768 -2.630  ...   24338.767  3.0 4s2 b3G     60237.810  2.0 4s6D7p 5D  7.53 -3.90 -7.06
278.5124 -0.849 -0.70  24118.819  4.0 4s2 b3G     60013.270  3.0 (4F)6p 3G  8.00 -3.41 -7.26
278.5779 -1.683 +0.00  26623.735  2.0 a2D)4s a3D  62509.750  3.0 8p 3D3G3F  8.28 -3.79 -7.03
279.0795 -2.199  ...   59532.970  4.0 s6d 3+[4+]  23711.456  4.0 5Dsp3P z7P  7.67 -3.91 -7.15
279.1453 -0.552  ...   24772.018  1.0 (2P)4s c3P  60585.090  2.0 3Psp1P 3P  8.58 -4.76 -7.25
279.1576 -2.773  ...   23783.619  5.0 4s2 b3G     59595.120  4.0 4s6D7p 7F  6.88 -4.11 -7.10
279.3095 -2.677  ...   26224.969  3.0 a2D)4s a3D  62016.990  3.0 (4F)7p 3G  8.12 -3.24 -7.09
279.5825 -1.901  ...   24118.819  4.0 4s2 b3G     59875.890  3.0 4s6D7p 5D  7.67 -3.29 -7.13
279.5890 -1.316  ...   26351.040  5.0 (2H)4s b3H  62107.290  5.0 3Gsp1P 3G  8.33 -2.99 -7.04
279.9276 -2.397  ...   23783.619  5.0 4s2 b3G     59496.620  4.0 4s6D7p 5D  7.42 -3.41 -7.08
280.1416 -2.539  ...   24118.819  4.0 4s2 b3G     59804.540  4.0 4s6D7p 5F  7.10 -4.56 -7.07
280.2177 -2.126  ...   24118.819  4.0 4s2 b3G     59794.850  3.0 3Dsp3P 1F  7.86 -4.18 -7.25
280.2297 -1.216  ...   24338.767  3.0 4s2 b3G     60013.270  3.0 (4F)6p 3G  8.00 -3.41 -7.26
280.4893 -1.346  ...   26224.969  3.0 a2D)4s a3D  61866.450  2.0 (4F)7p 3D  7.75 -3.45 -7.06
280.5999 -2.876 +0.20  19552.478  1.0 4s2 a3P     55179.910  0.0 (2P)4p 1S  8.38 -5.99 -7.74
280.7178 -0.967 -0.10  24118.819  4.0 4s2 b3G     59731.290  4.0 (4F)6p 3G  8.03 -3.71 -7.26
280.7875 -1.112  ...   24772.018  1.0 (2P)4s c3P  60375.650  1.0 (2F)4p 3D  8.52 -4.33 -7.37
280.8660 -1.840  ...   23783.619  5.0 4s2 b3G     59377.300  4.0 (4F)6p 5G  7.82 -3.96 -7.22
280.9145 -2.553  ...   26105.908  6.0 (2H)4s b3H  61693.440  5.0 (4F)7p 5G  7.69 -3.92 -7.08
280.9509 -2.915  ...   59294.380  3.0 4s4D4d 3G   23711.456  4.0 5Dsp3P z7P  8.29 -3.62 -7.37
281.0261 -0.635 -0.30  23783.619  5.0 4s2 b3G     59537.030  5.0 (4F)6p 3G  8.26 -4.27 -7.28
281.0993 -2.578  ...   24772.018  1.0 (2P)4s c3P  60336.160  1.0 4s6D7p 5F  7.42 -4.13 -7.07
281.2430 -2.044  ...   26224.969  3.0 a2D)4s a3D  61770.940  3.0 (4F)7p 5D  7.39 -2.63 -6.96
281.2893 -2.546 +0.20  24335.766  2.0 (2P)4s c3P  59875.890  3.0 4s6D7p 5D  7.67 -3.29 -7.13
281.3130 -2.644  ...   24338.767  3.0 4s2 b3G     59875.890  3.0 4s6D7p 5D  7.67 -3.29 -7.13
281.3358 -2.767  ...   23783.619  5.0 4s2 b3G     59317.860  4.0 4s6D7p 7D  7.22 -4.13 -7.11
281.7685 -2.756  ...   26627.609  4.0 (2H)4s b3H  62107.290  5.0 3Gsp1P 3G  8.33 -2.99 -7.04
281.7954 -2.590  ...   24118.819  4.0 4s2 b3G     59595.120  4.0 4s6D7p 7F  6.88 -4.11 -7.10
281.8021 -2.950  ...   64531.780  4.0 4s3H5s 5H   29056.324  3.0 5Dsp3P z5P  8.21 -4.00 -7.14
281.9322 -2.695  ...   24335.766  2.0 (2P)4s c3P  59794.850  3.0 3Dsp3P 1F  7.86 -4.18 -7.25
281.9560 -2.295 -0.40  24338.767  3.0 4s2 b3G     59794.850  3.0 3Dsp3P 1F  7.86 -4.18 -7.25
281.9607 -2.954  ...   59636.360  3.0 5d 5F5D3G   24180.862  3.0 5Dsp3P z7P  7.73 -3.56 -7.25
281.9783 -1.999  ...   26224.969  3.0 a2D)4s a3D  61678.260  4.0 (4F)7p 5F  7.62 -3.54 -7.02
282.0950 -1.946 -0.60  24574.655  4.0 (2G)4s a1G  60013.270  3.0 (4F)6p 3G  8.00 -3.41 -7.26
282.2168 -2.284 +0.10  26224.969  3.0 a2D)4s a3D  61648.300  4.0 (4F)7p 3G  7.95 -4.20 -7.12
282.4566 -2.806  ...   26623.735  2.0 a2D)4s a3D  62016.990  3.0 (4F)7p 3G  8.12 -3.24 -7.09
282.4624 -2.246  ...   24338.767  3.0 4s2 b3G     59731.290  4.0 (4F)6p 3G  8.03 -3.71 -7.26
282.5258 -2.314  ...   24118.819  4.0 4s2 b3G     59503.400  3.0 6p 5G5D3D  8.15 -3.43 -7.17
282.5800 -2.187  ...   24118.819  4.0 4s2 b3G     59496.620  4.0 4s6D7p 5D  7.42 -3.41 -7.08
282.7854 -2.153  ...   59532.970  4.0 s6d 3+[4+]  24180.862  3.0 5Dsp3P z7P  7.67 -3.91 -7.15
282.8630 -1.706  ...   26351.040  5.0 (2H)4s b3H  61693.440  5.0 (4F)7p 5G  7.69 -3.92 -7.08
283.1929 -1.202 -1.10  24574.655  4.0 (2G)4s a1G  59875.890  3.0 4s6D7p 5D  7.67 -3.29 -7.13
283.2149 -2.662  ...   61198.480  5.0 (2G)5s 3G   25899.989  4.0 5Dsp3P z5D  7.76 -3.84 -7.13
283.2248 -2.126  ...   26351.040  5.0 (2H)4s b3H  61648.300  4.0 (4F)7p 3G  7.95 -4.20 -7.12
283.3308 -1.190 -0.50  25091.599  0.0 (2P)4s c3P  60375.650  1.0 (2F)4p 3D  8.52 -4.33 -7.37
283.5363 -1.405 -0.50  24118.819  4.0 4s2 b3G     59377.300  4.0 (4F)6p 5G  7.82 -3.96 -7.22
283.6483 -2.863  ...   25091.599  0.0 (2P)4s c3P  60336.160  1.0 4s6D7p 5F  7.42 -4.13 -7.07
283.6994 -2.207 +0.00  24118.819  4.0 4s2 b3G     59357.030  5.0 (4F)6p 3G  8.26 -4.27 -7.28
283.7036 -1.258 -0.20  23783.619  5.0 4s2 b3G     59021.310  5.0 (4F)6p 5G  7.98 -4.37 -7.26
283.8445 -0.807 +0.10  24574.655  4.0 (2G)4s a1G  59794.850  3.0 3Dsp3P 1F  7.86 -4.18 -7.25
284.0151 -2.387 +0.40  24118.819  4.0 4s2 b3G     59317.860  4.0 4s6D7p 7D  7.22 -4.13 -7.11
284.2688 -1.193  ...   24335.766  2.0 (2P)4s c3P  59503.400  3.0 6p 5G5D3D  8.15 -3.43 -7.17
284.2931 -2.373  ...   24338.767  3.0 4s2 b3G     59503.400  3.0 6p 5G5D3D  8.15 -3.43 -7.17
284.3577 -1.884  ...   24574.655  4.0 (2G)4s a1G  59731.290  4.0 (4F)6p 3G  8.03 -3.71 -7.26
284.6002 -2.057  ...   26224.969  3.0 a2D)4s a3D  61351.660  3.0 (4F)7p 3D  7.90 -4.31 -7.07
284.9541 -1.934  ...   24335.766  2.0 (2P)4s c3P  59418.830  3.0 (4F)6p 5F  7.59 -4.10 -7.19
```

```
285.2176 -2.432  ...    26627.609  4.0 (2H)4s b3H  61678.260  4.0 (4F)7p 5F  7.62 -3.54 -7.02
285.3162 -2.659  ...    24338.767  3.0 4s2 b3G     59377.300  4.0 (4F)6p 5G  7.82 -3.96 -7.22
285.3474 -1.599  ...    26105.908  6.0 (2H)4s b3H  61140.620  5.0 (4F)7p 3G  7.98 -3.98 -7.12
285.4616 -1.398 -0.40   26627.609  4.0 (2H)4s b3H  61648.300  4.0 (4F)7p 3G  7.95 -4.20 -7.12
286.0703 -2.652 -0.10   23783.619  5.0 4s2 b3G     58729.800  4.0 (4F)6p 5D  7.68 -4.39 -7.23
286.1944 -2.820  ...    24772.018  1.0 (2P)4s c3P  59703.050  1.0 (4F)5p 5D  7.40 -3.86 -7.16
286.2131 -2.135 -0.60   24574.655  3.0 (2G)4s a1G  59503.400  3.0 6p 5G5D3D  8.15 -3.43 -7.17
286.4106 -2.144  ...    58616.110  3.0 (4F)5d 5P   23711.456  4.0 5Dsp3P z7P  8.00 -4.14 -7.38
286.4284 -2.791  ...    24118.819  4.0 4s2 b3G     59021.310  5.0 (4F)6p 5G  7.98 -4.37 -7.26
286.5440 -1.503 -0.20   26224.969  3.0 a2D)4s a3D  61113.380  4.0 (4F)7p 3F  7.84 -3.82 -7.11
287.2315 -2.307  ...    26351.040  5.0 (2H)4s b3H  61155.950  5.0 (4F)6p 5D  7.07 -4.17 -7.06
287.2502 -2.315  ...    24574.655  4.0 (2G)4s a1G  59377.300  4.0 (4F)6p 5G  7.82 -3.96 -7.22
287.3581 -0.669 -0.70   26351.040  5.0 (2H)4s b3H  61140.620  5.0 (4F)7p 3G  7.98 -3.98 -7.12
287.4176 -2.857  ...    24574.655  4.0 (2G)4s a1G  59357.030  5.0 (4F)6p 3G  8.26 -4.27 -7.28
287.5832 -2.331  ...    26351.040  5.0 (2H)4s b3H  61113.380  4.0 (4F)7p 3F  7.84 -3.82 -7.11
287.6924 -1.230 -0.60   26406.465  1.0 a2D)4s a3D  61155.620  1.0 3Psp1P 3P  8.52 -4.00 -7.19
287.8682 -2.649  ...    26623.735  2.0 a2D)4s a3D  61351.660  3.0 (4F)7p 5D  7.90 -4.31 -7.07
288.8370 -2.835 +00     25091.599  0.0 (2P)4s c3P  59703.050  1.0 (4F)5p 5D  7.40 -3.86 -7.16
288.8409 -2.699 -0.10   24118.819  4.0 4s2 b3G     58729.800  4.0 (4F)6p 5D  7.68 -4.39 -7.23
288.9432 -2.640  ...    58779.590  2.0 4s4D4d 3D   24180.862  3.0 5Dsp3P z7P  8.14 -4.56 -7.43
289.5026 -0.545  ...    26623.735  2.0 a2D)4s a3D  61155.620  1.0 3Psp1P 3P  8.52 -4.00 -7.19
289.6609 -2.062  ...    26627.609  4.0 (2H)4s b3H  61140.620  5.0 (4F)7p 3G  7.98 -3.98 -7.12
289.8897 -1.487 -0.40   26627.609  4.0 (2H)4s b3H  61113.380  4.0 (4F)7p 3F  7.84 -3.82 -7.11
290.1787 -1.534  ...    26623.735  2.0 a2D)4s a3D  61075.160  1.0 3Fsp1P 3D  8.23 -2.91 -7.06
290.9498 -0.312 -0.30   26224.969  3.0 a2D)4s a3D  60585.090  2.0 3Dsp1P 3P  8.58 -4.76 -7.25
292.4949 -2.086 -0.40   26406.465  1.0 a2D)4s a3D  60585.090  2.0 3Dsp1P 3P  8.58 -4.76 -7.25
293.0903 -2.439  ...    58616.110  3.0 (4F)5d 5P   24506.917  2.0 5Dsp3P z7P  8.00 -4.14 -7.38
293.7581 -2.509 +0.20   61198.480  5.0 (2G)5s 3G   27166.820  4.0 5Dsp3P z5F  7.78 -3.84 -7.13
293.9207 -2.147  ...    26224.969  3.0 a2D)4s a3D  60237.810  2.0 4s6D7p 5D  7.53 -3.90 -7.06
294.2984 -1.580  ...    26406.465  1.0 a2D)4s a3D  60375.650  1.0 (2F)4p 3D  8.52 -4.33 -7.37
294.3663 -2.044  ...    26623.735  2.0 a2D)4s a3D  60585.090  2.0 3Dsp1P 3P  8.58 -4.76 -7.25
294.6410 -2.939 +0.20   26406.465  1.0 a2D)4s a3D  60336.160  1.0 4s6D7p 5F  7.42 -4.13 -7.07
296.1930 -2.658  ...    26623.735  2.0 a2D)4s a3D  60375.650  1.0 (2F)4p 3D  8.52 -4.33 -7.37
296.2312 -1.994 -0.30   60087.260  2.0 4s4D4d 3P   26339.696  2.0 5Dsp3P z5D  7.92 -3.79 -7.18
296.3294 -2.921  ...    59636.360  3.0 5d 5F5D3G   25899.989  4.0 5Dsp3P z5D  7.84 -3.56 -7.25
297.0820 -2.760  ...    26224.969  3.0 a2D)4s a3D  59875.890  3.0 4s6D7p 5D  7.67 -3.29 -7.13
297.4076 -2.646  ...    26623.735  2.0 a2D)4s a3D  60237.810  2.0 4s6D7p 5D  7.53 -3.90 -7.06
297.4205 -1.637 -0.30   27543.003  1.0 (2P)4s a1P  61155.620  1.0 3Psp1P 3P  8.52 -4.00 -7.19
297.4624 -1.881 -0.30   60087.260  2.0 4s4D4d 3P   26479.361  1.0 5Dsp3P z5D  7.92 -3.79 -7.18
297.7992 -2.692  ...    26224.969  3.0 a2D)4s a3D  59794.850  3.0 3Dsp3P 1F  7.86 -4.18 -7.25
298.4544 -1.903  ...    59636.360  3.0 5d 5F5D3G   26140.179  3.0 5Dsp3P z5D  7.84 -3.56 -7.25
299.3642 -2.041 -0.40   59294.380  3.0 4s4D4d 3G   25899.989  4.0 5Dsp3P z5D  8.32 -3.62 -7.37
299.3785 -0.774 -0.40   59532.970  4.0 s6d 3+[4+]  26140.179  3.0 5Dsp3P z5D  7.79 -3.91 -7.15
299.4425 -2.233  ...    26627.609  4.0 (2H)4s b3H  60013.270  3.0 (4F)6p 3G  8.00 -3.41 -7.26
299.4910 -1.725 +00     26351.040  5.0 (2H)4s b3H  59731.290  4.0 (4F)6p 3G  8.03 -3.71 -7.26
300.1050 -2.527  ...    29371.812  3.0 4s2 b3D     62683.770  4.0 8p 3G5G5F  8.29 -3.24 -7.02
300.2409 -1.917  ...    59196.870  3.0 (4F)5d 5F   25899.989  4.0 5Dsp3P z5D  7.77 -3.95 -7.28
300.2429 -2.809  ...    59636.360  3.0 5d 5F5D3G   26339.696  2.0 5Dsp3P z5D  7.84 -3.56 -7.25
300.4074 -1.617  ...    26224.969  3.0 a2D)4s a3D  59503.400  3.0 6p 5G5D3D  8.15 -3.43 -7.17
300.6448 -2.616  ...    26623.735  2.0 a2D)4s a3D  59875.890  3.0 4s6D7p 5D  7.67 -3.29 -7.13
300.6541 -1.098 -0.40   26105.908  6.0 (2H)4s b3H  59357.030  5.0 (4F)6p 3G  8.26 -4.27 -7.28
301.1728 -2.000  ...    26224.969  3.0 a2D)4s a3D  59418.830  3.0 (4F)6p 5F  7.59 -4.10 -7.19
301.4145 -2.409  ...    26627.609  4.0 (2H)4s b3H  59794.850  3.0 3Dsp3P 1F  7.86 -4.18 -7.25
301.5331 -2.107 -0.40   59294.380  3.0 4s4D4d 3G   26140.179  3.0 5Dsp3P z5D  8.32 -3.62 -7.37
301.5439 -2.563  ...    29356.744  2.0 4s2 b3D     62509.750  3.0 8p 3D3G3F  8.28 -3.79 -7.03
301.6115 -2.754  ...    26351.040  5.0 (2H)4s b3H  59496.620  4.0 4s6D7p 5D  7.42 -3.41 -7.08
301.9933 -1.537 -0.20   26627.609  4.0 (2H)4s b3H  59731.290  4.0 (4F)6p 3G  8.03 -3.71 -7.26
302.4226 -1.836 -0.30   59196.870  3.0 (4F)5d 5F   26140.179  3.0 5Dsp3P z5D  7.76 -3.95 -7.28
302.6936 -2.175 +0.10   59366.790  2.0 (4F)5d 5F   26339.696  2.0 5Dsp3P z5D  7.72 -3.98 -7.25
302.7012 -1.789 -0.30   26351.040  5.0 (2H)4s b3H  59377.300  4.0 (4F)6p 5G  7.82 -3.96 -7.22
302.8871 -1.404 -0.30   26351.040  5.0 (2H)4s b3H  59357.030  5.0 (4F)6p 3G  8.26 -4.27 -7.28
303.2470 -2.649 +0.30   26351.040  5.0 (2H)4s b3H  59317.860  4.0 4s6D7p 7D  7.22 -4.13 -7.11
303.3020 -2.491  ...    59300.540  1.0 (4P)5s 3P   26339.696  2.0 5Dsp3P z5D  8.17 -4.12 -7.47
303.7207 -1.431  ...    26105.908  6.0 (2H)4s b3H  59021.310  5.0 (4F)6p 5G  7.98 -4.37 -7.26
303.9793 -2.411  ...    59366.790  2.0 (4F)5d 5F   26479.381  1.0 5Dsp3P z5D  7.71 -3.98 -7.25
304.0031 -2.964  ...    29798.846  4.0 4s2 b1G     62683.770  4.0 8p 3G5G5F  8.29 -3.24 -7.02
304.0509 -1.647  ...    26623.735  2.0 a2D)4s a3D  59503.400  3.0 6p 5G5D3D  8.15 -3.43 -7.17
304.0867 -2.797  ...    26627.609  4.0 (2H)4s b3H  59503.400  3.0 6p 5G5D3D  8.15 -3.43 -7.17
```

```
304.1494 -2.751 +0.30  26627.609  4.0 (2H)4s b3H   59496.620  4.0 4s6D7p 5D    7.42 -3.41 -7.08
304.2590 -2.255  ...    59196.870  3.0 (4F)5d 5F    26339.696  2.0 5Dsp3P z5D   7.76 -3.95 -7.28
304.8350 -2.168 -0.10  26623.735  2.0 a2D)4s a3D   59418.830  3.0 (4F)6p 5F    7.59 -4.10 -7.19
305.2576 -2.057 -0.20  26627.609  4.0 (2H)4s b3H   59377.300  4.0 (4F)6p 5G    7.82 -3.96 -7.22
305.4467 -2.671  ...    26627.609  4.0 (2H)4s b3H   59357.030  5.0 (4F)6p 3G    8.26 -4.27 -7.28
305.5709 -1.147  ...    58616.110  3.0 (4F)5d 5P    25899.989  4.0 5Dsp3P z5D   8.06 -4.14 -7.38
305.7910 -2.678  ...    60087.260  3.0 4s4D4d 3P    27394.691  3.0 5Dsp3P z5F   7.93 -3.79 -7.18
305.9997 -1.952 -0.20  26351.040  5.0 (2H)4s b3H   59021.310  5.0 (4F)6p 5G    7.98 -4.37 -7.26
306.1108 -2.888  ...    59532.970  4.0 s6d 3+[4+]  26874.550  5.0 5Dsp3P z5F   7.82 -3.91 -7.15
306.2890 -1.753  ...    58779.590  2.0 4s4D4d 3D    26140.179  3.0 5Dsp3P z5D   8.19 -4.56 -7.43
307.1209 -2.204  ...    28604.613  2.0 a2D)4s a1D   61155.620  1.0 3Psp1P 3P    8.52 -4.00 -7.19
307.5111 -2.544  ...    29356.744  2.0 4s2 b3D      61866.450  2.0 (4F)7p 3D    7.75 -3.45 -7.06
307.5572 -2.176  ...    26224.969  3.0 a2D)4s a3D   58729.800  4.0 (4F)6p 5D    7.68 -4.39 -7.23
307.6537 -2.441 -0.30  29371.812  3.0 4s2 b3D      61866.450  2.0 (4F)7p 3D    7.75 -3.45 -7.06
307.8309 -2.615  ...    58616.110  3.0 (4F)5d 5P    26140.179  3.0 5Dsp3P z5D   8.06 -4.14 -7.38
307.8820 -2.974 +0.70  28604.613  2.0 a2D)4s a1D   61075.160  1.0 3Fsp1P 3D    8.23 -2.91 -7.06
307.8915 -2.450  ...    59636.360  3.0 5d 5F5D3G    27166.820  4.0 5Dsp3P z5F   7.86 -3.56 -7.25
308.1729 -2.347  ...    58779.590  2.0 4s4D4d 3D    26339.696  2.0 5Dsp3P z5D   8.19 -4.56 -7.43
308.3533 -1.446 -0.70  60087.260  3.0 4s4D4d 3P    27666.348  1.0 5Dsp3P z5F   7.93 -3.79 -7.18
308.7548 -2.668 +0.40  26351.040  5.0 (2H)4s b3H   58729.800  4.0 (4F)6p 5D    7.68 -4.39 -7.23
308.8751 -1.049 -0.20  59532.970  4.0 s6d 3+[4+]   27166.820  4.0 5Dsp3P z5F   7.81 -3.91 -7.15
309.3098 -2.691  ...    28819.954  5.0 (2H)4s a1H   61140.620  5.0 (4F)7p 3G    7.98 -3.98 -7.12
309.6159 -1.611 +0.00  58628.410  1.0 4s4D4d 5P    26339.696  2.0 5Dsp3P z5D   8.14 -5.24 -7.33
309.7339 -1.858 -0.30  58616.110  3.0 (4F)5d 5P    26339.696  2.0 5Dsp3P z5D   8.06 -4.14 -7.38
310.0677 -1.330  ...    59636.360  3.0 5d 5F5D3G    27394.691  3.0 5Dsp3P z5F   7.86 -3.56 -7.25
310.1502 -1.933  ...    22946.816  1.0 (4P)4s b3P   55179.910  0.0 (2P)4p 1S    8.38 -5.99 -7.74
310.9612 -1.787 +0.00  58628.410  1.0 4s4D4d 5P    26479.381  1.0 5Dsp3P z5D   8.14 -5.24 -7.33
311.0652 -1.326  ...    59532.970  4.0 s6d 3+[4+]   27394.691  3.0 5Dsp3P z5F   7.81 -3.91 -7.15
311.1690 -2.592  ...    59294.380  3.0 4s4D4d 3G    27166.820  4.0 5Dsp3P z5F   8.33 -3.62 -7.37
311.6504 -0.758 -0.70  58628.410  1.0 4s4D4d 5P    26550.479  0.0 5Dsp3P z5D   8.14 -5.24 -7.33
311.6616 -2.537  ...    59636.360  3.0 5d 5F5D3G    27559.583  2.0 5Dsp3P z5F   7.85 -3.56 -7.25
312.1163 -1.956  ...    59196.870  3.0 (4F)5d 5F    27166.820  4.0 5Dsp3P z5F   7.79 -3.95 -7.28
312.6821 -1.897  ...    59366.790  2.0 (4F)5d 5F    27394.691  3.0 5Dsp3P z5F   7.74 -3.98 -7.25
312.9102 -0.890 -0.30  58428.170  0.0 4s6D4d 5D    26479.381  1.0 5Dsp3P z5D   8.22 -5.07 -7.40
313.3919 -2.215  ...    59294.380  3.0 4s4D4d 3G    27394.691  3.0 5Dsp3P z5F   8.33 -3.62 -7.37
314.3031 -1.912  ...    59366.790  2.0 (4F)5d 5F    27559.583  2.0 5Dsp3P z5F   7.74 -3.98 -7.25
314.3528 -1.626  ...    59196.870  3.0 (4F)5d 5F    27394.691  3.0 5Dsp3P z5F   7.78 -3.95 -7.28
314.3855 -2.366  ...    29356.744  2.0 4s2 b3D      61155.620  1.0 3Psp1P 3P    8.52 -4.00 -7.19
314.6609 -1.465  ...    28604.613  2.0 a2D)4s a1D   60375.650  1.0 (2F)4p 3D    8.52 -4.33 -7.37
314.8185 -2.756 +0.70  29320.026  1.0 4s2 b3D      61075.160  1.0 3Fsp1P 3D    8.23 -2.91 -7.06
314.9531 -2.293  ...    29371.812  3.0 4s2 b3D      61113.380  4.0 (4F)7p 3F    7.84 -3.82 -7.11
315.0203 -1.992  ...    59294.380  3.0 4s4D4d 3G    27559.583  2.0 5Dsp3P z5F   8.32 -3.62 -7.37
315.0526 -2.666  ...    28604.613  2.0 a2D)4s a1D   60336.160  1.0 4s6D7p 5F    7.42 -4.13 -7.07
315.1830 -2.541 +0.70  29356.744  2.0 4s2 b3D      61075.160  1.0 3Fsp1P 3D    8.23 -2.91 -7.06
316.7178 -2.743  ...    28604.613  2.0 a2D)4s a1D   60169.330  1.0 4s4F7p 5D    7.56 -3.76 -7.07
317.8803 -2.134 -0.40  58616.110  3.0 (4F)5d 5P    27166.820  4.0 5Dsp3P z5F   8.07 -4.14 -7.38
318.2915 -2.489  ...    28604.613  2.0 a2D)4s a1D   60013.270  3.0 (4F)6p 3G    8.00 -3.41 -7.26
319.6899 -1.729  ...    28604.613  2.0 a2D)4s a1D   59875.890  3.0 4s6D7p 5D    7.67 -3.29 -7.13
319.7534 -2.886  ...    29320.026  1.0 4s2 b3D      60585.090  2.0 3Psp1P 3P    8.58 -4.76 -7.25
320.1294 -1.911 -0.20  29356.744  2.0 4s2 b3D      60585.090  2.0 3Psp1P 3P    8.58 -4.76 -7.25
320.2004 -2.229 -0.40  58616.110  3.0 (4F)5d 5P    27394.691  3.0 5Dsp3P z5F   8.07 -4.14 -7.38
320.2149 -2.756  ...    58779.590  2.0 4s4D4d 3D    27559.583  2.0 5Dsp3P z5F   8.19 -4.56 -7.43
320.2839 -1.907  ...    29371.812  3.0 4s2 b3D      60585.090  2.0 3Psp1P 3P    8.58 -4.76 -7.25
320.5205 -1.360  ...    28604.613  2.0 a2D)4s a1D   59794.850  3.0 3Dsp3P 1F    7.86 -4.18 -7.25
321.1997 -0.705 +0.00  51143.920  0.0 4s6D4d e7F   20019.635  1.0 5Dsp3P z7D   8.31 -5.29 -7.55
321.2171 -2.436  ...    19552.478  1.0 4s2 a3P      50675.080  1.0 3Psp1P 1P    7.27 -6.11 -7.81
321.3137 -2.157  ...    58779.590  2.0 4s4D4d 3D    27666.348  1.0 5Dsp3P z5F   8.19 -4.56 -7.43
321.9005 -2.572  ...    58616.110  3.0 (4F)5d 5P    27559.583  2.0 5Dsp3P z5F   8.07 -4.14 -7.38
321.9099 -1.543 -0.40  29320.026  1.0 4s2 b3D      60375.650  1.0 (2F)4p 3D    8.52 -4.33 -7.37
322.8827 -1.411  ...    58628.410  1.0 4s4D4d 5P    27666.348  1.0 5Dsp3P z5F   8.15 -5.24 -7.33
323.5439 -2.750  ...    28604.613  2.0 a2D)4s a1D   59503.400  3.0 6p 5G5D3D    8.15 -3.43 -7.17
324.9845 -2.012 -0.20  58428.170  0.0 4s6D4d 5D    27666.348  1.0 5Dsp3P z5F   8.23 -5.07 -7.40
326.1008 -2.357  ...    29356.744  2.0 4s2 b3D      60013.270  3.0 (4F)6p 3G    8.00 -3.41 -7.26
326.5086 -2.406  ...    60087.260  3.0 4s4D4d 3P    29469.024  2.0 5Dsp3P z5P   7.96 -3.79 -7.18
326.6486 -2.507  ...    64300.510  6.0 4s3H5s 5H    33695.397  5.0 (4F)4p y5F   8.47 -4.39 -7.39
327.5688 -2.728  ...    29356.744  2.0 4s2 b3D      59875.890  3.0 4s6D7p 5D    7.67 -3.29 -7.13
327.7306 -2.596  ...    29371.812  3.0 4s2 b3D      59875.890  3.0 4s6D7p 5D    7.67 -3.29 -7.13
328.0256 -1.745  ...    59532.970  4.0 s6d 3+[4+]   29056.324  3.0 5Dsp3P z5P   7.84 -3.91 -7.15
```

```
328.6036 -2.267  ...   29371.812  3.0 4s2 b3D    59794.850  3.0 3Dsp3P 1F   7.86 -4.18 -7.25
328.7673 -2.314  ...   24772.018  1.0 (2P)4s c3P  55179.910  0.0 (2P)4p 1S   8.38 -5.99 -7.74
329.2916 -2.392  ...   29371.812  3.0 4s2 b3D    59731.290  4.0 (4F)6p 3G   8.03 -3.71 -7.26
329.3453 -2.327  ...   60087.260  2.0 4s4d4d 3P  29732.736  1.0 5Dsp3P z5P  7.96 -3.79 -7.18
329.8241 -2.507  ...   59366.790  2.0 (4F)5d 5F  29056.324  3.0 5Dsp3P z5P  7.77 -3.98 -7.25
330.6139 -2.035  ...   59294.380  3.0 4s4D4d 3G  29056.324  3.0 5Dsp3P z5P  8.33 -3.62 -7.37
330.8735 -2.744  ...   29798.936  4.0 4s2 b1G    60013.270  3.0 (4F)6p 3G   8.00 -3.41 -7.26
330.9974 -2.437  ...   64531.780  4.0 4s3H5s 5H  34328.752  3.0 (4F)4p y5F  8.45 -4.00 -7.14
331.6163 -2.221  ...   29356.744  2.0 4s2 b3D    59503.400  3.0 6p 5G5D3D   8.15 -3.43 -7.17
331.7822 -1.895 -0.40  29371.812  3.0 4s2 b3D    59503.400  3.0 6p 5G5D3D   8.15 -3.43 -7.17
332.3848 -2.449  ...   29798.936  4.0 4s2 b1G    59875.890  3.0 4s6D7p 5D   7.67 -3.29 -7.13
332.7160 -2.548  ...   29371.812  3.0 4s2 b3D    59418.830  3.0 (4F)6p 5F   7.59 -4.10 -7.19
333.2829 -2.201  ...   29798.936  4.0 4s2 b1G    59794.850  3.0 3Dsp3P 1F   7.86 -4.18 -7.25
334.3770 -2.768  ...   59366.790  2.0 (4F)5d 5F  29469.024  2.0 5Dsp3P z5P  7.78 -3.98 -7.25
335.3600 -2.806  ...   32873.632  4.0 3d8 c3F    62683.770  4.0 8p 3G5G5F   8.29 -3.24 -7.02
336.3401 -1.592  ...   58779.590  2.0 4s4D4d 3D  29056.324  3.0 5Dsp3P z5P  8.21 -4.56 -7.43
337.3292 -2.019  ...   32873.632  4.0 3d8 c3F    62509.750  3.0 8p 3D3G3F   8.28 -3.79 -7.03
338.2003 -1.878 -0.30  58616.110  3.0 (4F)5d 5P  29056.324  3.0 5Dsp3P z5P  8.09 -4.14 -7.38
338.6781 -1.157 -0.70  64300.510  6.0 4s3H5s 5H  34782.421  5.0 (4F)4p z5G  8.41 -4.39 -7.39
339.3856 -1.879  ...   64300.510  6.0 4s3H5s 5H  34843.957  6.0 (4F)4p z5G  8.44 -4.39 -7.39
341.4968 -1.952  ...   64531.780  4.0 4s3H5s 5H  35257.324  4.0 (4F)4p z5G  8.40 -4.00 -7.14
341.9734 -2.447  ...   32873.632  4.0 3d8 c3F    62107.290  5.0 3Gsp1P 3G   8.33 -2.99 -7.04
342.8445 -1.443 -0.10  58628.410  1.0 4s4D4d 5P  29469.024  2.0 5Dsp3P z5P  8.17 -5.24 -7.33
342.9891 -1.698 -0.60  58616.110  3.0 (4F)5d 5P  29469.024  2.0 5Dsp3P z5P  8.09 -4.14 -7.38
343.0330 -2.820  ...   32873.632  4.0 3d8 c3F    62016.990  3.0 (4F)7p 3G   8.12 -3.24 -7.09
343.5792 -1.690 -0.20  33412.717  3.0 3d8 c3F    62509.750  3.0 8p 3D3G3F   8.28 -3.79 -7.03
344.1727 -1.835  ...   58779.590  2.0 4s4D4d 3D  29732.736  1.0 5Dsp3P z5P  8.21 -4.56 -7.43
345.6669 -1.057 -0.10  64300.510  6.0 4s3H5s 5H  35379.208  5.0 (4F)4p z5G  8.41 -4.39 -7.39
345.6806 -1.251 -0.30  64531.780  4.0 4s3H5s 5H  35611.625  4.0 (4F)4p z5G  8.40 -4.00 -7.14
345.9539 -2.153 +0.00  32873.632  4.0 3d8 c3F    61770.940  3.0 (4F)7p 5D   7.39 -2.63 -6.96
345.9735 -0.617 -0.40  58628.410  1.0 4s4D4d 5P  29732.736  1.0 5Dsp3P z5P  8.17 -5.24 -7.33
346.2233 -2.913  ...   33412.717  3.0 3d8 c3F    62287.540  3.0 (4F)7p 5G   7.78 -4.05 -7.00
347.0671 -1.790 -0.30  32873.632  4.0 3d8 c3F    61648.300  4.0 (4F)7p 5F   7.62 -3.54 -7.02
347.4284 -2.087  ...   32873.632  4.0 3d8 c3F    61648.300  4.0 (4F)7p 3G   7.95 -4.20 -7.12
347.4432 -0.936 -0.20  26406.465  1.0 a2D)4s a3D  55179.910  0.0 (2P)4p 1S  8.38 -5.99 -7.74
347.5495 -1.351  ...   60087.260  2.0 4s4D4d 3P  31322.613  3.0 5Dsp3P z3D  7.87 -3.79 -7.18
347.5547 -2.438  ...   64531.780  4.0 4s3H5s 5H  35767.564  4.0 (4F)4p z5G  8.37 -4.00 -7.14
348.3878 -1.096 +0.00  58428.170  4.0 4s6D4d 5D  29732.736  1.0 5Dsp3P z5P  8.24 -5.07 -7.40
349.4981 -2.698  ...   33412.717  3.0 3d8 c3F    62016.990  3.0 (4F)7p 3G   8.12 -3.24 -7.09
350.5034 -2.434  ...   33765.306  2.0 3d8 c3F    62287.540  3.0 (4F)7p 5G   7.78 -4.05 -7.00
351.0475 -2.566  ...   32873.632  4.0 3d8 c3F    61351.660  3.0 (4F)7p 3D   7.90 -4.31 -7.07
351.3473 -2.240  ...   33412.717  3.0 3d8 c3F    61686.450  2.0 (4F)7p 3D   7.75 -3.45 -7.06
351.3636 -2.712  ...   64531.780  4.0 4s3H5s 5H  36079.372  3.0 (4F)4p z3G  8.36 -4.00 -7.14
352.0008 -1.260  ...   60087.260  2.0 4s4D4d 3P  31686.351  2.0 5Dsp3P z3D  7.87 -3.79 -7.18
352.5306 -2.971  ...   33412.717  3.0 3d8 c3F    61770.940  3.0 (4F)7p 5D   7.39 -2.63 -6.96
353.0844 -2.935  ...   59636.360  3.0 5d 5F5D3G  31322.613  3.0 5Dsp3P z3D  7.78 -3.56 -7.25
353.4784 -2.842  ...   60087.260  2.0 4s4D4d 3P  31805.071  3.0 5Dsp3P z3F  7.84 -3.79 -7.18
353.8601 -2.280  ...   33765.306  2.0 3d8 c3F    62016.990  3.0 (4F)7p 3G   8.12 -3.24 -7.09
354.0096 -1.033  ...   32873.632  4.0 3d8 c3F    61113.380  4.0 (4F)7p 3F   7.84 -3.82 -7.11
354.0619 -2.777  ...   33412.717  3.0 3d8 c3F    61648.300  4.0 (4F)7p 3G   7.95 -4.20 -7.12
354.1855 -1.744 -0.40  59532.970  4.0 s6d 3+[4+]  31307.245  4.0 5Dsp3P z3F  7.68 -3.91 -7.15
355.1392 -1.734 -0.10  60087.260  2.0 4s4D4d 3P  31937.325  1.0 5Dsp3P z3D  7.87 -3.79 -7.18
355.7558 -2.077 -0.30  33765.306  2.0 3d8 c3F    61866.450  2.0 (4F)7p 3D   7.75 -3.45 -7.06
356.4785 -1.619 -0.10  59366.790  2.0 (4F)5d 5F  31322.613  3.0 5Dsp3P z3D  7.63 -3.98 -7.25
357.2050 -2.201  ...   59294.380  3.0 4s4D4d 3G  31307.245  4.0 5Dsp3P z3F  8.29 -3.62 -7.37
357.6378 -2.091  ...   60087.260  2.0 4s4D4d 3P  32133.991  2.0 5Dsp3P z3F  7.84 -3.79 -7.18
357.6795 -1.930  ...   59636.360  3.0 5d 5F5D3G  31686.351  2.0 5Dsp3P z3D  7.78 -3.56 -7.25
357.8212 -1.980  ...   33412.717  3.0 3d8 c3F    61351.660  3.0 (4F)7p 3D   7.90 -4.31 -7.07
358.3325 -0.987 -0.30  51143.920  0.0 4s6D4d e7F  23244.838  1.0 5Dsp3P 1F  8.31 -5.29 -7.55
358.6516 -1.022  ...   59196.870  3.0 (4F)5d 5F  31322.613  3.0 5Dsp3P z3D  7.69 -3.95 -7.28
359.1348 -2.013 -0.50  22838.323  2.0 (4P)4s b3P  50675.080  1.0 3Psp3P 1P  7.27 -6.11 -7.81
359.2887 -1.671 -0.10  54304.210  0.0 (4F)4d 5D  26479.381  1.0 5Dsp3P 5D  8.05 -4.57 -7.51
360.5400 -1.889  ...   22946.816  1.0 (4P)4s b3P  50675.080  1.0 3Psp3P 1P  7.27 -6.11 -7.81
360.5447 -1.583  ...   59532.970  4.0 s6d 3+[4+]  31805.071  3.0 5Dsp3P z3F  7.68 -3.91 -7.15
360.8992 -2.701  ...   33412.717  3.0 3d8 c3F    61113.380  4.0 (4F)7p 3F   7.84 -3.82 -7.11
361.1629 -1.559 -0.40  59366.790  2.0 (4F)5d 5F  31686.351  2.0 5Dsp3P z3D  7.63 -3.98 -7.25
361.7318 -0.493 -0.30  27543.003  1.0 (2P)4s a1P  55179.910  0.0 (2P)4p 1S  8.38 -5.99 -7.74
361.9096 -2.497 +0.00  23051.750  0.0 (4P)4s b3P  50675.080  1.0 3Psp3P 1P  7.27 -6.11 -7.81
```

```
362.1102 -0.275 -0.20  59294.380  3.0 4s4D4d 3G   31686.351  2.0 5Dsp3P z3D  8.30 -3.62 -7.37
363.3937 -1.948  ...   59196.870  3.0 (4F)5d 5F   31686.351  2.0 5Dsp3P z3D  7.69 -3.95 -7.28
363.5014 -0.593 -0.30  59636.360  3.0 5d 5F5D3G   32133.991  2.0 5Dsp3P z3F  7.74 -3.56 -7.25
363.6741 -0.172 -0.40  59294.380  3.0 4s4D4d 3G   31805.071  3.0 5Dsp3P z3D  8.29 -3.62 -7.37
364.1024 -0.641 -0.80  58779.590  2.0 4s4D4d 3D   31322.613  3.0 5Dsp3P z3D  8.16 -4.56 -7.43
364.9688 -1.804  ...   59196.870  3.0 (4F)5d 5F   31805.071  3.0 5Dsp3P z3D  7.65 -3.95 -7.28
365.3500 -1.356  ...   59300.540  1.0 (4P)5s 3P   31937.325  1.0 5Dsp3P z3D  8.14 -4.12 -7.47
366.0639 -2.412  ...   33765.306  2.0 3d8 c3F     61075.160  1.0 3Fsp1P 3D   8.23 -2.91 -7.06
366.0772 -1.063 -0.30  58616.110  3.0 (4F)5d 5P   31307.245  4.0 5Dsp3P z3F  8.01 -4.14 -7.38
366.2833 -0.422 -0.20  58616.110  3.0 (4F)5d 5P   31322.613  3.0 5Dsp3P z3D  8.02 -4.14 -7.38
367.0997 -2.588 +0.00  59366.790  2.0 (4F)5d 5F   32133.991  2.0 5Dsp3P z3F  7.59 -3.98 -7.25
367.9161 -1.956 -0.20  33412.717  3.0 3d8 c3F     60585.090  2.0 3Psp1P 3P   8.58 -4.76 -7.25
367.9950 -2.469  ...   59300.540  1.0 (4P)5s 3P   32133.991  2.0 5Dsp3P z3F  8.13 -4.12 -7.47
368.0784  0.098 -0.40  59294.380  3.0 4s4D4d 3G   32133.991  2.0 5Dsp3P z3F  8.29 -3.62 -7.37
368.0978 -2.839  ...   61198.480  5.0 (2G)5s 3G   34039.516  4.0 (4F)4p y5D  8.27 -3.84 -7.13
368.9907 -0.523  ...   58779.590  2.0 4s4D4d 3D   31686.351  2.0 5Dsp3P z3D  8.16 -4.56 -7.43
369.4047 -2.042  ...   59196.870  3.0 (4F)5d 5F   32133.991  2.0 5Dsp3P z3F  7.64 -3.95 -7.28
370.6148 -1.264  ...   58779.590  2.0 4s4D4d 3D   31805.071  3.0 5Dsp3P z3F  8.15 -4.56 -7.43
372.2274 -2.881  ...   32873.632  4.0 3d8 c3F     59731.290  4.0 (4F)6p 3G   8.03 -3.71 -7.26
372.4409 -2.019  ...   58779.590  2.0 4s4D4d 3D   31937.325  1.0 5Dsp3P z3D  8.16 -4.56 -7.43
372.7531 -2.473  ...   33765.306  2.0 3d8 c3F     60585.090  2.0 3Psp1P 3P   8.58 -4.76 -7.25
375.1899 -2.815  ...   58779.590  2.0 4s4D4d 3D   32133.991  2.0 5Dsp3P z3F  8.15 -4.56 -7.43
375.2988 -1.850 -0.20  54304.210  0.0 (4F)4d 5D   27666.348  1.0 5Dsp3P z5F  8.06 -4.57 -7.51
375.4129 -2.523 -0.40  32873.632  4.0 3d8 c3F     59503.400  3.0 6p 5G5D3D  8.15 -3.43 -7.17
375.5085 -2.967  ...   32873.632  4.0 3d8 c3F     59496.620  4.0 4s6D7p 5D   7.42 -3.41 -7.08
375.6869 -1.676 -0.40  33765.306  2.0 3d8 c3F     60375.650  1.0 (2F)4p 3D   8.52 -4.33 -7.37
376.6768 -2.677  ...   59636.360  3.0 5d 5F5D3G   33095.941  4.0 (4F)4p y5D  8.40 -3.56 -7.25
377.1991 -2.766  ...   32873.632  4.0 3d8 c3F     59377.300  4.0 (4F)6p 5G   7.82 -3.96 -7.22
377.4878 -1.608  ...   32873.632  4.0 3d8 c3F     59357.030  5.0 (4F)6p 3G   8.26 -4.27 -7.28
377.7763 -2.956  ...   33412.717  3.0 3d8 c3F     59875.890  3.0 4s6D7p 5D   7.67 -3.29 -7.13
379.5528 -1.665 -0.50  24335.766  2.0 (2P)4s c3P  50675.080  1.0 3Psp3P 1P   7.27 -6.11 -7.81
379.8520 -1.802  ...   33412.717  3.0 3d8 c3F     59731.290  4.0 (4F)6p 3G   8.03 -3.71 -7.26
380.3272 -1.872  ...   60087.260  2.0 4s4D4d 3P   33801.572  2.0 (4F)4p y5D  8.42 -3.79 -7.18
380.8738 -1.954  ...   33765.306  2.0 3d8 c3F     60013.270  3.0 (4F)6p 3G   8.00 -3.41 -7.26
382.3346 -2.001 -0.30  32873.632  4.0 3d8 c3F     59021.310  5.0 (4F)6p 5G   7.98 -4.37 -7.26
382.4422 -0.806  ...   60087.260  2.0 4s4D4d 3P   33946.933  2.0 5Dsp3P z3P  7.88 -3.79 -7.18
382.6045 -2.000  ...   59636.360  3.0 5d 5F5D3G   33507.123  3.0 (4F)4p y5D  8.40 -3.56 -7.25
384.1245 -2.267  ...   59532.970  4.0 s6d 3+[4+]  33507.123  3.0 (4F)4p y5D  8.39 -3.91 -7.15
384.4159 -2.058  ...   33412.717  3.0 3d8 c3F     59418.830  3.0 (4F)6p 5F   7.59 -4.10 -7.19
385.0308 -2.047 -0.30  33412.717  3.0 3d8 c3F     59377.300  4.0 (4F)6p 5G   7.82 -3.96 -7.22
385.9143 -2.743  ...   33412.717  3.0 3d8 c3F     59317.860  4.0 4s6D7p 7D   7.22 -4.13 -7.11
385.9453 -2.369 +0.00  24772.018  1.0 (2P)4s c3P  50675.080  1.0 3Psp3P 1P   7.27 -6.11 -7.81
386.5897 -2.465 +0.10  29320.026  1.0 4s2 b3D     55179.910  0.0 (2P)4p 1S   8.38 -5.99 -7.74
386.5930 -2.211  ...   59366.790  2.0 (4F)5d 5F   33507.123  3.0 (4F)4p y5D  8.37 -3.98 -7.25
386.9653 -1.227  ...   59636.360  3.0 5d 5F5D3G   33801.572  2.0 (4F)4p y5D  8.40 -3.56 -7.25
388.4076 -2.664  ...   34636.792  2.0 4s2 b1D     60375.650  1.0 (2F)4p 3D   8.52 -4.33 -7.37
388.4191 -2.875  ...   33765.306  2.0 3d8 c3F     59503.400  3.0 6p 5G5D3D  8.15 -3.43 -7.17
388.6261 -0.832  ...   60087.260  2.0 4s4D4d 3P   34362.873  1.0 5Dsp3P z3P  7.87 -3.79 -7.18
388.8710 -1.195 -0.30  36975.588  4.0 (2F)4s d3F  62683.770  4.0 8p 3G5G5F  8.29 -3.24 -7.02
389.1501 -1.252 +0.00  59196.870  3.0 (4F)5d 5F   33507.123  3.0 (4F)4p y5D  8.38 -3.95 -7.28
389.8220 -2.731  ...   61724.840  3.0 (2G)5s 3G   36079.372  3.0 (4F)4p z3G  8.18 -4.91 -7.51
389.9380 -0.948 -0.60  37045.934  4.0 (2F)4s d3F  62683.770  4.0 8p 3G5G5F  8.29 -3.24 -7.02
390.5625 -2.493  ...   59636.360  3.0 5d 5F5D3G   34039.516  4.0 (4F)4p y5F  8.30 -3.56 -7.25
390.7665 -2.314 -0.40  25091.599  0.0 (2P)4s c3P  50675.080  1.0 3Psp3P 1P   7.27 -6.11 -7.81
391.0457 -1.526 -0.30  59366.790  2.0 (4F)5d 5F   33801.572  2.0 (4F)4p y5D  8.37 -3.98 -7.25
391.5213 -0.944 -0.20  36975.588  3.0 (2F)4s d3F  62509.750  3.0 8p 3D3G3F  8.28 -3.79 -7.03
391.7360 -1.487 -0.50  58616.110  3.0 (4F)5d 5P   33095.941  4.0 (4F)4p y5D  8.47 -4.14 -7.38
392.0617 -1.661  ...   59300.540  1.0 (4P)5s 3P   33801.572  2.0 (4F)4p y5D  8.52 -4.12 -7.47
392.1465 -2.870  ...   59532.970  4.0 s6d 3+[4+]  34039.516  4.0 (4F)4p y5F  8.28 -3.91 -7.15
392.6030 -2.203  ...   37045.934  4.0 (2F)4s d3F  62509.750  3.0 8p 3D3G3F  8.28 -3.79 -7.03
393.2819 -2.396  ...   59366.790  2.0 (4F)5d 5F   33946.933  2.0 5Dsp3P z3P  7.65 -3.98 -7.25
393.6622 -2.639  ...   59196.870  3.0 (4F)5d 5F   33801.572  2.0 (4F)4p y5D  8.38 -3.95 -7.28
394.3096 -0.798 -0.50  59300.540  1.0 (4P)5s 3P   33946.933  2.0 5Dsp3P z3P  8.15 -4.12 -7.47
394.4054 -1.095  ...   59294.380  3.0 4s4D4d 3G   33946.933  2.0 5Dsp3P z3P  8.30 -3.62 -7.37
394.4131 -1.515  ...   36940.590  2.0 (2F)4s d3F  62287.540  3.0 (4F)7p 5G   7.78 -4.05 -7.00
394.9585 -2.401  ...   36975.588  3.0 (2F)4s d3F  62287.540  3.0 (4F)7p 5G   7.78 -4.05 -7.00
395.0263 -2.020 -0.40  59636.360  3.0 5d 5F5D3G   34328.752  3.0 (4F)4p y5F  8.30 -3.56 -7.25
395.4040 -2.815 +0.40  59300.540  1.0 (4P)5s 3P   34017.103  1.0 (4F)4p y5D  8.52 -4.12 -7.47
```

```
395.5756 -2.789  ...   58779.590  2.0 4s4D4d 3D    33507.123  3.0 (4F)4p y5D  8.53 -4.56 -7.43
395.8513 -2.857  ...   59294.380  3.0 4s4D4d 3G    34039.516  4.0 (4F)4p y5F  8.53 -3.62 -7.37
395.9286 -2.091 -0.30  59196.870  3.0 (4F)5d 5F    33946.933  2.0 5Dsp3P z3P  7.70 -3.95 -7.28
396.6468 -1.809  ...   59532.970  4.0 s6d 3+[4+]   34238.752  3.0 (4F)4p y5F  8.28 -3.91 -7.15
397.0450 -2.607  ...   59300.540  1.0 (4P)5s 3P    34121.603  0.0 (4F)4p y5D  8.53 -4.12 -7.47
397.3857 -1.413  ...   59196.870  3.0 (4F)5d 5F    34039.516  4.0 (4F)4p y5F  8.27 -3.95 -7.28
398.1512 -1.697 -0.40  58616.110  3.0 (4F)5d 5P    33507.123  3.0 (4F)4p y5D  8.48 -4.14 -7.38
398.4660 -0.914 -0.30  59636.360  3.0 5d 5F5D3G    34547.211  2.0 (4F)4p y5F  8.30 -3.56 -7.25
398.6686 -0.998  ...   36940.590  2.0 (2F)4s d3F   62016.990  3.0 (4F)7p 3G   8.12 -3.24 -7.09
398.9079 -0.818 -0.10  37045.934  4.0 (2F)4s d3F   62107.290  5.0 3Gsp1P 3G   8.33 -2.99 -7.04
399.2258 -1.816  ...   36975.588  3.0 (2F)4s d3F   62016.990  3.0 (4F)7p 3G   8.12 -3.24 -7.09
399.2794 -1.496  ...   59366.790  2.0 (4F)5d 5F    34328.752  3.0 (4F)4p y5F  8.26 -3.98 -7.25
399.8243 -2.425 -0.40  59366.790  2.0 (4F)5d 5F    34362.873  1.0 5Dsp3P z3P  7.63 -3.98 -7.25
400.2389 -2.153  ...   58779.590  2.0 4s4D4d 3D    33801.572  2.0 (4F)4p y5D  8.53 -4.56 -7.43
400.8865 -0.661 +0.00  59300.540  1.0 (4P)5s 3P    34362.873  1.0 5Dsp3P z3P  8.14 -4.12 -7.47
402.0077 -1.242 -0.30  59196.870  3.0 (4F)5d 5F    34328.752  3.0 (4F)4p y5F  8.27 -3.95 -7.28
402.5818 -0.377 -0.90  58779.590  2.0 4s4D4d 3D    33946.933  2.0 5Dsp3P z3P  8.17 -4.56 -7.43
402.6761 -2.356 -0.10  58628.410  1.0 4s4D4d 5P    33801.572  2.0 (4F)4p y5D  8.51 -5.24 -7.33
402.7939 -1.584 -0.20  59366.790  2.0 (4F)5d 5F    34547.211  2.0 (4F)4p y5F  8.26 -3.98 -7.25
402.8757 -1.579 +0.00  58616.110  3.0 (4F)5d 5P    33801.572  2.0 (4F)4p y5D  8.47 -4.14 -7.38
403.5558 -2.336  ...   61935.470  4.0 (2G)5s 1G    37162.746  3.0 (4F)4p y3F  8.25 -5.13 -7.54
403.7226 -2.641  ...   58779.590  2.0 4s4D4d 3D    34017.103  1.0 (4F)4p y5D  8.53 -4.56 -7.43
403.8720 -2.569  ...   59300.540  1.0 (4P)5s 3P    34547.211  2.0 (4F)4p y5F  8.44 -4.12 -7.47
403.9173 -2.715 -0.30  59532.970  4.0 s6d 3+[4+]   34782.421  5.0 (4F)4p z5G  8.17 -3.91 -7.15
404.0088 -0.572 -0.40  59300.540  1.0 (4P)5s 3P    34555.597  0.0 5Dsp3P z3P  8.14 -4.12 -7.47
404.7002 -2.369 +0.40  36975.588  3.0 (2F)4s d3F   61678.260  4.0 (4F)7p 5F   7.62 -3.54 -7.02
405.0477 -2.502 +0.10  58628.410  1.0 4s4D4d 5P    33946.933  2.0 5Dsp3P z3P  8.12 -5.24 -7.33
405.1916 -1.067  ...   36975.588  3.0 (2F)4s d3F   61648.300  4.0 (4F)7p 3G   7.95 -4.20 -7.12
405.2497 -0.523 -0.50  58616.110  3.0 (4F)5d 5P    33946.933  2.0 5Dsp3P z3P  8.03 -4.14 -7.38
405.3259 -1.500 -0.20  51143.920  0.0 4s6D4d e7F  26479.381  1.0 5Dsp3P z5D  8.34 -5.29 -7.55
405.4945 -2.705  ...   61340.460  4.0 (2G)5s 3G    36686.176  4.0 (4F)4p y3F  8.25 -4.61 -7.49
405.5706 -2.599  ...   59196.870  3.0 (4F)5d 5F    34547.211  2.0 (4F)4p y5F  8.27 -3.95 -7.28
405.6060 -1.387 -0.20  37045.934  4.0 (2F)4s d3F   61663.440  5.0 (4F)7p 5G   7.69 -3.92 -7.08
405.8560 -1.799 +0.20  37045.934  4.0 (2F)4s d3F   61678.260  4.0 (4F)7p 5F   7.62 -3.54 -7.02
406.2026 -2.604 +0.50  58628.410  1.0 4s4D4d 5P    34017.103  1.0 (4F)4p y5D  8.51 -5.24 -7.33
406.3502 -2.045  ...   37045.934  4.0 (2F)4s d3F   61648.300  4.0 (4F)7p 3G   7.95 -4.20 -7.12
406.7763 -1.622 -0.30  58616.110  3.0 (4F)5d 5P    34039.516  4.0 (4F)4p y5F  8.39 -4.14 -7.38
407.0165 -2.970  ...   61724.840  3.0 (2G)5s 3G    37162.746  3.0 (4F)4p y3F  8.25 -4.91 -7.51
407.8432 -1.890 +0.40  61198.480  5.0 (2G)5s 3G    36686.176  4.0 (4F)4p y3F  8.21 -3.84 -7.13
407.9347 -1.761 +0.10  58628.410  1.0 4s4D4d 5P    34121.603  0.0 (4F)4p y5D  8.52 -5.24 -7.33
408.8685 -2.655  ...   58779.590  2.0 4s4D4d 3D    34328.752  3.0 (4F)4p y5F  8.45 -4.56 -7.43
409.4399 -1.095 -0.70  58779.590  2.0 4s4D4d 3D    34362.873  1.0 5Dsp3P z3P  8.16 -4.56 -7.43
409.5347 -1.895 +0.00  58428.170  0.0 4s6D4d 5D    34017.103  1.0 (4F)4p y5D  8.55 -5.07 -7.40
411.3096 -2.143  ...   37045.934  4.0 (2F)4s d3F   61351.660  3.0 (4F)7p 3D   7.90 -4.31 -7.07
411.7024 -2.963  ...   38602.260  3.0 (2F)4s 1F    62884.800  3.0 3Gsp1P 3F   8.34 -3.05 -7.06
411.8193 -1.589 -0.50  59532.970  4.0 s6d 3+[4+]   35257.324  4.0 (4F)4p z5G  8.20 -3.91 -7.15
411.9386 -2.262 -0.10  26406.465  4.0 a2D)4s a3D   50675.080  1.0 3Psp3P 1P   7.27 -6.11 -7.81
411.9909 -1.991 +0.00  58628.410  1.0 4s4D4d 5P    34362.873  1.0 5Dsp3P z3P  8.12 -5.24 -7.33
413.0438 -1.997  ...   61724.840  3.0 (2G)5s 3G    37521.161  2.0 (4F)4p y3F  8.25 -4.91 -7.51
413.4874 -1.664 -0.20  61340.460  4.0 (2G)5s 3G    37162.746  3.0 (4F)4p y3F  8.25 -4.61 -7.49
413.8974 -2.187 -0.10  59532.970  4.0 s6d 3+[4+]   35379.208  5.0 (4F)4p z3G  8.18 -3.91 -7.15
414.1713 -2.166  ...   36975.588  3.0 (2F)4s d3F   61113.380  4.0 (4F)7p 3F   7.84 -3.82 -7.11
414.2266 -2.736  ...   36940.590  2.0 (2F)4s d3F   61075.160  1.0 3Fsp1P 3D   8.23 -2.91 -7.06
414.6485 -2.531 +0.50  37045.934  4.0 (2F)4s d3F   61155.950  5.0 (4F)7p 5F   7.07 -4.17 -7.06
414.9123 -1.073 -0.20  37045.934  4.0 (2F)4s d3F   61140.620  5.0 (4F)7p 3G   7.99 -3.98 -7.12
415.0371 -2.867  ...   58779.590  2.0 4s4D4d 3D    34692.148  1.0 (4F)4p y5F  8.45 -4.56 -7.43
415.1393 -2.900  ...   38602.260  3.0 (2F)4s 1F    62683.770  4.0 8p 3G5G5F   8.29 -3.24 -7.02
415.2893 -2.592 +0.10  58628.410  1.0 4s4D4d 5P    34555.597  0.0 5Dsp3P z3P  8.11 -5.24 -7.33
415.3819 -1.766 +0.10  37045.934  4.0 (2F)4s d3F   61113.380  4.0 (4F)7p 3F   7.84 -3.82 -7.11
415.4190 -2.515 +0.00  58428.170  0.0 4s6D4d 5D    34362.873  1.0 5Dsp3P z3P  8.20 -5.07 -7.40
415.6600 -2.496 +0.00  26623.735  2.0 a2D)4s a3D   50675.080  1.0 3Psp3P 1P   7.27 -6.11 -7.81
416.1204 -1.159  ...   59636.360  3.0 5d 5F5D3G    35611.625  3.0 (4F)4p z5G  8.23 -3.56 -7.25
417.6012 -2.030 -0.30  59196.870  3.0 (4F)5d 5F    35257.324  4.0 (4F)4p z5G  8.19 -3.95 -7.28
417.9189 -1.619 -0.40  59532.970  4.0 s6d 3+[4+]   35611.625  3.0 (4F)4p z5G  8.21 -3.91 -7.15
418.8390 -2.953  ...   59636.360  3.0 5d 5F5D3G    35767.564  4.0 (4F)4p z3G  8.18 -3.56 -7.25
420.4037 -1.885  ...   59636.360  3.0 5d 5F5D3G    35856.402  2.0 (4F)4p z3G  8.23 -3.56 -7.25
420.6612 -1.292  ...   59532.970  4.0 s6d 3+[4+]   35767.564  4.0 (4F)4p z3G  8.16 -3.91 -7.15
420.8425 -2.113 -0.40  59366.790  2.0 (4F)5d 5F    35611.625  3.0 (4F)4p z5G  8.18 -3.98 -7.25
```

```
422.1293 -2.718 +0.40  59294.380  3.0 4s4D4d 3G    35611.625  3.0 (4F)4p z5G   8.49 -3.62 -7.37
422.8123 -2.666 +0.10  36940.590  2.0 (2F)4s d3F   60585.090  2.0 3Psp1P 3P    8.58 -4.76 -7.25
423.4390 -1.730 -0.50  36975.588  3.0 (2F)4s d3F   60585.090  2.0 3Psp1P 3P    8.58 -4.76 -7.25
423.8745 -2.394  ...   59196.870  3.0 (4F)5d 5F    35611.625  3.0 (4F)4p z5G   8.20 -3.95 -7.28
424.3830 -2.012  ...   59636.360  3.0 5d 5F5D3G    36079.372  3.0 (4F)4p z3G   8.16 -3.56 -7.25
425.2242 -2.806  ...   59366.790  2.0 (4F)5d 5F    35856.402  2.0 (4F)4p z5G   8.19 -3.98 -7.25
425.8186 -2.397 -0.30  51143.920  0.0 4s6D4d e7F   27666.348  1.0 5Dsp3P z5F   8.34 -5.29 -7.55
426.5910 -1.565  ...   36940.590  2.0 (2F)4s d3F   60375.650  1.0 (2F)4p 3D    8.52 -4.33 -7.37
426.6958 -1.896  ...   59196.870  3.0 (4F)5d 5F    35767.564  4.0 (4F)4p z3G   8.14 -3.95 -7.28
429.2957 -2.450  ...   59366.790  2.0 (4F)5d 5F    36079.372  3.0 (4F)4p z3G   8.11 -3.98 -7.25
430.6347 -2.367 +0.20  59294.380  3.0 4s4D4d 3G    36079.372  3.0 (4F)4p z3G   8.45 -3.62 -7.37
432.1786 -1.592 -0.20  27543.003  1.0 (2P)4s a1P   50675.080  1.0 3Psp3P 1P    7.27 -6.11 -7.81
433.2912 -1.292 -0.30  36940.590  2.0 (2F)4s d3F   60013.270  3.0 (4F)6p 3G    8.00 -3.41 -7.26
433.9494 -2.521  ...   36975.588  3.0 (2F)4s d3F   60013.270  3.0 (4F)6p 3G    8.00 -3.41 -7.26
435.6039 -2.825  ...   59636.360  3.0 5d 5F5D3G    36686.176  4.0 (4F)4p y3F   8.24 -3.56 -7.25
435.8866 -2.492  ...   36940.590  2.0 (2F)4s d3F   59875.890  3.0 4s6D7p 5D    7.67 -3.29 -7.13
436.0917 -2.209 -0.70  60087.260  2.0 4s4D4d 3P    37162.746  3.0 (4F)4p y3F   8.27 -3.79 -7.18
437.4323 -2.237  ...   36940.590  2.0 (2F)4s d3F   59794.850  3.0 3Dsp3P 1F    7.86 -4.18 -7.25
437.5752 -1.850 -0.50  59532.970  4.0 s6d 3+[4+]   36686.176  4.0 (4F)4p y3F   8.22 -3.91 -7.15
437.9172 -2.676  ...   36975.588  3.0 (2F)4s d3F   59804.540  4.0 4s6D7p 5F    7.10 -4.56 -7.07
438.1032 -2.561  ...   36975.588  3.0 (2F)4s d3F   59794.850  3.0 3Dsp3P 1F    7.86 -4.18 -7.25
439.1281 -2.319 -0.10  59532.970  4.0 s6d 3+[4+]   36766.966  3.0 5Dsp1P y5P   8.35 -3.91 -7.15
439.3269 -1.099 -0.40  36975.588  3.0 (2F)4s d3F   59731.290  4.0 (4F)6p 3G    8.03 -3.71 -7.26
439.4579 -2.572  ...   37045.934  4.0 (2F)4s d3F   59794.850  3.0 3Dsp3P 1F    7.86 -4.18 -7.25
441.9717 -2.648  ...   36975.588  3.0 (2F)4s d3F   59595.120  4.0 4s6D7p 7F    6.88 -4.11 -7.10
442.3571 -2.903  ...   59366.790  2.0 (4F)5d 5F    36766.966  3.0 5Dsp1P y5P   8.33 -3.98 -7.25
443.0828 -2.286  ...   36940.590  2.0 (2F)4s d3F   59503.400  3.0 6p 5G5D3D    8.15 -3.43 -7.17
443.7712 -2.536  ...   36975.588  3.0 (2F)4s d3F   59503.400  3.0 6p 5G5D3D    8.15 -3.43 -7.17
443.7790 -2.819  ...   59294.380  3.0 4s4D4d 3G    36766.966  3.0 5Dsp1P y5P   8.57 -3.62 -7.37
443.9048 -2.195 -0.30  36975.588  3.0 (2F)4s d3F   59496.620  4.0 4s6D7p 5D    7.42 -3.41 -7.08
444.1086 -1.193 -0.10  59196.870  3.0 (4F)5d 5F    36686.176  4.0 (4F)4p y3F   8.21 -3.95 -7.28
444.7389 -2.743  ...   59636.360  3.0 5d 5F5D3G    37157.566  2.0 5Dsp1P y5P   8.35 -3.56 -7.25
444.8414 -2.934  ...   59636.360  3.0 5d 5F5D3G    37162.746  3.0 (4F)4p y3F   8.24 -3.56 -7.25
445.1613 -2.287  ...   37045.934  4.0 (2F)4s d3F   59503.400  3.0 6p 5G5D3D    8.15 -3.43 -7.17
445.4434 -2.722  ...   36975.588  3.0 (2F)4s d3F   59418.830  3.0 (4F)6p 5F    7.59 -4.10 -7.19
445.8673 -1.489 -0.10  67716.750  6.0 (2G)4d 1I    45294.846  5.0 3Hsp3P y3G   8.08 -5.15 -7.51
446.2692 -1.449 -0.40  36975.588  3.0 (2F)4s d3F   59377.300  4.0 (4F)6p 5G    7.82 -3.96 -7.22
446.8440 -2.655  ...   37045.934  4.0 (2F)4s d3F   59418.830  3.0 (4F)6p 5F    7.59 -4.10 -7.19
446.8974 -1.991 -0.70  59532.970  4.0 s6d 3+[4+]   37162.746  3.0 (4F)4p y3F   8.22 -3.91 -7.15
447.4565 -2.498  ...   36975.588  3.0 (2F)4s d3F   59317.860  4.0 4s6D7p 7D    7.22 -4.13 -7.11
448.0817 -0.928 -0.30  37045.934  4.0 (2F)4s d3F   59357.030  5.0 (4F)6p 3G    8.26 -4.27 -7.28
449.1186 -1.880 -0.30  67687.990  5.0 (2G)4d 3I    45428.402  4.0 3Hsp3P y3G   8.09 -4.84 -7.51
450.2421 -1.319 -0.70  59366.790  2.0 (4F)5d 5F    37162.746  3.0 (4F)4p y3F   8.19 -3.98 -7.25
451.2320 -2.944  ...   38602.260  3.0 (2F)4s 1F    60757.596  4.0 3Hsp1P t3H   8.54 -5.36 -7.63
451.7152 -1.448 +0.00  59294.380  3.0 4s4D4d 3G    37162.746  3.0 (4F)4p y3F   8.50 -3.62 -7.37
452.0509 -1.993 +0.00  59636.360  3.0 5d 5F5D3G    37521.161  2.0 (4F)4p y3F   8.24 -3.56 -7.25
452.9672 -1.534 -0.20  28604.613  2.0 a2D)4s a1D   50675.080  1.0 3Psp3P 1P    7.27 -6.11 -7.81
453.7143 -2.991  ...   59196.870  3.0 (4F)5d 5F    37162.746  3.0 (4F)4p y3F   8.21 -3.95 -7.28
454.1574 -2.612  ...   58779.590  2.0 4s4D4d 3D    36766.966  3.0 5Dsp1P y5P   8.50 -4.56 -7.43
454.9273 -1.402 -0.10  37045.934  4.0 (2F)4s d3F   59021.310  5.0 (4F)6p 5G    7.98 -4.37 -7.26
456.2450 -1.992 -0.50  60087.260  2.0 4s4D4d 3P    38175.355  3.0 (4F)4p y3D   8.36 -3.79 -7.18
459.0213 -2.677 +0.00  59300.540  1.0 (4P)5s 3P    37521.161  2.0 (4F)4p y3F   8.40 -4.12 -7.47
459.1512 -1.119 +0.00  59294.380  3.0 4s4D4d 3G    37521.161  2.0 (4F)4p y3F   8.49 -3.62 -7.37
459.5523 -2.481 +0.20  36975.588  3.0 (2F)4s d3F   58729.800  4.0 (4F)6p 5D    7.68 -4.39 -7.23
462.0133 -2.049 +0.00  40871.412  2.0 3d8 3P       62509.750  3.0 8p 3D3G3F    8.28 -3.79 -7.03
462.4081 -0.439 -1.00  64531.780  4.0 4s3H5s 5H    42911.917  5.0 3Hsp3P y5G   8.30 -4.00 -7.14
463.4727 -2.746  ...   38602.260  3.0 (2F)4s 1F    60172.464  3.0 3Fsp1P 3F    8.47 -4.81 -7.40
463.8605 -2.693  ...   67687.990  5.0 (2G)4d 3I    46135.820  5.0 3Hsp3P z3I   8.03 -4.84 -7.51
464.1208 -0.216 -0.60  64531.780  4.0 4s3H5s 5H    42991.697  5.0 3Hsp3P z5I   8.19 -4.00 -7.14
464.6369 -1.300 -0.10  64300.510  6.0 4s3H5s 5H    42784.352  6.0 3Hsp3P y5G   8.33 -4.39 -7.39
464.7960 -1.592  ...   64531.780  4.0 4s3H5s 5H    43022.985  4.0 3Hsp3P y5G   8.29 -4.00 -7.14
465.6175 -2.192 +0.30  58828.410  1.0 4s4D4d 5P    37157.566  2.0 5Dsp1P y5P   8.47 -5.24 -7.33
465.8310 -2.701  ...   59636.360  3.0 5d 5F5D3G    38175.355  3.0 (4F)4p y3D   8.33 -3.56 -7.25
465.8844 -2.664  ...   58616.110  3.0 (4F)5d 5P    37157.566  2.0 5Dsp1P y5P   8.43 -4.14 -7.38
465.9969 -2.669  ...   58616.110  3.0 (4F)5d 5P    37162.746  3.0 (4F)4p y3F   8.34 -4.14 -7.38
466.6604 -0.181 -0.60  64531.780  4.0 4s3H5s 5H    43108.917  4.0 3Hsp3P z5I   8.18 -4.00 -7.14
466.9149 -1.660 -0.30  51143.920  0.0 4s6D4d e7F   29732.736  1.0 5Dsp3P z5P   8.36 -5.29 -7.55
466.9187 -2.413  ...   38602.260  3.0 (2F)4s 1F    60013.270  3.0 (4F)6p 3G    8.00 -3.41 -7.26
```

```
466.9577 -1.837 -0.70   60087.260  2.0 4s4D4d 3P   38678.039  2.0 (4F)4p y3D   8.36 -3.79 -7.18
467.2321  0.011 -0.10   64300.510  6.0 4s3H5s 5H   42903.861  6.0 3Hsp3P z5I   8.24 -4.39 -7.39
467.2836 -0.721 +0.00   64531.780  4.0 4s3H5s 5H   43137.487  3.0 3Hsp3P y5G   8.30 -4.00 -7.14
467.4081 -0.149 -0.50   64300.510  6.0 4s3H5s 5H   42911.917  5.0 3Hsp3P y5G   8.33 -4.39 -7.39
467.8140 -2.171  ...    58779.590  2.0 4s4D4d 3D   37409.555  1.0 5Dsp1P y5P   8.48 -4.56 -7.43
468.0861 -2.035  ...    59532.970  4.0 s6d 3+[4+]  38175.355  3.0 (4F)4p y3D   8.32 -3.91 -7.15
468.9485 -2.317 +0.00   29356.744  2.0 4s2 b3D    50675.080  1.0 3Psp3P 1P    7.27 -6.11 -7.81
469.1581 -1.835  ...    64300.510  6.0 4s3H5s 5H   42991.697  5.0 3Hsp3P z5I   8.23 -4.39 -7.39
469.9340 -2.048  ...    38602.260  3.0 (2F)4s 1F   59875.890  3.0 4s6D7p 5D   7.67 -3.29 -7.13
469.9340 -2.048  ...    38602.260  3.0 (2F)4s 1F   59875.890  3.0 4s6D7p 5D   7.67 -3.29 -7.13
471.1471 -1.377 +0.00   58628.410  1.0 4s4D4d 5P   37409.555  1.0 5Dsp1P y5P   8.46 -5.24 -7.33
471.4368 -0.478  ...    64531.780  4.0 4s3H5s 5H   43325.964  3.0 3Hsp3P z5H   8.18 -4.00 -7.14
471.7311 -1.777 -0.10   38602.260  3.0 (2F)4s 1F   59794.850  3.0 3Dsp3P 1F    7.86 -4.18 -7.25
471.7568 -1.093 -0.10   59366.790  2.0 (4F)5d 5F   38175.355  3.0 (4F)4p y3D   8.30 -3.98 -7.25
473.3743 -2.629  ...    59294.380  3.0 4s4D4d 3G   38175.355  3.0 (4F)4p y3D   8.55 -3.62 -7.37
474.0466 -0.334 -0.20   64531.780  4.0 4s3H5s 5H   43442.705  4.0 3Hsp3P z5H   8.16 -4.00 -7.14
474.4384 -0.126 +0.00   64531.780  4.0 4s3H5s 5H   43460.121  5.0 3Hsp3P z5H   8.15 -4.00 -7.14
475.5701 -1.117 +0.00   59196.870  3.0 (4F)5d 5F   38175.355  3.0 (4F)4p y3D   8.31 -3.95 -7.28
475.6357 -1.870 +0.00   54828.170  0.0 4s6D4d 5D   37409.555  1.0 5Dsp1P y5P   8.50 -5.07 -7.40
476.1699 -2.562  ...    40871.412  2.0 3d8 3P     61866.450  2.0 (4F)7p 3D    7.76 -3.45 -7.06
476.5245 -0.082 -1.40   64300.510  6.0 4s3H5s 5H   43321.096  6.0 3Hsp3P z5H   8.19 -4.39 -7.39
477.0041 -2.765  ...    59636.360  3.0 5d 5F5D3G   38678.039  2.0 (4F)4p y3D   8.33 -3.56 -7.25
478.3091 -2.963  ...    38602.260  3.0 (2F)4s 1F   59503.400  3.0 6p 5G5D3D   8.15 -3.43 -7.17
478.3091 -2.963  ...    38602.260  3.0 (2F)4s 1F   59503.400  3.0 6p 5G5D3D   8.15 -3.43 -7.17
478.3460 -2.140  ...    40871.412  2.0 3d8 3P     61770.940  3.0 (4F)7p 5D    7.41 -2.63 -6.96
479.7034 -0.478 -0.90   64300.510  6.0 4s3H5s 5H   43460.121  5.0 3Hsp3P z5H   8.20 -4.39 -7.39
480.6615 -1.729 +0.00   67687.990  4.0 3Fsp3P x3F  46889.142  4.0 3Fsp3P x3F  8.22 -4.84 -7.51
482.1549 -2.744  ...    67716.750  6.0 (2G)4d 1I   46982.320  6.0 3Hsp3P z3H   8.10 -5.15 -7.51
482.7614 -0.360 -0.10   67716.750  6.0 (2G)4d 1I   47008.371  5.0 3Hsp3P z3H   8.13 -5.15 -7.51
483.2194 -2.276  ...    59366.790  2.0 (4F)5d 5F   38678.039  2.0 (4F)4p y3D   8.30 -3.98 -7.25
483.2361 -2.174  ...    41178.412  1.0 3d8 3P     61866.450  2.0 (4F)7p 3D    7.75 -3.45 -7.06
483.4328 -2.729  ...    67687.990  5.0 (2G)4d 3I   47008.371  5.0 3Hsp3P z3H   8.14 -4.84 -7.51
484.7718 -2.058 +0.00   59300.540  1.0 (4P)5s 3P   38678.039  2.0 (4F)4p y3D   8.47 -4.12 -7.47
484.9167 -1.633  ...    59294.380  3.0 4s4D4d 3G   38678.039  2.0 (4F)4p y3D   8.55 -3.62 -7.37
485.2016 -1.335 -0.50   58779.590  2.0 4s4D4d 3D   38175.355  3.0 (4F)4p y3D   8.48 -4.56 -7.43
485.2060 -2.460 -0.40   61198.480  2.0 (2G)5s 3G   40594.432  4.0 5Dsp1P x5F  8.81 -3.84 -7.13
485.7374  0.007  ...    67687.990  5.0 (2G)4d 3I   47106.484  4.0 3Hsp3P z3H   8.13 -4.84 -7.51
487.2211 -1.805  ...    59196.870  3.0 (4F)5d 5F   38678.039  2.0 (4F)4p y3D   8.31 -3.95 -7.28
487.4486 -2.123  ...    64531.780  4.0 4s3H5s 5H   44022.525  4.0 3Fsp3P w5F  8.20 -4.00 -7.14
488.1390 -1.655 -0.90   40871.412  2.0 3d8 3P     61351.660  3.0 (4F)7p 3D    7.90 -4.31 -7.07
489.0821 -1.154  ...    58616.110  3.0 (4F)5d 5P   38175.355  3.0 (4F)4p y3D   8.42 -4.14 -7.38
490.7556 -2.316 -0.30   59366.790  2.0 (4F)5d 5F   38995.736  1.0 (4F)4p y3D   8.30 -3.98 -7.25
490.8876 -1.854 -0.10   64531.780  4.0 4s3H5s 5H   44166.206  3.0 3Fsp3P v5D   8.24 -4.00 -7.14
491.1819 -2.833  ...    67716.750  6.0 (2G)4d 1I   47363.376  6.0 3Gsp3P w5G  8.10 -5.15 -7.51
492.3568 -2.327 -0.10   59300.540  1.0 (4P)5s 3P   38995.736  1.0 (4F)4p y3D   8.47 -4.12 -7.47
492.5578 -0.969  ...    67716.750  6.0 (2G)4d 1I   47420.228  5.0 3Gsp3P w5G  8.23 -5.15 -7.51
492.7623 -2.971  ...    64531.780  4.0 4s3H5s 5H   44243.685  5.0 3Fsp3P w5F  8.28 -4.00 -7.14
492.7863 -0.758 -0.20   54304.210  0.0 (4F)4d 5D   34017.103  1.0 (4F)4p y5D  8.48 -4.57 -7.51
492.8568 -2.322 -0.50   40871.412  2.0 3d8 3P     61155.620  1.0 3Psp1P 3P    8.52 -4.00 -7.19
493.2750 -2.385  ...    66293.980  5.0 (2H)5s 1H   46026.971  6.0 3Hsp3P z3I   7.83 -5.22 -7.54
494.0487 -0.628 -0.20   67687.990  5.0 (2G)4d 3I   47452.717  4.0 (2G)4p z1G  8.08 -4.84 -7.51
496.9606 -2.445  ...    64531.780  4.0 4s3H5s 5H   44415.074  4.0 3Fsp3P w5F  8.27 -4.00 -7.14
497.1106 -2.571  ...    67716.750  6.0 (2G)4d 1I   47606.114  5.0 3Gsp3P v5F  8.18 -5.15 -7.51
497.3353 -1.706 -0.50   58779.590  2.0 4s4D4d 3D   38678.039  2.0 (4F)4p y3D   8.48 -4.56 -7.43
497.4246 -0.367 -0.50   67687.990  5.0 (2G)4d 3I   47590.048  4.0 3Gsp3P w5G  8.15 -4.84 -7.51
498.4443 -1.523 -0.20   64300.510  6.0 4s3H5s 5H   44243.685  5.0 3Fsp3P w5F  8.31 -4.39 -7.39
500.3498 -1.943 -0.20   64531.780  4.0 4s3H5s 5H   44551.335  3.0 3Fsp3P w5F  8.29 -4.00 -7.14
500.4309 -2.915  ...    41178.412  1.0 3d8 3P     61155.620  1.0 3Psp1P 3P    8.52 -4.00 -7.19
501.3311 -2.443  ...    54304.210  0.0 (4F)4d 5D   34362.873  1.0 5Dsp3P z3P  8.02 -4.57 -7.51
501.8400 -2.010  ...    41234.505  0.0 3d8 3P     61155.620  1.0 3Psp1P 3P    8.52 -4.00 -7.19
502.1916 -2.943  ...    59532.970  4.0 s6d 3+[4+]  39625.804  4.0 5Dsp1P x5D  8.70 -3.91 -7.15
502.9823 -1.012 +0.30   67687.990  5.0 (2G)4d 3I   47812.118  4.0 3Gsp3P x3G  8.13 -4.84 -7.51
505.9832 -1.520 +0.20   67687.990  5.0 (2G)4d 3I   47929.997  4.0 3Gsp3P v5F  8.26 -4.84 -7.51
507.1206 -2.877  ...    40871.412  2.0 3d8 3P     60585.090  2.0 3Psp1P 3P    8.58 -4.76 -7.25
508.2835 -2.932  ...    59294.380  3.0 4s4D4d 3G   39625.804  4.0 5Dsp1P x5D  8.81 -3.62 -7.37
509.5821 -2.896  ...    67716.750  6.0 (2G)4d 1I   48098.293  6.0 3Hsp3P 1I   8.00 -5.15 -7.51
509.7482 -1.115 -0.20   54304.210  0.0 (4F)4d 5D   34692.148  1.0 (4F)4p y5F  8.39 -4.57 -7.51
510.3302 -2.751  ...    67687.990  5.0 (2G)4d 3I   48098.293  6.0 3Hsp3P 1I   8.02 -4.84 -7.51
```

```
511.0236 -2.119  ...    59532.970  4.0 s6d 3+[4+]    39969.853  3.0 5Dsp1P x5D   8.69 -3.91 -7.15
513.0600 -1.504  ...    67716.750  6.0 (2G)4d 1I     48231.280  5.0 3Gsp3P y5H   8.04 -5.15 -7.51
513.8184 -2.800  ...    67687.990  5.0 (2G)4d 3I     48231.280  5.0 3Gsp3P y5H   8.05 -4.84 -7.51
515.1430 -1.833 -0.20   41178.412  1.0 3d8 3P        60585.090  2.0 3Psp1P 3P    8.58 -4.76 -7.25
515.1869 -2.276  ...    59636.360  3.0 5d 5F5D3G     40231.336  2.0 5Dsp1P x5D   8.70 -3.56 -7.25
517.0756  0.317 -0.10   67716.750  6.0 (2G)4d 1I     48382.603  5.0 (2G)4p z1H   8.23 -5.15 -7.51
517.2907 -1.338  ...    67687.990  5.0 (2G)4d 3I     48361.882  4.0 3Gsp3P y5H   8.05 -4.84 -7.51
517.3330 -2.928  ...    59294.380  3.0 4s4D4d 3G     39969.853  3.0 5Dsp1P x5D   8.81 -3.62 -7.37
517.6777 -2.426  ...    66293.980  6.0 (2H)5s 1H     46982.320  6.0 3Hsp3P z3H   7.94 -5.22 -7.54
517.8459 -1.225 +0.00   67687.990  5.0 (2G)4d 3I     48382.603  5.0 (2G)4p z1H   8.24 -4.84 -7.51
518.3770 -1.397  ...    66293.980  6.0 (2H)5s 1H     47008.371  5.0 3Hsp3P z3H   7.99 -5.22 -7.54
519.9567 -2.318 +0.10   59196.870  3.0 (4F)5d 5F     39969.853  3.0 5Dsp1P x5D   8.69 -3.95 -7.28
520.7633 -2.084  ...    41178.412  1.0 3d8 3P        60375.650  1.0 (2F)4p 3D    8.52 -4.33 -7.37
521.0277 -1.983 +0.00   66293.980  6.0 (2H)5s 1H     47106.484  4.0 3Hsp3P z3H   7.97 -5.22 -7.54
522.2894 -2.610  ...    41234.505  0.0 3d8 3P        60375.650  1.0 (2F)4p 3D    8.52 -4.33 -7.37
522.4447 -2.568  ...    59196.870  3.0 (4F)5d 5F     40231.336  2.0 5Dsp1P x5D   8.68 -3.98 -7.25
522.7518 -2.511  ...    43163.326  4.0 4s6D5s e7D    62287.540  3.0 (4F)7p 5G    8.32 -4.05 -7.00
524.0971 -2.514  ...    43434.627  3.0 4s6D5s e7D    62509.750  3.0 8p 3D3G3F    8.53 -3.79 -7.03
526.0125 -1.510 -0.30   64300.510  6.0 4s3H5s 5H     45294.846  5.0 3Hsp3P y3G   8.25 -4.39 -7.39
526.0454 -2.926  ...    40871.412  2.0 3d8 3P        59875.890  3.0 4s6D7p 5D    7.68 -3.29 -7.13
526.4379 -2.022 +0.00   58616.110  3.0 (4F)5d 5P     39625.804  4.0 5Dsp1P x5D   8.74 -4.14 -7.38
526.5725 -0.953  ...    67687.990  5.0 (2G)4d 3I     48702.535  4.0 3Hsp3P y1G   8.19 -4.84 -7.51
527.0346 -1.451  ...    64531.780  4.0 4s3H5s 5H     45562.974  3.0 3Hsp3P y3G   8.20 -4.00 -7.14
527.1256 -2.553  ...    59196.870  3.0 (4F)5d 5F     40231.336  2.0 5Dsp1P x5D   8.69 -3.95 -7.28
527.2162 -2.768  ...    59366.790  2.0 (4F)5d 5F     40404.518  1.0 5Dsp1P x5D   8.68 -3.98 -7.25
527.3723 -1.930 -0.40   60087.260  2.0 4s4D4d 3D     41130.599  1.0 5Dsp1P x5F   8.82 -3.79 -7.18
527.8770 -1.612 -0.10   59532.970  4.0 s6d 3+[4+]    40594.432  4.0 5Dsp1P x5F   8.81 -3.91 -7.15
528.0982 -2.965  ...    66293.980  6.0 (2H)5s 1H     47363.376  6.0 3Gsp3P w5G   7.94 -5.22 -7.54
528.2982 -2.620  ...    40871.412  2.0 3d8 3P        59794.850  3.0 3Dsp3P 1F    7.87 -4.18 -7.25
529.6890 -2.158  ...    66293.980  6.0 (2H)5s 1H     47420.228  5.0 3Gsp3P w5G   8.12 -5.22 -7.54
531.4917 -1.901 -0.10   58779.590  2.0 4s4D4d 3D     39969.853  3.0 5Dsp1P x5D   8.77 -4.56 -7.43
531.6072 -1.695 -0.20   64531.780  4.0 4s3H5s 5H     45726.130  5.0 3Fsp3P x5G   8.30 -4.00 -7.14
531.9309 -1.851 -0.10   59636.360  3.0 5d 5F5D3G     40842.154  3.0 5Dsp1P x5F   8.82 -3.56 -7.25
534.6519 -0.892 -0.10   64531.780  4.0 4s3H5s 5H     45833.223  4.0 3Fsp3P x5G   8.31 -4.00 -7.14
534.8352 -0.900  ...    64300.510  6.0 4s3H5s 5H     45608.361  6.0 3Fsp3P x5G   8.33 -4.39 -7.39
534.8734 -1.540 -0.40   59532.970  4.0 s6d 3+[4+]    40842.154  3.0 5Dsp1P x5F   8.81 -3.91 -7.15
536.1516 -1.985  ...    58616.110  3.0 (4F)5d 5P     39969.853  3.0 5Dsp1P x5D   8.74 -4.14 -7.38
536.5622 -2.242  ...    40871.412  2.0 3d8 3P        59503.400  3.0 6p 5G5D3D    8.15 -3.43 -7.17
536.9564 -2.854  ...    59636.360  3.0 5d 5F5D3G     41018.051  2.0 5Dsp1P x5F   8.81 -3.56 -7.25
536.9571 -0.432  ...    64531.780  4.0 4s3H5s 5H     45913.497  3.0 3Fsp3P x5G   8.31 -4.00 -7.14
537.2650 -2.257  ...    43163.326  4.0 4s6D5s e7D    61770.940  3.0 (4F)7p 5D    8.24 -2.63 -6.96
537.4145 -2.727 +0.40   59196.870  3.0 (4F)5d 5F     40594.432  4.0 5Dsp1P x5F   8.81 -3.95 -7.28
537.8507 -2.945  ...    61724.840  3.0 (2G)5s 3G     43137.487  3.0 3Hsp3P y5G   8.08 -4.91 -7.51
538.0898 -1.491 +0.50   67687.990  5.0 (2G)4d 3I     49108.896  4.0 (2G)4p w3F   8.25 -4.84 -7.51
538.2263 -0.252 -0.20   64300.510  6.0 4s3H5s 5H     45726.130  5.0 3Fsp3P x5G   8.33 -4.39 -7.39
538.9844 -1.798 +0.10   58779.590  2.0 4s4D4d 3D     40231.336  2.0 5Dsp1P x5D   8.77 -4.56 -7.43
539.0087 -2.820  ...    40871.412  2.0 3d8 3P        59418.830  3.0 (4F)6p 5F    7.61 -4.10 -7.19
539.6716 -2.963  ...    59366.790  2.0 (4F)5d 5F     40842.154  3.0 5Dsp1P x5F   8.80 -3.98 -7.25
540.9206 -2.567  ...    66293.980  6.0 (2H)5s 1H     47812.118  4.0 3Fsp3P x3G   7.97 -5.22 -7.54
541.3107 -2.300  ...    41234.505  0.0 3d8 3P        59703.050  1.0 (4F)6p 5D    7.41 -3.86 -7.16
541.5780 -2.656  ...    66293.980  6.0 (2H)5s 1H     47834.550  5.0 3Fsp3P x3G   7.92 -5.22 -7.54
541.7894 -2.664  ...    59294.380  3.0 4s4D4d 3G     40842.154  3.0 5Dsp1P x5F   8.90 -3.62 -7.37
542.1828 -2.129  ...    66293.980  6.0 (2H)5s 1H     47855.143  6.0 3Gsp3P y5H   7.79 -5.22 -7.54
543.3604 -2.665  ...    61724.840  3.0 (2G)5s 3G     43325.964  3.0 3Hsp3P x5D   7.85 -4.91 -7.51
544.0643 -2.236 +0.10   58779.590  2.0 4s4D4d 3D     40404.518  1.0 5Dsp1P x5D   8.76 -4.56 -7.43
544.8362 -1.158 -0.40   38602.260  3.0 (2F)4s 1F     56951.301  4.0 (2H)4p v1G   8.25 -6.21 -7.73
545.6254 -2.895  ...    64300.510  6.0 4s3H5s 5H     45978.008  7.0 3Hsp3P z3I   8.20 -4.39 -7.39
546.8166 -2.122  ...    67716.750  6.0 (2G)4d 1I     49434.163  6.0 (2G)4p y3H   8.28 -5.15 -7.51
547.0038 -2.577  ...    59294.380  3.0 4s4D4d 3G     41018.051  2.0 5Dsp1P x5F   8.90 -3.62 -7.37
547.6175 -1.672  ...    67716.750  6.0 (2G)4d 1I     49460.902  5.0 (2G)4p v3G   8.30 -5.15 -7.51
548.5778 -2.399 +0.20   58628.410  1.0 4s4D4d 5P     40404.518  1.0 5Dsp1P x5D   8.76 -5.24 -7.33
549.4281 -0.545 -1.20   66293.980  6.0 (2H)5s 1H     48098.293  6.0 3Hsp3P 1I    7.80 -5.22 -7.54
550.3657 -2.367  ...    64300.510  6.0 4s3H5s 5H     46135.820  5.0 3Hsp3P z3I   8.20 -4.39 -7.39
551.2021 -2.101 +0.20   58628.410  1.0 4s4D4d 5P     40491.284  0.0 5Dsp1P x5D   8.76 -5.24 -7.33
551.9570  0.340 -0.40   67716.750  6.0 (2G)4d 1I     49604.427  5.0 (2G)4p y3H   8.30 -5.15 -7.51
552.8348 -1.648  ...    67687.990  5.0 (2G)4d 3I     49604.427  5.0 (2G)4p y3H   8.31 -4.84 -7.51
553.5529 -1.454  ...    67687.990  5.0 (2G)4d 3I     49627.884  4.0 (2G)4p v3G   8.31 -4.84 -7.51
554.6725 -2.632 +0.50   58428.170  0.0 4s6D4d 5D     40404.518  1.0 5Dsp1P x5D   8.78 -5.07 -7.40
```

```
556.6073  0.342 -0.60   67687.990  5.0 (2G)4d 3I    49726.990  4.0 (2G)4p y3H   8.31 -4.84 -7.51
557.5044 -2.454  ...     66293.980  5.0 (2H)5s 1H    48361.882  4.0 3Gsp3P y5H   7.85 -5.22 -7.54
558.1494 -2.395  ...     66293.980  5.0 (2H)5s 1H    48382.603  5.0 (2G)4p z1H   8.12 -5.22 -7.54
558.5742 -2.979  ...     61340.460  4.0 (2G)5s 3G    43442.705  4.0 3Hsp3P z5H   7.82 -4.61 -7.49
558.9851 -1.484 +00      58779.590  2.0 4s4D4d 3D    40894.990  2.0 3Psp3P z5S   8.23 -4.56 -7.43
560.9793 -2.056 -0.40    38602.260  3.0 (2F)4s 1F    56423.283  4.0 1Isp3P u3H   8.23 -5.45 -7.73
562.0591 -2.989  ...     64531.780  4.0 4s3H5s 5H    46744.993  3.0 (4P)4p u5D   8.40 -4.00 -7.14
562.8580 -2.503  ...     58779.590  2.0 4s4D4d 3D    41018.051  2.0 5Dsp1P x5F   8.87 -4.56 -7.43
564.1418 -1.048  ...     58616.110  3.0 (4F)5d 5P    40894.990  2.0 3Psp3P z5S   8.11 -4.14 -7.38
567.6901 -2.689 -0.10    58628.410  1.0 4s4D4d 5P    41018.051  2.0 5Dsp1P x5F   8.86 -5.24 -7.33
568.3004 -0.700 -0.30    66293.980  5.0 (2H)5s 1H    48702.535  4.0 3Hsp3P y1G   8.06 -5.22 -7.54
570.7941 -2.898  ...     64531.780  4.0 4s3H5s 5H    47017.188  3.0 (4P)4p w3D   8.47 -4.00 -7.14
571.3416 -2.279 -0.10    58628.410  1.0 4s4D4d 5P    41130.599  1.0 5Dsp1P x5F   8.86 -5.24 -7.33
571.4145 -1.000 -0.50    38602.260  3.0 (2F)4s 1F    56097.836  3.0 3Gsp3P 1F    8.04 -6.13 -7.78
573.2661 -2.218  ...     64531.780  4.0 4s3H5s 5H    47092.712  3.0 3Fsp3P x3F   8.26 -4.00 -7.14
573.5550 -2.967  ...     44677.006  4.0 4s6D5s e5D   62107.290  5.0 3Gsp1P 3G    8.49 -2.99 -7.04
576.5419 -2.815  ...     44677.006  4.0 4s6D5s e5D   62016.990  3.0 (4F)7p 3G    8.36 -3.24 -7.09
577.2675 -2.460  ...     64300.510  6.0 4s3H5s 5H    46982.320  6.0 3Hsp3P z3H   8.26 -4.39 -7.39
577.9556 -2.374 +0.00    58428.170  0.0 4s6D4d 5D    41130.599  1.0 5Dsp1P x5F   8.88 -5.07 -7.40
578.1371 -1.994 -0.30    64300.510  6.0 4s3H5s 5H    47008.371  6.0 3Hsp3P z3H   8.28 -4.39 -7.39
580.3498 -2.248  ...     45061.329  3.0 4s6D5s e5D   62287.540  3.0 (4F)7p 5G    8.19 -4.05 -7.00
581.6252 -2.503  ...     38602.260  3.0 (2F)4s 1F    55790.696  3.0 1Gsp3P s3G   8.06 -6.09 -7.73
581.7387 -1.179 -0.30    66293.980  5.0 (2H)5s 1H    49108.896  4.0 (2G)4p w3F   8.14 -5.22 -7.54
584.8407 -2.942  ...     44677.006  4.0 4s6D5s e5D   61770.940  3.0 (4F)7p 5D    8.08 -2.63 -6.96
585.5233 -1.714 +0.30    67687.990  5.0 (2G)4d 3I    50613.983  4.0 3Fsp3P x1G   8.15 -4.84 -7.51
587.5044 -2.896  ...     44677.006  4.0 4s6D5s e5D   61693.440  5.0 (4F)7p 5G    8.16 -3.92 -7.08
588.0289 -2.033  ...     44677.006  4.0 4s6D5s e5D   61678.260  4.0 (4F)7p 5F    8.14 -3.54 -7.02
589.6101 -2.848  ...     45061.329  3.0 4s6D5s e5D   62016.990  3.0 (4F)7p 3G    8.36 -3.24 -7.09
589.6795 -2.962  ...     45333.875  2.0 4s6D5s e5D   62287.540  3.0 (4F)7p 5G    8.20 -4.05 -7.00
590.2551 -2.695  ...     64300.510  6.0 4s3H5s 5H    47363.376  6.0 3Gsp3P w5G   8.26 -4.39 -7.39
591.2101 -2.870 -0.10    33765.306  2.0 3d8 c3F      50675.080  1.0 3Psp3P 1P    7.27 -6.03 -7.81
592.2431 -1.619  ...     64300.510  6.0 4s3H5s 5H    47420.228  5.0 3Gsp3P w5G   8.35 -4.39 -7.39
592.9620 -2.592  ...     66293.980  5.0 (2H)5s 1H    49434.163  6.0 (2G)4p y3H   8.19 -5.22 -7.54
593.5276 -2.412 -0.50    38602.260  3.0 (2F)4s 1F    55446.008  4.0 1Gsp3P 3H    7.80 -6.19 -7.79
593.7112 -1.452 -0.40    64531.780  4.0 4s3H5s 5H    47693.239  3.0 3Gsp3P w5G   8.23 -4.00 -7.14
593.9039 -2.551  ...     66293.980  5.0 (2H)5s 1H    49460.902  5.0 (2G)4p v3G   8.21 -5.22 -7.54
597.9326 -1.508 -0.30    64531.780  4.0 4s3H5s 5H    47812.118  4.0 3Psp3P x3G   8.24 -4.00 -7.14
598.2923 -2.968  ...     45061.329  3.0 4s6D5s e5D   61770.940  3.0 (4F)7p 5D    8.09 -2.63 -6.96
598.7359 -2.006  ...     64531.780  4.0 4s3H5s 5H    47834.550  5.0 3Psp3P x3G   8.21 -4.00 -7.14
598.8375 -1.390 -0.20    64300.510  6.0 4s3H5s 5H    47606.114  5.0 3Gsp3P v5F   8.31 -4.39 -7.39
599.0113 -1.865  ...     66293.980  5.0 (2H)5s 1H    49604.427  5.0 (2G)4p y3H   8.21 -5.22 -7.54
599.8544 -1.992  ...     66293.980  5.0 (2H)5s 1H    49627.884  4.0 (2G)4p v3G   8.21 -5.22 -7.54
600.7726 -2.502  ...     61935.470  4.0 (2G)5s 1G    45294.846  5.0 3Hsp3P y3G   7.90 -5.13 -7.54
601.6292 -2.913  ...     45061.329  3.0 4s6D5s e5D   61678.260  4.0 (4F)7p 5F    8.14 -3.54 -7.02
602.1782 -1.770  ...     64531.780  4.0 4s3H5s 5H    47929.997  4.0 3Psp3P v5F   8.34 -4.00 -7.14
603.4428 -2.520  ...     66293.980  5.0 (2H)5s 1H    49726.990  4.0 (2G)4p y3H   8.21 -5.22 -7.54
605.5264 -2.744  ...     59532.970  4.0 s6d 3+[4+]   43022.985  4.0 3Hsp3P y5G   8.02 -3.91 -7.15
605.6334 -1.336 -0.10    61935.470  4.0 (2G)5s 1G    45428.402  4.0 3Hsp3P y3G   7.90 -5.13 -7.54
605.8192 -2.993  ...     42815.855  4.0 4s6D5s e7D   59317.860  4.0 4s6D7p 7D    8.22 -4.13 -7.11
606.0106 -1.889 -0.60    44677.006  4.0 4s6D5s e5D   61173.800  4.0 (4F)7p 5D    8.08 -3.66 -7.03
606.6670 -2.075 -0.30    44677.006  4.0 4s6D5s e5D   61155.950  5.0 (4F)7p 5F    8.03 -4.17 -7.06
607.1454 -1.981 -0.20    64300.510  6.0 4s3H5s 5H    47834.550  5.0 3Fsp3P x3G   8.25 -4.39 -7.39
607.9057 -0.697  ...     64300.510  6.0 4s3H5s 5H    47855.143  6.0 3Gsp3P y5H   8.19 -4.39 -7.39
608.0574 -2.950  ...     43434.627  3.0 4s6D5s e7D   59875.890  3.0 4s6D7p 5D    8.29 -3.29 -7.13
608.0760 -2.988  ...     59300.540  1.0 (4P)5s 3P    42859.778  2.0 3Psp3P x5P   8.22 -4.12 -7.47
608.2127 -2.445 -0.10    45333.875  2.0 4s6D5s e5D   61770.940  3.0 (4F)7p 5D    8.09 -2.63 -6.96
608.7560 -2.799  ...     43633.533  2.0 4s6D5s e7D   60055.930  3.0 4s6D7p 5F    8.22 -4.07 -7.07
609.2585 -2.338  ...     64531.780  4.0 4s3H5s 5H    48122.928  3.0 3Gsp3P v5F   8.41 -4.00 -7.14
610.2055 -2.652  ...     67687.990  5.0 (2G)4d 3I    51304.604  4.0 (2D)4p v3F   8.18 -4.84 -7.51
610.6114 -2.344 -0.30    61935.470  4.0 (2G)5s 1G    45562.974  3.0 3Hsp3P y3G   7.89 -5.13 -7.54
610.7077 -2.305 +0.10    43434.627  3.0 4s6D5s e7D   59804.540  4.0 4s6D7p 5F    8.21 -4.56 -7.07
611.1787 -2.692  ...     44509.152  1.0 4s6D5s e5D   61866.450  2.0 (4F)7p 3D    8.19 -3.45 -7.06
611.7194 -0.786 -0.20    67716.750  4.0 (2G)4d 1I    51373.910  5.0 (2H)4p u3G   8.18 -5.15 -7.51
611.8230 -2.975  ...     43163.326  4.0 4s6D5s e7D   59503.400  3.0 6p 5G5D3D   8.46 -3.43 -7.17
612.0770 -2.233 -0.20    43163.326  4.0 4s6D5s e7D   59496.620  4.0 4s6D7p 5D    8.25 -3.41 -7.08
612.7978 -2.164  ...     67687.990  5.0 (2G)4d 3I    51373.910  5.0 (2H)4p u3G   8.19 -4.84 -7.51
613.3084 -1.740 -0.20    64531.780  4.0 4s3H5s 5H    48231.280  5.0 3Gsp3P y5H   8.18 -4.00 -7.14
613.4612 -1.958 -0.10    61724.840  3.0 (2G)5s 3G    45428.402  4.0 3Hsp3P y3G   7.90 -4.91 -7.51
```

```
613.6912 -2.882  ...    45061.329  3.0 4s6D5s e5D  61351.660  3.0 (4F)7p 3D   8.24 -4.31 -7.07
614.1234 -1.700 -0.20   67687.990  5.0 (2G)4d 3I   51409.124  4.0 3Gsp3P x3H  8.07 -4.84 -7.51
615.0060 -2.578  ...    43163.326  4.0 4s6D5s e7D  59418.830  3.0 (4F)6p 5F   8.28 -4.10 -7.19
615.3337 -2.034 -0.20   58779.590  2.0 4s4D4d 3D   42532.741  3.0 3Psp3P x5P  8.20 -4.56 -7.43
616.5813 -2.239  ...    43163.326  4.0 4s6D5s e7D  59377.300  4.0 (4F)6p 5G   8.34 -3.96 -7.22
616.7859 -1.922 -0.60   38602.260  3.0 (2F)4s 1F   54810.856  4.0 3Gsp3P w1G  7.82 -6.22 -7.78
617.0287 -2.403  ...    64300.510  4.0 4s3H5s 5H   48098.293  6.0 3Hsp3P 1I   8.19 -4.39 -7.39
617.3531 -2.650  ...    43163.326  4.0 4s6D5s e7D  59357.030  5.0 (4F)6p 3G   8.52 -4.27 -7.28
618.2620 -1.143 -0.30   64531.780  4.0 4s3H5s 5H   48361.882  4.0 3Gsp3P y5H  8.17 -4.00 -7.14
618.4650 -2.473 -0.20   60087.260  2.0 4s4D4d 3P   43922.668  3.0 3Psp3P w5D  8.35 -3.79 -7.18
618.5693 -0.625 -0.10   61724.840  3.0 (2G)5s 3G   44562.974  3.0 3Hsp3P y3G  7.90 -4.91 -7.51
618.6218 -1.573 +0.20   43434.627  3.0 4s6D5s e7D  59595.120  4.0 4s6D7p 7F   8.20 -4.11 -7.10
618.8500 -1.690 -0.60   43163.326  4.0 4s6D5s e7D  59317.860  4.0 4s6D7p 7D   8.22 -4.13 -7.11
620.4656 -2.093  ...    45061.329  3.0 4s6D5s e5D  61173.800  4.0 (4F)7p 5D   8.09 -3.66 -7.03
620.8596 -2.579  ...    61935.470  4.0 (2G)5s 1G   45833.223  4.0 3Fsp3P x5G  8.08 -5.13 -7.54
621.4645 -0.758 +0.10   67716.750  6.0 (2G)4d 1I   51630.178  5.0 3Hsp3P 1H   8.21 -5.15 -7.51
621.5883 -1.375 -0.30   58616.110  3.0 (4F)5d 5P   42532.741  3.0 3Psp3P x5P  8.07 -4.14 -7.38
622.1352 -1.605  ...    64300.510  4.0 4s3H5s 5H   48231.280  5.0 3Gsp3P y5H  8.22 -4.39 -7.39
622.1529 -2.850  ...    43434.627  3.0 4s6D5s e7D  59503.400  3.0 6p 5G5D3D  8.46 -3.43 -7.17
622.5776 -2.013  ...    67687.990  5.0 (2G)4d 3I   51630.178  5.0 3Hsp3P 1H   8.22 -4.84 -7.51
622.6442 -1.968  ...    64531.780  4.0 4s3H5s 5H   48475.686  3.0 3Gsp3P y5H  8.17 -4.00 -7.14
623.0509 -2.103  ...    61340.460  4.0 (2G)5s 3G   45294.846  5.0 3Hsp3P y3G  7.91 -4.61 -7.49
624.0547 -2.527  ...    67687.990  5.0 (2G)4d 3I   51668.186  4.0 (2H)4p u3G  8.24 -4.84 -7.51
624.1334 -2.477 -0.30   45333.875  2.0 4s6D5s e5D  61351.660  3.0 (4F)7p 3D   8.25 -4.31 -7.07
627.9744 -1.523  ...    58779.590  2.0 4s4D4d 3D   42859.778  2.0 3Psp3P x5P  8.23 -4.56 -7.43
628.2805 -0.582 -0.30   61340.460  4.0 (2G)5s 3G   45428.402  4.0 3Hsp3P y3G  7.90 -4.61 -7.49
628.6132 -0.566 -0.10   61198.480  5.0 (2G)5s 3G   45294.846  5.0 3Hsp3P y3G  7.81 -3.84 -7.13
632.2825 -1.896  ...    61724.840  3.0 (2G)5s 3G   45913.497  3.0 3Fsp3P x5G  8.09 -4.91 -7.51
633.6393 -1.811 -0.10   61340.460  4.0 (2G)5s 3G   44562.974  3.0 3Hsp3P y3G  7.90 -4.61 -7.49
633.9370 -1.789 -0.10   61198.480  5.0 (2G)5s 3G   45428.402  4.0 3Hsp3P y3G  7.80 -3.84 -7.13
633.9951 -2.697  ...    58628.410  1.0 4s4D4d 5P   42859.778  2.0 3Psp3P x5P  8.19 -5.24 -7.33
636.7436 -1.783 -0.20   58779.590  2.0 4s4D4d 3D   43079.023  1.0 3Psp3P x5P  8.17 -4.56 -7.43
637.5790 -0.739 -0.10   66293.980  5.0 (2H)5s 1H   50613.983  4.0 3Fsp3P x1G  8.01 -5.22 -7.54
638.9452 -2.976  ...    45509.152  1.0 4s6D5s e5D  61155.620  1.0 3Psp1P 3P   8.63 -4.00 -7.19
639.4076 -2.958 -0.50   38602.260  3.0 (2F)4s 1F   54237.415  4.0 3Gsp3P t3G  8.04 -5.45 -7.71
640.4256 -2.959  ...    59532.970  4.0 s6d 3+[4+]  43922.668  3.0 3Psp3P w5D  8.30 -3.91 -7.15
641.2547 -2.970  ...    61198.480  5.0 (2G)5s 3G   45608.361  6.0 3Fsp3P x5G  8.01 -3.84 -7.13
644.6820 -1.770 -0.60   61340.460  4.0 (2G)5s 3G   45833.223  4.0 3Fsp3P x5G  8.09 -4.61 -7.49
646.1357 -1.938 -0.50   61198.480  5.0 (2G)5s 3G   45726.130  5.0 3Fsp3P x5G  8.02 -3.84 -7.13
647.3167 -2.944  ...    59366.790  2.0 (4F)5d 5F   43922.668  3.0 3Psp3P w5D  8.28 -3.98 -7.25
648.0367 -2.604  ...    61340.460  4.0 (2G)5s 3G   45913.497  3.0 3Fsp3P x5G  8.09 -4.61 -7.49
650.6391 -2.545  ...    61198.480  5.0 (2G)5s 3G   45833.223  4.0 3Fsp3P x5G  8.02 -3.84 -7.13
653.1660 -1.759 -0.30   47377.955  4.0 (4F)5s e5F  62683.770  4.0 8p 3G5G5F  8.41 -3.24 -7.02
654.6629 -1.967 +0.00   66293.980  5.0 (2H)5s 1H   51023.162  6.0 3Gsp3P x3H  7.85 -5.22 -7.54
656.6218 -2.795  ...    66293.980  5.0 (2H)5s 1H   51068.718  5.0 3Gsp3P x3H  7.88 -5.22 -7.54
657.0807 -2.122  ...    61935.470  4.0 (2G)5s 1G   46720.842  4.0 (4P)4p u5D  8.36 -5.13 -7.54
657.5123 -2.344  ...    61340.460  4.0 (2G)5s 3G   46135.820  5.0 3Hsp3P z3I  7.82 -4.61 -7.49
658.4716 -2.509 -0.40   38602.260  3.0 (2F)4s 1F   53784.750  3.0 (4F)5p 5D   7.98 -4.76 -7.59
658.9482 -2.758  ...    61198.480  5.0 (2G)5s 3G   46026.971  6.0 3Hsp3P z3I  7.69 -3.84 -7.13
660.6777 -2.477  ...    47377.955  4.0 (4F)5s e5F  62509.750  3.0 8p 3D3G3F  8.40 -3.79 -7.03
661.3415 -1.204 -0.10   58616.110  3.0 (4F)5d 5P   43499.505  4.0 3Psp3P w5D  8.48 -4.14 -7.38
662.7261 -2.996  ...    59636.360  3.0 5d 5F5D3G   44551.335  3.0 3Fsp3P w5F  8.04 -3.56 -7.25
664.4305 -1.174 -0.50   61935.470  4.0 (2G)5s 1G   46889.142  4.0 3Fsp3P x3F  8.09 -5.13 -7.54
666.3050 -2.955  ...    61724.840  3.0 (2G)5s 3G   46720.842  4.0 (4P)4p u5D  8.36 -4.91 -7.51
666.9550 -1.869 +0.00   66293.980  5.0 (2H)5s 1H   51304.604  4.0 (2D)4p v3F  8.04 -5.22 -7.54
668.6249 -2.476  ...    45061.329  3.0 4s6D5s e5D  60013.270  3.0 (4F)6p 3G   8.29 -3.41 -7.26
669.7376 -1.260 -0.10   61935.470  4.0 (2G)5s 1G   47008.371  5.0 3Hsp3P z3H  7.97 -5.13 -7.54
670.0531 -1.872  ...    66293.980  5.0 (2H)5s 1H   51373.910  5.0 (2H)4p u3G  8.05 -5.22 -7.54
670.5170 -2.650  ...    67716.750  6.0 (2G)4d 1I   52807.001  5.0 (4F)5p 5D   8.19 -4.76 -7.51
670.7786 -2.323 -0.20   45333.875  2.0 4s6D5s e5D  60237.810  2.0 4s6D7p 5D   8.12 -3.90 -7.06
671.6383 -2.898  ...    66293.980  5.0 (2H)5s 1H   51409.124  4.0 3Gsp3P x3H  7.88 -5.22 -7.54
672.4641 -1.738 +0.30   60087.260  2.0 4s4D4d 3P   46223.681  3.0 3Psp3P x3D  8.11 -3.79 -7.18
672.4746 -2.979  ...    64300.510  4.0 4s3H5s 5H   49434.163  6.0 (2G)4p y3H  8.39 -4.39 -7.39
672.9012 -1.379 +0.00   58779.590  2.0 4s4D4d 3D   43922.668  3.0 3Psp3P w5D  8.47 -4.56 -7.43
673.5433 -1.168 -0.30   61935.470  4.0 (2G)5s 1G   47092.712  3.0 3Fsp3P x3F  8.00 -5.13 -7.54
673.8638 -1.699 -0.10   61724.840  3.0 (2G)5s 3G   46889.142  4.0 3Fsp3P x3F  8.09 -4.91 -7.51
673.8749 -1.818 -0.10   45333.875  2.0 4s6D5s e5D  60169.330  1.0 4s4F7p 5D   8.12 -3.76 -7.07
674.1688 -1.979 +0.10   61935.470  4.0 (2G)5s 1G   47106.484  4.0 3Hsp3P z3H  7.96 -5.13 -7.54
```

```
674.2588 -2.194 +0.40   45509.152  1.0 4s6D5s e5D   60336.160  1.0 4s6D7p 5F   8.09 -4.13 -7.07
674.2867 -1.900 -0.50   44677.006  4.0 4s6D5s e5D   59503.400  3.0 6p 5G5D3D   8.37 -3.43 -7.17
674.5952 -1.484  ...    44677.006  4.0 4s6D5s e5D   59496.620  4.0 4s6D7p 5D   8.09 -3.41 -7.08
674.6800 -2.750  ...    67716.750  6.0 (2G)4d 1I    52898.998  5.0 (2H)4p y3I  8.32 -5.15 -7.51
674.8253 -1.666 -0.20   45061.329  3.0 4s6D5s e5D   59875.890  3.0 4s6D7p 5D   8.16 -3.29 -7.13
675.2416 -2.377  ...    60087.260  2.0 4s4D4d 3P    45281.833  2.0 3Psp3P x3D  8.12 -3.79 -7.18
676.0041 -2.477 -0.20   59300.540  1.0 (4P)5s 3P    44511.812  2.0 (4P)4p y5S  8.33 -4.12 -7.47
678.0911 -1.629 -0.50   45061.329  3.0 4s6D5s e5D   59804.540  4.0 4s6D7p 5F   8.04 -4.56 -7.07
678.1549 -2.906  ...    44677.006  4.0 4s6D5s e5D   59418.830  3.0 (4F)6p 5F   8.13 -4.10 -7.19
678.1894 -2.487  ...    44595.086  0.0 4s6D5s e5D   60036.160  1.0 4s6D7p 5F   8.09 -4.13 -7.07
678.5371 -2.179  ...    45061.329  3.0 4s6D5s e5D   59794.850  3.0 3Dsp3P 1F   8.23 -4.18 -7.25
678.7299 -2.680  ...    47377.955  4.0 (4F)5s e5F   62107.290  5.0 3Gsp1P 3G   8.44 -2.99 -7.04
678.7611 -1.761 -0.20   45509.152  1.0 4s6D5s e5D   60237.810  2.0 4s6D7p 5D   8.12 -3.90 -7.06
679.0298 -1.267 -0.50   47960.940  1.0 (4P)5s e3F   62683.770  4.0 8p 3G5G5F   8.40 -3.24 -7.02
679.0656 -1.656 -0.40   45333.875  2.0 4s6D5s e5D   60055.930  3.0 4s6D7p 5F   8.06 -4.07 -7.07
680.3880 -2.066 -0.40   58616.110  3.0 (4F)5d 5P    43922.668  3.0 3Psp3P w5D  8.41 -4.14 -7.38
681.7632 -1.517 +0.00   66293.980  5.0 (2H)5s 1H    51630.178  5.0 3Hsp3P 1H   8.09 -5.22 -7.54
682.8318 -2.052 +0.00   44677.006  4.0 4s6D5s e5D   59317.860  4.0 4s6D7p 7D   8.05 -4.13 -7.11
683.0366 -2.375 -0.40   59300.540  1.0 (4P)5s 3P    44664.075  2.0 3Fsp3P v5D  8.37 -4.12 -7.47
683.2391 -2.459  ...    61724.840  3.0 (2G)5s 3G    47092.712  3.0 3Fsp3P x3F  8.00 -4.91 -7.51
683.5349 -0.915 -0.20   66293.980  5.0 (2H)5s 1H    51668.186  4.0 (2H)4p u3G  8.12 -5.22 -7.54
683.8237 -2.415  ...    61340.460  4.0 (2G)5s 3G    46720.842  4.0 (4P)4p u5D  8.36 -4.61 -7.49
683.8827 -0.361 -0.10   61724.840  3.0 (2G)5s 3G    47106.484  4.0 3Hsp3P z3H  7.96 -4.91 -7.51
684.1154 -1.746  ...    58779.590  2.0 4s4D4d 3D    44166.206  3.0 3Fsp3P v5D  8.24 -4.56 -7.43
684.9552 -2.682  ...    61340.460  4.0 (2G)5s 3G    46744.993  3.0 (4P)4p u5D  8.24 -4.61 -7.49
685.0436 -1.053 -0.40   58616.110  3.0 (4F)5d 5P    44022.525  4.0 3Fsp3P v5D  8.08 -4.14 -7.38
685.2703 -2.766  ...    61724.840  3.0 (2G)5s 3G    47136.084  2.0 (4P)4p w5D  8.31 -4.91 -7.51
687.1518 -2.046  ...    47960.940  4.0 (4F)5s e3F   62509.750  3.0 8p 3D3G3F   8.39 -3.79 -7.03
687.7814 -2.857  ...    60087.260  2.0 4s4D4d 3P    45551.767  1.0 3Psp3P x3D  8.13 -3.79 -7.18
687.8619 -2.376  ...    45061.329  3.0 4s6D5s e5D   59595.120  4.0 4s6D7p 7F   8.02 -4.11 -7.10
688.1442 -0.733 -0.10   61724.840  3.0 (2G)5s 3G    47197.010  2.0 (3F)sp x3F  7.97 -4.91 -7.51
688.7410 -2.134  ...    61935.470  4.0 (2G)5s 1G    47420.228  5.0 3Gsp3P w5G  8.11 -5.13 -7.54
689.7439 -1.531 -0.40   58779.590  2.0 4s4D4d 3D    44285.454  2.0 3Fsp3P w5F  8.26 -4.56 -7.43
690.2860 -0.413 -0.20   61935.470  4.0 (2G)5s 1G    47452.717  4.0 (2G)4p z1G  7.88 -5.13 -7.54
690.5299 -1.319  ...    61198.480  5.0 (2G)5s 3G    46720.842  4.0 (4P)4p u5D  8.33 -3.84 -7.13
691.7876 -1.977  ...    61340.460  4.0 (2G)5s 3G    46889.142  4.0 3Fsp3P x3F  8.09 -4.61 -7.49
691.8553 -1.907  ...    58616.110  3.0 (4F)5d 5P    44166.206  3.0 3Fsp3P v5D  8.13 -4.14 -7.38
692.1006 -2.408 -0.30   58628.410  1.0 4s4D4d 5P    44183.628  2.0 3Fsp3P w5D  8.35 -5.24 -7.33
692.6904 -1.447 -0.30   58616.110  3.0 (4F)5d 5P    44183.628  2.0 3Fsp3P w5D  8.30 -4.14 -7.38
694.0933 -2.192  ...    67716.750  6.0 (2G)4d 1I    53313.438  5.0 3Gsp3P y1H  8.04 -5.15 -7.51
694.5913 -2.277  ...    47377.955  4.0 (4F)5s e5F   61770.940  3.0 (4F)7p 5D   7.94 -2.63 -6.96
695.7783 -2.187  ...    58779.590  2.0 4s4D4d 3D    44411.160  1.0 3Psp3P w5D  8.27 -4.56 -7.43
696.8943 -1.075 -0.70   61935.470  4.0 (2G)5s 1G    47590.048  4.0 3Gsp3P w5G  7.98 -5.13 -7.54
697.0141 -2.909  ...    58628.410  1.0 4s4D4d 5P    44285.454  2.0 3Fsp3P w5F  8.23 -5.24 -7.33
697.5426 -0.215 -0.20   61340.460  4.0 (2G)5s 3G    47008.371  5.0 3Hsp3P z3H  7.98 -4.61 -7.49
697.6124 -2.054 -0.10   58616.110  3.0 (4F)5d 5P    44285.454  2.0 3Fsp3P w5F  8.16 -4.14 -7.38
697.9720 -2.339  ...    61340.460  4.0 (2G)5s 3G    47017.188  3.0 (4P)4p w3D  8.34 -4.61 -7.49
698.3517 -1.386 -0.10   47377.955  4.0 (4F)5s e5F   61693.440  5.0 (4F)7p 5G   8.05 -3.92 -7.08
698.6517 -0.553 -0.10   61198.480  5.0 (2G)5s 3G    46889.142  4.0 3Fsp3P x3F  8.03 -3.84 -7.13
699.0930 -1.518 -0.20   47377.955  4.0 (4F)5s e5F   61678.260  4.0 (4F)7p 5F   8.02 -3.54 -7.02
700.4734 -1.428  ...    61724.840  3.0 (2G)5s 3G    47452.717  4.0 (2G)4p z1G  7.89 -4.91 -7.51
700.6867 -1.714 +0.40   58779.590  2.0 4s4D4d 3D    44511.812  2.0 (4P)4p y5S  8.35 -4.56 -7.43
700.9975 -2.779  ...    47755.537  3.0 (4F)5s e5F   62016.990  3.0 (4F)7p 3G   8.29 -3.24 -7.09
701.5182 -1.668 +0.00   48036.673  2.0 (4F)5s e5F   62287.540  3.0 (4F)7p 5G   8.09 -4.05 -7.00
701.6718 -0.505 -0.20   61340.460  4.0 (2G)5s 3G    47092.712  3.0 3Fsp3P x3F  8.01 -4.61 -7.49
701.9436 -1.916 +0.20   61935.470  4.0 (2G)5s 1G    47693.239  3.0 3Gsp3P w5G  7.95 -5.13 -7.54
702.3507 -0.702 +0.00   61340.460  4.0 (2G)5s 3G    47106.484  4.0 3Hsp3P z3H  7.97 -4.61 -7.49
702.6331 -1.986  ...    58779.590  2.0 4s4D4d 3D    44551.335  3.0 3Fsp3P w5F  8.29 -4.56 -7.43
703.1770 -2.887  ...    58628.410  1.0 4s4D4d 5P    44411.160  1.0 3Psp3P w5D  8.23 -5.24 -7.33
703.2309 -0.348 +0.00   61198.480  5.0 (2G)5s 3G    46982.320  6.0 3Hsp3P z3H  7.84 -3.84 -7.13
703.9799 -1.718  ...    58616.110  3.0 (4F)5d 5P    44415.074  4.0 3Fsp3P w5F  8.17 -4.14 -7.38
704.5220 -1.065 +0.00   61198.480  5.0 (2G)5s 3G    47008.371  5.0 3Hsp3P z3H  7.89 -3.84 -7.13
705.6067 -1.396 -0.50   47005.506  5.0 (4F)5s e5F   61173.800  4.0 (4F)7p 5D   7.94 -3.66 -7.03
706.4239 -1.465 -0.50   48531.865  3.0 (4F)5s e3F   62683.770  4.0 8p 3G5G5F   8.40 -3.24 -7.02
706.4968 -1.444 -0.30   47005.506  5.0 (4F)5s e5F   61155.950  5.0 (4F)7p 5F   7.87 -4.17 -7.06
706.7013 -1.017 -0.10   47960.940  4.0 (4F)5s e3F   62107.290  5.0 3Gsp1P 3G   8.43 -2.99 -7.04
706.7133 -2.580  ...    59366.790  2.0 (4F)5d 5F    45220.681  3.0 3Fsp3P x3D  7.99 -3.98 -7.25
707.2630 -2.907  ...    47005.506  5.0 (4F)5s e5F   61140.620  5.0 (4F)7p 3G   8.20 -3.98 -7.12
```

```
707.2791 -0.882 -0.40   61724.840   3.0 (2G)5s 3G   47590.048   4.0 3Gsp3P w5G   7.99 -4.91 -7.51
707.8520 -2.194 -0.30   61935.470   4.0 (2G)5s 1G   47812.118   4.0 3Fsp3P x3G   7.96 -5.13 -7.54
708.1907 -2.864  ...    58628.410   1.0 4s4D4d 5P   44511.812   2.0 (4P)4p y5S   8.32 -5.24 -7.33
708.2450 -2.562  ...    58779.590   2.0 4s4D4d 3D   44664.075   2.0 3Fsp3P v5D   8.38 -4.56 -7.43
708.4760 -2.029 +0.40   47755.537   3.0 (4F)5s e5F  61866.450   2.0 (4F)7p 3D    8.07 -3.45 -7.06
708.8083 -1.850  ...    58616.110   3.0 (4F)5d 5P   44511.812   2.0 (4P)4p y5S   8.26 -4.14 -7.38
708.9616 -2.596  ...    61935.470   4.0 (2G)5s 1G   47834.221   3.0 3Fsp3P x3G   7.85 -5.13 -7.54
708.9781 -2.060 -0.20   61935.470   4.0 (2G)5s 1G   47834.550   5.0 3Fsp3P x3G   7.91 -5.13 -7.54
709.4271 -1.319 -0.20   61198.480   5.0 (2G)5s 3G   47106.484   4.0 3Hsp3P z3H   7.88 -3.84 -7.13
709.9982 -2.878  ...    38602.260   3.0 (2F)4s 1F   52682.920   2.0 (2D)4p t3D   8.07 -5.56 -7.69
710.8001 -2.078  ...    58616.110   3.0 (4F)5d 5P   44551.335   3.0 3Psp3P w5F   8.19 -4.14 -7.38
711.4061 -2.380  ...    44677.006   4.0 4s6D5s e5D  58729.800   4.0 (4F)6p 5D    8.16 -4.39 -7.23
712.4806 -1.591 -0.10   61724.840   3.0 (2G)5s 3G   47693.239   3.0 3Gsp3P w5G   7.96 -4.91 -7.51
713.1290 -2.995  ...    58779.590   2.0 4s4D4d 3D   44760.746   1.0 3Psp3P v5D   8.46 -4.56 -7.43
713.1360 -0.904 -0.30   59300.540   1.0 (4P)5s 3P   45281.833   2.0 3Psp3P x3D   8.30 -4.12 -7.47
713.2223 -2.991  ...    58428.170   4.0 4s6D4d 5D   44411.160   1.0 3Psp3P w5D   8.30 -5.07 -7.40
713.4495 -2.477  ...    59294.380   3.0 4s4D4d 3G   45281.833   2.0 3Psp3P x3D   8.41 -3.62 -7.37
715.0942 -2.673  ...    48036.673   2.0 (4F)5s e5F  62016.990   3.0 (4F)7p 3G    8.29 -3.24 -7.09
715.2187 -1.178 -0.10   48531.865   3.0 (4F)5s e3F  62509.750   3.0 8p 3D3G3F    8.39 -3.79 -7.03
715.4326 -1.526 -0.40   47377.955   4.0 (4F)5s e5F  61351.660   3.0 (4F)7p 3D    8.15 -4.31 -7.07
716.5438 -2.236  ...    58616.110   3.0 (4F)5d 5P   44664.075   2.0 3Fsp3P v5D   8.30 -4.14 -7.38
717.5457 -2.156 -0.40   44677.006   4.0 4s6D5s e5D  58609.560   5.0 (4F)6p 5F    8.05 -3.24 -7.23
718.1809 -0.704  ...    61340.460   4.0 (2G)5s 3G   47420.228   5.0 3Gsp3P w5G   8.11 -4.61 -7.49
718.5686 -1.484 +0.40   61724.840   3.0 (2G)5s 3G   47812.118   4.0 3Fsp3P x3G   7.97 -4.91 -7.51
719.2422 -1.272 -0.10   54304.210   3.0 (4F)4d 5D   40404.518   1.0 5Dsp1P x5D   8.74 -4.57 -7.51
719.5530 -0.881 +0.00   61724.840   3.0 (2G)5s 3G   47831.153   2.0 3Gsp3P w5G   7.99 -4.91 -7.51
719.6009 -2.315 +0.60   47755.537   3.0 (4F)5s e5F  61648.300   4.0 (4F)7p 3G    8.18 -4.20 -7.12
719.7120 -1.727  ...    61724.840   3.0 (2G)5s 3G   47834.221   3.0 3Fsp3P x3G   7.86 -4.91 -7.51
719.8610 -1.332  ...    61340.460   4.0 (2G)5s 3G   47452.717   4.0 (2G)4p z1G   7.89 -4.61 -7.49
721.1701 -2.988  ...    66293.980   5.0 (2H)5s 1H   52431.447   6.0 (2H)4p w3H   8.27 -5.22 -7.54
722.5999 -0.522 -0.30   61198.480   5.0 (2G)5s 3G   47363.376   6.0 3Gsp3P w5G   7.84 -3.84 -7.13
723.7802 -2.258  ...    61935.470   4.0 (2G)5s 1G   48122.928   3.0 3Gsp3P v5F   8.24 -5.13 -7.54
723.9135 -1.716  ...    47960.940   4.0 (4F)5s e3F  61770.940   3.0 (4F)7p 5D    7.90 -2.63 -6.96
724.7089 -1.937  ...    61724.840   3.0 (2G)5s 3G   47929.997   4.0 3Gsp3P v5F   8.15 -4.91 -7.51
725.4674 -1.649 -0.20   66293.980   5.0 (2H)5s 1H   52513.560   6.0 (2H)4p y3I   8.24 -5.22 -7.54
725.5815 -2.863  ...    61198.480   5.0 (2G)5s 3G   47420.228   5.0 3Gsp3P w5G   8.05 -3.84 -7.13
726.4033 -2.145  ...    47377.955   4.0 (4F)5s e5F  61140.620   5.0 (4F)7p 3G    8.20 -3.98 -7.12
726.7724 -2.929  ...    48531.865   3.0 (4F)5s e3F  62287.540   3.0 (4F)7p 5G    8.06 -4.05 -7.00
727.0506 -1.524 -0.20   61340.460   4.0 (2G)5s 3G   47590.048   4.0 3Gsp3P w5G   7.99 -4.61 -7.49
727.1373 -1.305 -0.30   59300.540   1.0 (4P)5s 3P   45551.767   1.0 3Psp3P x3D   8.30 -4.12 -7.47
727.2965 -1.418  ...    61198.480   5.0 (2G)5s 3G   47452.717   4.0 (2G)4p z1G   7.78 -3.84 -7.13
727.8439 -2.493  ...    47377.955   4.0 (4F)5s e5F  61113.380   4.0 (4F)7p 3F    8.11 -3.82 -7.11
727.9001 -2.232 -0.10   51143.920   0.0 4s6D4d e7F  37409.555   1.0 5Dsp1P y5P   8.57 -5.29 -7.55
727.9011 -2.268 -0.10   61340.460   4.0 (2G)5s 3G   47606.114   5.0 3Gsp3P v5F   8.05 -4.61 -7.49
727.9482 -2.360  ...    67716.750   6.0 (2G)4d 1I   53983.293   5.0 3Gsp3P t3G   8.37 -5.15 -7.51
729.5028 -1.424  ...    61935.470   4.0 (2G)5s 1G   48231.280   5.0 3Gsp3P y5H   7.84 -5.13 -7.54
730.7453 -0.590 -0.20   66293.980   5.0 (2H)5s 1H   52613.090   5.0 (2H)4p w3H   8.22 -5.22 -7.54
731.4092 -2.211  ...    45061.329   3.0 4s6D5s e5D  58729.800   4.0 (4F)6p 5D    8.16 -4.39 -7.23
732.5481 -1.272 -0.10   61340.460   4.0 (2G)5s 3G   47693.239   3.0 3Gsp3P w5G   7.96 -4.61 -7.49
734.6361 -1.420 -0.40   61198.480   5.0 (2G)5s 3G   47590.048   4.0 3Gsp3P w5G   7.91 -3.84 -7.13
735.5045 -1.697  ...    61198.480   5.0 (2G)5s 3G   47606.114   5.0 3Gsp3P v5F   7.98 -3.84 -7.13
736.1004 -2.435  ...    48928.388   2.0 (4F)5s e3F  62509.750   3.0 8p 3D3G3F    8.39 -3.79 -7.03
736.5220 -2.822  ...    61935.470   4.0 (2G)5s 1G   48361.882   4.0 3Gsp3P y5H   7.83 -5.13 -7.54
737.3194 -1.380 -0.40   58779.590   2.0 4s4D4d 3D   45220.681   3.0 3Psp3P x3D   8.30 -4.56 -7.43
737.6481  0.089 -0.20   61935.470   4.0 (2G)5s 1G   48382.603   5.0 (2G)4p z1H   8.11 -5.13 -7.54
738.9853 -1.442 +0.10   61340.460   4.0 (2G)5s 3G   47812.118   4.0 3Fsp3P x3G   7.97 -4.61 -7.49
739.1550 -2.524  ...    66293.980   5.0 (2H)5s 1H   52768.743   4.0 (2H)4p w3H   8.26 -5.22 -7.54
740.1947 -1.698 -0.30   61340.460   4.0 (2G)5s 3G   47834.221   3.0 3Fsp3P x3G   7.86 -4.61 -7.49
740.2127 -1.171 +0.10   61340.460   4.0 (2G)5s 3G   47834.550   5.0 3Fsp3P x3G   7.92 -4.61 -7.49
740.6598 -1.632 -0.10   58779.590   2.0 4s4D4d 3D   45281.833   2.0 3Psp3P x3D   8.31 -4.56 -7.43
741.2517 -1.475 -0.30   66293.980   5.0 (2H)5s 1H   52807.001   5.0 (4F)5p 3G    8.07 -4.76 -7.51
741.3059 -2.002 -0.60   61724.840   3.0 (2G)5s 3G   48238.827   2.0 3Gsp3P v5F   8.25 -4.91 -7.51
741.3536 -2.937  ...    48531.865   3.0 (4F)5s e3F  62016.990   3.0 (4F)7p 3G    8.27 -3.24 -7.09
743.2579 -2.664  ...    67687.990   5.0 (2G)4d 3I   54237.415   4.0 3Gsp3P t3G   8.33 -4.84 -7.51
744.9404 -2.602  ...    61724.840   3.0 (2G)5s 3G   48304.643   2.0 (4P)4p x3P   8.49 -4.91 -7.51
745.4811 -2.006  ...    61340.460   4.0 (2G)5s 3G   47929.997   4.0 3Gsp3P v5F   8.15 -4.61 -7.49
746.2930 -2.834  ...    59532.970   4.0 s6d 3+[4+]  46137.097   3.0 (4P)4p w5P   8.58 -3.91 -7.15
746.3178 -2.453  ...    58616.110   3.0 (4F)5d 5P   45220.681   3.0 3Psp3P x3D   8.21 -4.14 -7.38
```

```
746.5802 -1.412 -0.20   47960.940  4.0 (4F)5s e3F   61351.660  3.0 (4F)7p 3D   8.13 -4.31 -7.07
746.8072 -2.972  ...     67687.990  5.0 (2G)4d 3I    54301.340  4.0 3Dsp3P 5D   8.27 -4.84 -7.51
746.8233 -1.786  ...     61198.480  5.0 (2G)5s 3G    47812.118  4.0 3Fsp3P x3G  7.88 -3.84 -7.13
748.0769 -1.512 +0.20    61198.480  5.0 (2G)5s 3G    47834.550  5.0 3Gsp3P x3G  7.82 -3.84 -7.13
748.1313 -1.702 -0.30    61724.840  3.0 (2G)5s 3G    48361.882  4.0 3Gsp3P y5H  7.84 -4.91 -7.51
748.2865 -1.736 +0.20    60087.260  2.0 4s4D4d 3P    46727.074  2.0 3Psp3P y3P  8.16 -3.79 -7.18
748.3444 -2.170  ...     48928.388  2.0 (4F)5s e3F   62287.540  3.0 (4F)7p 5G   8.06 -4.05 -7.00
749.2314 -2.955  ...     61198.480  5.0 (2G)5s 3G    47855.143  6.0 3Gsp3P y5H  7.63 -3.84 -7.13
749.2915 -1.648 +0.20    60087.260  2.0 4s4D4d 3P    46744.993  3.0 (4P)4p u5D  8.26 -3.79 -7.18
749.7231 -2.272  ...     48531.865  3.0 (4F)5s e3F   61866.450  2.0 (4F)7p 3D   8.04 -3.45 -7.06
753.4582 -1.716 +0.00    61198.480  5.0 (2G)5s 3G    47929.997  4.0 3Gsp3P v5F  8.09 -3.84 -7.13
754.5574 -1.207  0.00    61724.840  3.0 (2G)5s 3G    48475.686  3.0 3Gsp3P y5H  7.84 -4.91 -7.51
755.4822 -1.035 +0.40    61935.470  4.0 (2G)5s 1G    48702.535  4.0 3Hsp3P y1G  8.05 -5.13 -7.54
755.7742 -2.909  ...     58779.590  2.0 4s4D4d 3D    45551.767  1.0 3Psp3P x3D  8.31 -4.56 -7.43
756.3626 -1.576 -0.30    61340.460  4.0 (2G)5s 3G    48122.928  3.0 3Gsp3P v5F  8.25 -4.61 -7.49
757.3414  0.302 -0.20    66293.980  5.0 (2H)5s 1H    53093.529  6.0 (2H)4p z1I  8.23 -5.22 -7.54
757.4394 -2.384  ...     60087.260  2.0 4s4D4d 3P    46888.517  2.0 (4P)4p u3D  8.24 -3.79 -7.18
758.2043 -1.579  ...     60087.260  2.0 4s4D4d 3P    46901.832  1.0 3Psp3P y3P  8.29 -3.79 -7.18
758.8844 -1.672 +0.00    54304.210  0.0 (4F)4d 5D    41130.599  1.0 5Dsp1P x5F  8.84 -4.57 -7.51
759.8262 -2.584  ...     59294.380  3.0 4s4D4d 3G    46137.097  3.0 (4P)4p w5P  8.73 -3.62 -7.37
760.1059 -2.862  ...     47960.940  4.0 (4F)5s e3F   61113.380  4.0 (4F)7p 3F   8.09 -3.82 -7.11
762.1925 -2.200  ...     48531.865  3.0 (4F)5s e3F   61648.300  4.0 (4F)7p 3G   8.16 -4.20 -7.12
762.6143 -1.229  ...     61340.460  4.0 (2G)5s 3G    48231.280  5.0 3Gsp3P y5H  7.85 -4.61 -7.49
762.7966 -1.971  ...     38602.260  3.0 (2F)4s 1F    51708.307  2.0 (2P)4p y1D  7.99 -5.22 -7.77
763.1378 -1.674  ...     61198.480  5.0 (2G)5s 3G    48098.293  6.0 3Hsp3P 1I   7.65 -3.84 -7.13
763.8133 -1.719  ...     48928.388  2.0 (4F)5s e3F   62016.990  3.0 (4F)7p 3G   8.27 -3.24 -7.09
764.8962 -1.971  ...     60087.260  2.0 4s4D4d 3P    47017.188  3.0 (4P)4p w3D  8.35 -3.79 -7.18
767.7019 -1.680  ...     61724.840  3.0 (2G)5s 3G    48702.535  4.0 3Hsp3P y1G  8.05 -4.91 -7.51
769.7887 -2.672  ...     59300.540  1.0 (4P)5s 3P    46313.537  2.0 (4P)4p w5P  8.66 -4.12 -7.47
770.1719 -1.514 -0.10    66293.980  5.0 (2H)5s 1H    53313.438  5.0 3Gsp3P y1H  7.87 -5.22 -7.54
770.2884 -1.269 -0.30    61340.460  4.0 (2G)5s 3G    48361.882  4.0 3Gsp3P y5H  7.84 -4.61 -7.49
770.9643 -1.500 -0.20    61198.480  5.0 (2G)5s 3G    48231.280  5.0 3Gsp3P y5H  7.73 -3.84 -7.13
771.0865 -1.968  ...     66293.980  5.0 (2H)5s 1H    53328.835  4.0 (4F)5p 3F   8.03 -4.80 -7.49
771.5202 -0.413 -0.10    61340.460  4.0 (2G)5s 3G    48382.603  5.0 (2G)4p z1H  8.12 -4.61 -7.49
772.7006 -2.944  ...     48928.388  2.0 (4F)5s e3F   61866.450  2.0 (4F)7p 3D   8.04 -3.45 -7.06
774.3787 -1.939  ...     60087.260  2.0 4s4D4d 3P    47177.234  1.0 (4P)4p u5D  8.31 -3.79 -7.18
775.5722 -2.420  ...     59300.540  1.0 (4P)5s 3P    46410.381  1.0 (4P)4p w5P  8.63 -4.12 -7.47
779.4169 -0.573 -1.40    61935.470  4.0 (2G)5s 1G    49108.896  4.0 (2G)4p w3F  8.13 -5.13 -7.54
779.8291 -2.719  ...     48531.865  3.0 (4F)5s e3F   61351.660  3.0 (4F)7p 3D   8.13 -4.31 -7.07
781.0940 -2.364  ...     47005.506  3.0 (4F)5s e5F   59804.540  4.0 4s6D7p 5F   7.87 -4.56 -7.07
781.7694 -2.536  ...     59532.970  4.0 s6d 3+[4+]   46744.993  3.0 (4P)4p u5D  8.21 -3.91 -7.15
787.2022 -2.124  ...     59300.540  1.0 (4P)5s 3P    46600.818  1.0 3Psp3P z3S  8.25 -4.12 -7.47
787.6449 -1.147 -0.30    61935.470  4.0 (2G)5s 1G    49242.886  3.0 (2G)4p w3F  8.15 -5.13 -7.54
788.5526 -2.121 -0.40    47377.955  4.0 (4F)5s e5F   60055.930  3.0 4s6D7p 5F   7.89 -4.07 -7.07
789.2001 -2.491  ...     60087.260  2.0 4s4D4d 3P    47419.687  2.0 3Psp3P 1D   8.00 -3.79 -7.18
790.7657 -2.065 -0.30    58779.590  2.0 4s4D4d 3D    46137.097  3.0 (4P)4p w5P  8.68 -4.56 -7.43
790.9394 -2.085  ...     59366.790  2.0 (4F)5d 5F    46727.074  2.0 3Psp3P y3P  8.06 -3.98 -7.25
791.0515 -0.702 -0.50    61340.460  4.0 (2G)5s 3G    48702.535  4.0 3Hsp3P y1G  8.05 -4.61 -7.49
791.6733 -1.655 +0.10    59300.540  1.0 (4P)5s 3P    46672.540  0.0 3Psp3P y3P  8.27 -4.12 -7.47
792.0623 -2.475  ...     59366.790  2.0 (4F)5d 5F    46744.993  3.0 (4P)4p u5D  8.18 -3.98 -7.25
793.4937 -2.508 -0.30    38602.260  3.0 (2F)4s 1F    51201.289  2.0 (2D)4p v3F  7.54 -6.14 -7.81
794.0743 -2.311  ...     61724.840  3.0 (2G)5s 3G    49135.023  3.0 3Psp3P v3D  7.98 -4.91 -7.51
794.0871 -2.775  ...     47005.506  5.0 (4F)5s e5F   59595.120  4.0 4s6D7p 7F   7.84 -4.11 -7.10
794.5983 -2.917  ...     48531.865  3.0 (4F)5s e3F   61113.380  4.0 (4F)7p 3F   8.09 -3.82 -7.11
795.1069 -1.243 +0.00    59300.540  1.0 (4P)5s 3P    46727.074  2.0 3Psp3P y3P  8.32 -4.12 -7.47
795.4967 -2.755  ...     59294.380  3.0 4s4D4d 3G    46727.074  2.0 3Psp3P y3P  8.43 -3.62 -7.37
797.7601 -2.890  ...     60087.260  2.0 4s4D4d 3P    47555.610  1.0 (4P)4p y3S  8.16 -3.79 -7.18
799.7624 -2.966  ...     59636.360  3.0 5d 5F5D3G    47136.084  2.0 (4P)4p w5D  8.30 -3.56 -7.25
800.0396 -1.194  ...     61198.480  5.0 (2G)5s 3G    48702.535  4.0 3Hsp3P y1G  7.98 -3.84 -7.13
800.3490 -1.787  ...     47005.506  3.0 (4F)5s e5F   59496.620  4.0 4s6D7p 5D   7.94 -3.41 -7.08
800.9159 -2.075  ...     47755.537  3.0 (4F)5s e5F   60237.810  2.0 4s6D7p 5D   7.98 -3.90 -7.06
800.9193 -1.246 -0.90    61724.840  3.0 (2G)5s 3G    49242.886  3.0 3Psp3P v3D  7.99 -4.91 -7.51
800.9363 -1.008 -0.40    61724.840  3.0 (2G)5s 3G    49242.886  3.0 (2G)4p w3F  8.15 -4.91 -7.51
801.1251 -1.141 -0.80    58616.110  3.0 (4F)5d 5P    46137.097  3.0 (4P)4p w5P  8.64 -4.14 -7.38
801.4106 -2.059  ...     61935.470  4.0 (2G)5s 1G    49460.902  5.0 (2G)4p v3G  8.20 -5.13 -7.54
801.9580 -1.667 -0.30    58779.590  2.0 4s4D4d 3D    46137.097  3.0 (4P)4p w5P  8.67 -4.56 -7.43
802.4543 -0.280 -0.20    61935.470  4.0 (2G)5s 1G    49477.127  3.0 3Fsp3P 1F   7.94 -5.13 -7.54
803.5170 -1.416  ...     66293.980  5.0 (2H)5s 1H    53852.114  4.0 (4F)5p 5G   8.13 -4.29 -7.54
```

```
804.5051 -2.625  ...   47377.955  4.0 (4F)5s e5F   59804.540  4.0 4s6D7p 5F   7.87 -4.56 -7.07
805.1329 -2.422  ...   47377.955  4.0 (4F)5s e5F   59794.850  3.0 3Dsp3P 1F   8.13 -4.18 -7.25
805.4489 -1.028 -0.20  59300.540  1.0 (4P)5s 3P    46888.517  2.0 (4P)4p u3D  8.38 -4.12 -7.47
806.3139 -1.833 +0.00  59300.540  1.0 (4P)5s 3P    46901.832  1.0 3Psp3P y3P  8.41 -4.12 -7.47
808.2369 -1.155 -0.50  58779.590  2.0 4s4D4d 3D    46410.381  1.0 (4P)4p w5P  8.64 -4.56 -7.43
810.7385 -0.296 -0.60  61935.470  4.0 (2G)5s 1G    49604.427  5.0 (2G)4p y3H  8.20 -5.13 -7.54
811.8030 -2.102  ...   58628.410  1.0 4s4D4d 5P    46313.537  2.0 (4P)4p w5P  8.65 -5.24 -7.33
811.9691 -2.252  ...   47005.506  5.0 (4F)5s e5F   59317.860  4.0 4s6D7p 7D   7.89 -4.13 -7.11
812.0791 -2.108  ...   66293.980  5.0 (2H)5s 1H    53983.293  5.0 3Gsp3P t3G  8.30 -5.22 -7.54
812.2837 -0.684 -0.10  61935.470  4.0 (2G)5s 1G    49627.884  4.0 (2G)4p v3G  8.20 -5.13 -7.54
812.6147 -1.776  ...   58616.110  3.0 (4F)5d 5P    46313.537  2.0 (4P)4p w5P  8.63 -4.14 -7.38
812.8185 -2.290  ...   48036.673  2.0 (4F)5s e5F   60336.160  1.0 4s6D7p 5F   7.95 -4.13 -7.07
813.3329 -0.428 +0.00  61724.840  3.0 (2G)5s 3G    49433.131  2.0 (2G)4p w3F  8.15 -4.91 -7.51
813.6492 -2.659  ...   67716.750  6.0 (2G)4d 1I    55429.819  5.0 1Gsp3P 3G   8.30 -5.15 -7.51
814.3472 -1.801  ...   66293.980  5.0 (2H)5s 1H    54017.581  4.0 (4F)5p 3G   8.09 -4.72 -7.49
816.6367 -0.647 +0.30  67687.990  5.0 (2G)4d 3I    55446.008  4.0 1Gsp3P 3H   8.22 -4.84 -7.51
817.6218 -2.221  ...   48928.388  2.0 (4F)5s e3F   61155.620  1.0 3Psp1P 3P   8.58 -4.00 -7.19
818.2377 -1.645 +0.00  58628.410  1.0 4s4D4d 5P    46410.381  1.0 (4P)4p w5P  8.62 -5.24 -7.33
818.2955 -2.850  ...   47377.955  4.0 (4F)5s e5F   59595.120  4.0 4s6D7p 7F   7.84 -4.11 -7.10
818.8777 -1.742  ...   61935.470  4.0 (2G)5s 1G    49726.990  4.0 (2G)4p y3H  8.20 -5.13 -7.54
819.0818 -1.286 +0.40  61340.460  4.0 (2G)5s 3G    49135.023  3.0 3Fsp3P v3D  7.98 -4.61 -7.49
820.0392 -0.888  ...   67716.750  6.0 (2G)4d 1I    55525.562  5.0 1Gsp3P 3H   8.17 -5.15 -7.51
821.8412 -1.021 -0.10  59300.540  1.0 (4P)5s 3P    47136.084  2.0 (4P)4p w5D  8.45 -4.12 -7.47
821.9783 -2.726  ...   67687.990  5.0 (2G)4d 3I    55525.562  5.0 1Gsp3P 3H   8.18 -4.84 -7.51
823.0377 -1.570  ...   48928.388  2.0 (4F)5s e3F   61075.160  1.0 3Fsp1P 3D   8.35 -2.91 -7.06
823.9953 -2.147  ...   48036.673  2.0 (4F)5s e5F   60169.330  1.0 4s4F7p 5D   7.99 -3.76 -7.07
824.4854 -1.823 -0.20  47377.955  4.0 (4F)5s e5F   59503.400  3.0 6p 5G5D3D   8.31 -3.43 -7.17
824.8317 -1.987  ...   47755.537  3.0 (4F)5s e5F   59875.890  3.0 4s6D7p 5D   8.04 -3.29 -7.13
824.9466 -1.895  ...   47377.955  4.0 (4F)5s e5F   59496.620  4.0 4s6D7p 5D   7.94 -3.41 -7.08
825.2074 -2.695  ...   48221.324  1.0 (4F)5s e5F   60336.160  1.0 4s6D7p 5F   7.95 -4.13 -7.07
825.9782 -2.634  ...   59300.540  1.0 (4P)5s 3P    47197.010  2.0 (3F)sp x3F  8.22 -4.12 -7.47
826.3849 -0.453 -0.40  61340.460  4.0 (2G)5s 3G    49242.886  3.0 (2G)4p w3F  8.15 -4.61 -7.49
826.3988 -2.918  ...   59294.380  3.0 4s4D4d 3G    47197.010  2.0 (3F)sp x3F  8.36 -3.62 -7.37
826.4271 -2.786  ...   61724.840  3.0 (2G)5s 3G    49627.884  4.0 (2G)4p v3G  8.21 -4.91 -7.51
826.5614 -2.590  ...   47960.940  4.0 (4F)5s e3F   60055.930  3.0 4s6D7p 5F   7.86 -4.07 -7.07
826.9310 -0.589 -0.30  61198.480  5.0 (2G)5s 3G    49108.896  4.0 (2G)4p w3F  8.08 -3.84 -7.13
827.2529 -1.624  ...   61935.470  4.0 (2G)5s 1G    49850.590  3.0 (2G)4p v3G  8.22 -5.13 -7.54
829.1957 -1.139  ...   66293.980  5.0 (2H)5s 1H    54237.415  4.0 3Gsp3P t3G  8.23 -5.22 -7.54
829.4255 -2.901  ...   48531.865  3.0 (4F)5s e3F   60585.090  2.0 3Psp1P 3P   8.64 -4.76 -7.25
829.4743 -1.572  ...   58779.590  2.0 4s4D4d 3D    46727.074  2.0 3Psp3P y3P  8.33 -4.56 -7.43
829.4871 -2.242  ...   47960.940  4.0 (4F)5s e3F   60013.270  3.0 (4F)6p 3G   8.19 -3.41 -7.26
829.7161 -2.339  ...   47755.537  3.0 (4F)5s e5F   59804.540  4.0 4s6D7p 5F   7.88 -4.56 -7.07
830.2762 -2.913  ...   47377.955  4.0 (4F)5s e5F   59418.830  3.0 (4F)6p 5F   8.00 -4.10 -7.19
830.3839 -2.300  ...   47755.537  3.0 (4F)5s e5F   59794.850  3.0 3Dsp3P 1F   8.13 -4.18 -7.25
830.7093 -2.502 +0.00  58779.590  2.0 4s4D4d 3D    46744.993  3.0 (4P)4p u5D  8.40 -4.56 -7.43
831.1295 -1.091 -0.30  59300.540  1.0 (4P)5s 3P    47272.027  1.0 (4P)4p w3D  8.37 -4.12 -7.47
831.7696 -2.747  ...   48036.673  2.0 (4F)5s e5F   60055.930  3.0 4s6D7p 5F   7.90 -4.07 -7.07
831.8712 -2.254  ...   58428.170  0.0 4s6D4d 5D    46410.381  1.0 (4P)4p w5P  8.65 -5.07 -7.40
832.2913 -2.053 -0.20  38602.260  3.0 (2F)4s 1F    50613.983  4.0 3Fsp3P x1G  7.61 -6.17 -7.80
833.1498 -2.886  ...   47377.955  4.0 (4F)5s e5F   59377.300  4.0 (4F)6p 5G   8.11 -3.96 -7.22
833.2537 -0.059 -0.60  61724.840  3.0 (2G)5s 3G    49726.990  4.0 (2G)4p y3H  8.21 -4.91 -7.51
833.6156 -1.275  ...   66293.980  5.0 (2H)5s 1H    54301.340  4.0 3Dsp3P 5D   8.16 -5.22 -7.54
834.1736 -2.533 -0.10  38602.260  3.0 (2F)4s 1F    50586.878  3.0 (2G)4p z1F  7.78 -6.14 -7.80
834.7323 -1.959  ...   48036.673  2.0 (4F)5s e5F   60013.270  3.0 (4F)6p 3G   8.21 -3.41 -7.26
834.7911 -1.908 +0.10  47755.537  3.0 (4F)5s e5F   59731.290  4.0 (4F)6p 3G   8.23 -3.71 -7.26
836.7298 -1.783  ...   48221.324  1.0 (4F)5s e5F   60169.330  1.0 4s4F7p 5D   7.99 -3.76 -7.07
836.7930 -2.691  ...   59366.790  2.0 (4F)5d 5F    47419.687  2.0 3Psp3P 1D   7.84 -3.98 -7.25
837.2975 -1.656  ...   47377.955  4.0 (4F)5s e5F   59317.860  4.0 4s6D7p 7D   7.89 -4.13 -7.11
838.4274 -2.226  ...   60087.260  2.0 4s4D4d 3P    48163.446  2.0 s2p v5P   8.48 -3.79 -7.18
839.0512 -1.973  ...   47960.940  4.0 (4F)5s e3F   59875.890  3.0 4s6D7p 5D   8.01 -3.29 -7.13
840.0110 -2.661  ...   58628.410  1.0 4s4D4d 5P    46727.074  2.0 3Psp3P y3P  8.30 -5.24 -7.33
840.4395 -0.705 -0.20  58616.110  3.0 (4F)5d 5P    46720.842  4.0 (4P)4p u5D  8.43 -4.14 -7.38
840.8800 -1.952  ...   58616.110  3.0 (4F)5d 5P    46727.074  2.0 3Psp3P y3P  8.25 -4.14 -7.38
841.4592 -1.479 -0.20  59300.540  1.0 (4P)5s 3P    47419.687  2.0 3Psp3P 1D   8.22 -4.12 -7.47
841.5509 -1.438  ...   61340.460  4.0 (2G)5s 3G    49460.902  5.0 (2G)4p v3G  8.21 -4.61 -7.49
841.6784 -1.208 -0.30  58779.590  2.0 4s4D4d 3D    46901.832  1.0 3Psp3P y3P  8.42 -4.56 -7.43
841.9271 -0.231 -0.20  61724.840  3.0 (2G)5s 3G    49850.590  3.0 (2G)4p v3G  8.22 -4.91 -7.51
842.1493 -1.846 +0.30  58616.110  3.0 (4F)5d 5P    46744.993  3.0 (4P)4p u5D  8.33 -4.14 -7.38
```

```
842.7019 -0.996 -0.30   61340.460  4.0 (2G)5s 3G    49477.127  3.0 3Fsp3P 1F    7.95 -4.61 -7.49
843.7630 -2.875  ...    60087.260  2.0 4s4D4d 3P    48238.847  2.0 3Gsp3P v5F   8.27 -3.79 -7.18
844.3923 -2.913  ...    47755.537  3.0 (4F)5s e5F   59595.120  4.0 4s6D7p 7F    7.85 -4.11 -7.10
844.4184 -2.271  ...    48036.673  2.0 (4F)5s e5F   59875.890  3.0 4s6D7p 5D    8.04 -3.29 -7.13
844.7971 -1.949 -0.30   47960.940  4.0 (4F)5s e3F   59794.850  3.0 3Dsp3P 1F    8.11 -4.18 -7.25
846.4229 -2.194  ...    59366.790  2.0 (4F)5d 5F    47555.610  1.0 (4P)4p y3S   8.06 -3.98 -7.25
846.5381 -0.758  ...    67716.750  6.0 (2G)4d 1I    55907.178  5.0 (2H)4p 1H    8.45 -5.15 -7.51
847.4123 -2.702  ...    60087.260  2.0 4s4D4d 3P    48289.871  1.0 s2p v5P      8.53 -3.79 -7.18
848.4865 -2.428  ...    67687.990  5.0 (2G)4d 3I    55905.536  4.0 1Gsp3P 3G    8.33 -4.84 -7.51
848.6048 -2.493  ...    67687.990  5.0 (2G)4d 3I    55907.178  5.0 (2H)4p 1H    8.45 -4.84 -7.51
849.7946 -0.165  ...    61198.480  3.0 (2G)5s 3G    49434.163  6.0 (2G)4p y3H   8.13 -3.84 -7.13
849.9330 -0.693 -0.20   58779.590  2.0 4s4D4d 3D    47017.188  3.0 (4P)4p w3D   8.47 -4.56 -7.43
850.9848 -1.789 -0.40   47755.537  3.0 (4F)5s e5F   59503.400  3.0 6p 5G5D3D    8.31 -3.43 -7.17
851.1973 -1.072  ...    59300.540  1.0 (4P)5s 3P    47555.610  1.0 (4P)4p y3S   8.32 -4.12 -7.47
851.4762 -1.960  ...    47755.537  3.0 (4F)5s e5F   59496.620  4.0 4s6D7p 5D    7.95 -3.41 -7.08
851.5626 -2.316 +0.00   58628.410  1.0 4s4D4d 5P    46888.517  2.0 (4P)4p u3D   8.36 -5.24 -7.33
851.7305 -0.259 -0.40   61198.480  5.0 (2G)5s 3G    49460.902  5.0 (2G)4p v3G   8.16 -3.84 -7.13
851.7976 -2.729  ...    60087.260  2.0 4s4D4d 3P    48350.606  1.0 3Gsp3P v5F   8.28 -3.79 -7.18
851.8426 -0.245 -0.30   61340.460  4.0 (2G)5s 3G    49604.427  5.0 (2G)4p y3H   8.21 -4.61 -7.49
852.5011 -1.523 +0.00   58616.110  3.0 (4F)5d 5P    46889.142  4.0 3Fsp3P x3F   8.21 -4.14 -7.38
852.5295 -2.293  ...    58628.410  1.0 4s4D4d 5P    46901.832  1.0 3Psp3P y3P   8.40 -5.24 -7.33
852.6956 -1.029 -0.60   47005.506  5.0 (4F)5s e5F   58729.800  4.0 (4F)6p 5D    8.04 -4.39 -7.23
853.5486 -0.310 -0.90   61340.460  4.0 (2G)5s 3G    49627.884  4.0 (2G)4p v3G   8.21 -4.61 -7.49
854.0322 -2.454  ...    48531.865  3.0 (4F)5s e3F   60237.810  2.0 4s6D7p 5D    7.95 -3.90 -7.06
855.4255 -2.734  ...    58779.590  2.0 4s4D4d 3D    47092.712  3.0 3Fsp3P x3F   8.26 -4.56 -7.43
856.9287 -1.769 +0.00   48036.673  2.0 (4F)5s e5F   59703.050  1.0 (4F)6p 5D    7.94 -3.86 -7.16
857.1553 -1.480 +0.10   47755.537  3.0 (4F)5s e5F   59418.830  3.0 (4F)6p 5F    8.01 -4.10 -7.19
857.6400 -2.716  ...    48928.388  2.0 (4F)5s e3F   60585.090  2.0 3Psp1P 3P    8.64 -4.76 -7.25
858.6120 -1.188 -0.30   58779.590  2.0 4s4D4d 3D    47136.084  2.0 (4P)4p w5D   8.45 -4.56 -7.43
858.6231 -0.737 -0.20   47377.955  4.0 (4F)5s e5F   59021.310  5.0 (4F)6p 5G    8.20 -4.37 -7.26
858.8893 -2.391  ...    61198.480  3.0 (2G)5s 3G    49558.734  4.0 4s6D5p n7D   7.76 -3.84 -7.13
860.2183 -0.974 -0.20   47755.537  3.0 (4F)5s e5F   59377.300  5.0 (4F)6p 5G    8.11 -3.96 -7.22
861.5311 -0.853 -0.10   47005.506  5.0 (4F)5s e5F   58609.560  5.0 (4F)6p 5F    7.89 -3.24 -7.23
861.6114 -2.675  ...    59532.970  4.0 s6d 3+[4+]   47929.997  4.0 3Gsp3P v5F   8.11 -3.91 -7.15
861.6572 -2.696  ...    58779.590  2.0 4s4D4d 3D    47177.234  1.0 (4P)4p w3D   8.44 -4.56 -7.43
861.9123 -1.330  ...    58616.110  3.0 (4F)5d 5P    47017.188  3.0 (4P)4p w3D   8.41 -4.14 -7.38
862.2743 -1.367 +0.60   61198.480  5.0 (2G)5s 3G    49604.427  5.0 (2G)4p y3H   8.16 -3.84 -7.13
864.0224 -1.386 -0.20   61198.480  5.0 (2G)5s 3G    49627.884  4.0 (2G)4p v3G   8.16 -3.84 -7.13
864.3369 -2.065  ...    59532.970  4.0 s6d 3+[4+]   47966.585  3.0 s2p v5P     8.65 -3.91 -7.15
864.6406 -1.863 +0.50   47755.537  3.0 (4F)5s e5F   59317.860  4.0 4s6D7p 7D    7.90 -4.13 -7.11
866.1285 -1.241  ...    47960.940  4.0 (4F)5s e3F   59503.400  3.0 6p 5G5D3D    8.29 -3.43 -7.17
866.6376 -2.391  ...    47960.940  4.0 (4F)5s e3F   59496.620  4.0 4s6D7p 5D    7.91 -3.41 -7.08
867.5613 -2.864  ...    58616.110  3.0 (4F)5d 5P    47092.712  3.0 3Fsp3P x3F   8.15 -4.14 -7.38
868.7551 -2.488  ...    58779.590  2.0 4s4D4d 3D    47272.027  1.0 (4P)4p w3D   8.39 -4.56 -7.43
869.8174 -2.866  ...    64300.510  4.0 4s3H5s 5H    52807.001  5.0 (4F)5p 3G    8.32 -4.39 -7.39
869.9069 -2.440  ...    58628.410  1.0 4s4D4d 5P    47136.084  2.0 (4P)4p w5D   8.43 -5.24 -7.33
870.0929 -1.281  ...    61340.460  4.0 (2G)5s 3G    49850.590  3.0 (2G)4p v3G   8.22 -4.61 -7.49
870.6040 -0.606 +0.00   66293.980  5.0 (2H)5s 1H    54810.856  4.0 3Gsp3P w1G   8.11 -5.22 -7.54
870.7344 -2.903  ...    48531.865  3.0 (4F)5s e3F   60013.270  3.0 (4F)6p 3G    8.19 -3.41 -7.26
870.8390 -2.558  ...    58616.110  3.0 (4F)5d 5P    47136.084  2.0 (4P)4p w5D   8.39 -4.14 -7.38
871.4870 -2.463  ...    61198.480  5.0 (2G)5s 3G    49726.990  4.0 (2G)4p y3H   8.16 -3.84 -7.13
871.8490 -1.971  ...    48036.673  2.0 (4F)5s e5F   59503.400  3.0 6p 5G5D3D    8.31 -3.43 -7.17
872.5214 -1.776 +0.60   47960.940  4.0 (4F)5s e3F   59418.830  3.0 (4F)6p 5F    7.98 -4.10 -7.19
872.5984 -2.794  ...    58628.410  1.0 4s4D4d 5P    47171.531  0.0 (4P)4p u5D   8.50 -5.24 -7.33
873.0330 -2.619  ...    58628.410  1.0 4s4D4d 5P    47177.234  1.0 (4P)4p u5D   8.41 -5.24 -7.33
873.3315 -1.464 -0.10   48928.388  2.0 (4F)5s e3F   60375.650  1.0 (2F)4p 3D    8.59 -4.33 -7.37
875.6954 -2.685  ...    47960.940  4.0 (4F)5s e3F   59377.300  4.0 (4F)6p 5G    8.09 -3.96 -7.22
877.1435 -2.663  ...    59636.360  3.0 5d 5F5D3G    48238.847  2.0 3Gsp3P v5F   8.23 -3.56 -7.25
877.2530 -0.599 -0.30   47960.940  4.0 (4F)5s e3F   59357.030  5.0 (4F)6p 3G    8.37 -4.27 -7.28
877.7568 -1.909  ...    51294.220  3.0 4s4D5s e3D   62683.770  4.0 8p 3G5G5F    8.54 -3.24 -7.02
880.2787 -2.658  ...    47960.940  4.0 (4F)5s e3F   59317.860  4.0 4s6D7p 7D    7.86 -4.13 -7.11
880.9247 -0.494 -0.30   61935.470  3.0 (2G)5s 1G    50586.878  3.0 (2G)4p z1F   8.08 -5.13 -7.54
881.2793 -2.190  ...    48531.865  3.0 (4F)5s e3F   59875.890  3.0 4s6D7p 5D    8.01 -3.29 -7.13
882.0520 -1.064 -0.40   67716.750  6.0 (2G)4d 1I    56382.662  5.0 1Isp3P u3H   8.43 -5.15 -7.51
882.5420 -2.666  ...    59294.380  3.0 4s4D4d 3G    47966.585  3.0 s2p v5P      8.77 -3.62 -7.37
883.0337 -1.021 -0.60   61935.470  3.0 (2G)5s 1G    50613.983  4.0 3Fsp3P x1G   7.99 -5.13 -7.54
883.9758 -2.949  ...    48928.388  2.0 (4F)5s e3F   60237.810  2.0 4s6D7p 5D    7.95 -3.90 -7.06
886.8574 -2.300  ...    48531.865  3.0 (4F)5s e3F   59804.540  4.0 4s6D7p 5F    7.83 -4.56 -7.07
```

```
887.4847 -0.877 -0.20   67687.990  5.0 (2G)4d 3I    56423.283  4.0 1Isp3P u3H   8.44 -4.84 -7.51
887.6204 -2.267  ...     48531.865  3.0 (4F)5s e3F   59794.850  3.0 3Dsp3P 1F    8.10 -4.18 -7.25
890.6092 -2.836  ...     54304.210  0.0 (4F)4d 5D    43079.023  1.0 3Psp3P x5P   8.04 -4.57 -7.51
890.7050 -2.492  ...     58779.590  2.0 4s4D4d 3D    47555.610  1.0 (4P)4p y3S   8.33 -4.56 -7.43
890.8542 -2.922  ...     51461.670  4.0 4s6D4d f5F   62683.770  4.0 8p 3G5G5F    8.48 -3.24 -7.02
892.6579 -0.897 -0.10   48531.865  3.0 (4F)5s e3F   59731.290  4.0 (4F)6p 3G    8.21 -3.71 -7.26
897.5839 -2.351  ...     61724.840  3.0 (2G)5s 3G    50586.878  3.0 (2G)4p z1F   8.08 -4.91 -7.51
897.6539 -1.360  ...     59300.540  1.0 (4P)5s 3P    48163.446  2.0 s2p v5P     8.57 -4.12 -7.47
898.1506 -2.549  ...     59294.380  3.0 4s4D4d 3G    48163.446  2.0 s2p v5P     8.63 -3.62 -7.37
```